\documentclass[twocolumn, twocolappendix]{aastex631}

\usepackage{graphicx}	
\usepackage{amsmath}	
\usepackage{soul}
\usepackage{enumerate}
\usepackage{bbm}
\usepackage[table]{xcolor}
\usepackage{multirow}
\usepackage{hhline}

\newcommand{\dd}{\mbox{$\>\mathrm{d}$}}
\newcommand{\Msun}{\>{\rm M_{\odot}}}
\newcommand{\kpc}{\>{\rm kpc}}
\newcommand{\kms}{\,\mathrm{km/s}}

\definecolor{ForestGreen}{RGB}{34,139,34}

\defcitealias{Watkins2010}{WEA10}
\defcitealias{Erkal2020}{EBP20}
\defcitealias{Garavito-Camargo2019}{GC19}

\begin{document}
\title[The meaning of Milky Way mass estimates]{The meaning of Milky Way mass estimates in the presence of the LMC}

\author[0000-0002-0740-1507]{Leandro {Beraldo e Silva}}
\affiliation{Observatório Nacional, Rio de Janeiro - RJ, 20921-400,
  Brasil}
\affiliation{Department of Astronomy and Steward Observatory,
  University of Arizona, 933 North Cherry Ave., Tucson, AZ 85721, USA}
\affiliation{Department of Astronomy and Astrophysics, University of Michigan, Ann Arbor, MI, USA}
\author[0000-0003-0715-2173]{Gurtina Besla}
\affiliation{Department of Astronomy and Steward Observatory, University of Arizona, 933 North Cherry Ave., Tucson, AZ 85721, USA}
\author[0000-0001-7107-1744]{Nicol\'as Garavito-Camargo}\thanks{NASA NFHP Einstein Fellow}
\affiliation{Department of Astronomy, University of Maryland, College Park, MD 20742, USA}
\affiliation{Steward Observatory, University of Arizona, 933 North Cherry Avenue, Tucson, AZ 85721, USA}
\affiliation{Instituto de Astrof\'isica, Pontificia Universidad Cat\'olica de Chile, Av. Vicu\~na Mackenna 4860, 782-0436 Macul, Santiago, Chile}

\author[0000-0002-6257-2341]{Monica Valluri}
\affiliation{Department of Astronomy and Astrophysics, University of Michigan, Ann Arbor, MI, USA}
\author[0000-0003-2594-8052]{Kathryne J. Daniel}
\affiliation{Department of Astronomy and Steward Observatory, University of Arizona, 933 North Cherry Ave., Tucson, AZ 85721, USA}

\author[0000-0002-9820-1219]{Ekta~Patel}
\affiliation{Department of Astrophysics and Planetary Sciences, Villanova University,  800 E. Lancaster Ave, Villanova, PA 19085, USA}

\author[0000-0003-2644-135X]{Sergey E. Koposov}
\affiliation{Institute for Astronomy, University of Edinburgh, Royal Observatory, Blackford Hill, Edinburgh EH9 3HJ, UK}
\affiliation{Institute of Astronomy, University of Cambridge, Madingley Road, Cambridge CB3 0HA, UK}
\author[0000-0002-1343-134X]{Laura L. Watkins}
\affiliation{AURA for the European Space Agency (ESA), Space Telescope Science Institute, 3700 San Martin Drive, Baltimore, MD 21218, USA}
\author[0000-0002-5689-8791]{Amanda Bystr\"om}
\affil{Institute for Astronomy, University of Edinburgh, Royal Observatory, Blackford Hill, Edinburgh EH9 3HJ, UK}
\author[0000-0001-6146-2645]{Alis J. Deason}
\affiliation{Institute for Computational Cosmology, Department of Physics, University of Durham, South Road, Durham DH1 3LE, UK}
\author[0000-0003-3922-7336]{Chervin F. P. Laporte}
\affiliation{LIRA, Observatoire de Paris, Universit\'e PSL, Sorbonne Universit\'e, Universit\'e Paris Cit\'e, CY Cergy Paris Universit\'e, CNRS, 92190 Meudon, France}
\affiliation{Institut de Ci\`{e}ncies del Cosmos (ICCUB), Universitat de Barcelona, Mart\'{i} i Franqu\`{e}s 1, E-08028 Barcelona, Spain}
\affiliation{Kavli IPMU (WPI), UTIAS, The University of Tokyo, Kashiwa, Chiba 277-8583, Japan}
\correspondingauthor{Leandro {Beraldo e Silva}}
\email{lberaldoesilva@on.br}

\begin{abstract}
The LMC's recent arrival perturbs the observed kinematics of the Milky Way's (MW) stellar halo, complicating equilibrium-based MW mass estimates at Galactocentric distances $r\gtrsim 30$ kpc. Using N-body MW+LMC simulations, we show that the MW stellar halo can be used to disentangle the total MW+LMC, current LMC, and {\it pre-infall} MW mass profiles, by exploiting the LMC's orbital path and the LMC-induced motion of the MW's center of mass (``reflex motion"). Tracers in regions crossed by the LMC longest ago (Galactic coordinates $\ell >0^\circ, b>0^\circ$) are approximately relaxed and trace the total MW+LMC mass. Since the reflex motion is primarily latitudinal, a radial-velocity-only estimator ($M_{v_r}$) recovers the \emph{total} MW+LMC mass with $\lesssim 5\%$ errors across $30<r/\kpc<200$, without any reflex-motion correction. Subtracting mass estimates in the region of the LMC's future orbit ($\ell < 0^\circ, b >0^\circ$) from those in its recent-orbit region ($\ell > 0^\circ, b <0^\circ$), which maximizes the LMC-induced velocity dipole, recovers the LMC's current mass profile with $10-20\%$ errors, but this requires tracers at $r\gtrsim 100-120$ kpc. Subtracting this from the total MW+LMC mass recovers the MW's {\it pre-LMC infall} mass at $r\gtrsim 80$ kpc with $5-10$\% errors. 
Applying $M_{v_r}$ to DESI-MWS DR1 BHBs at $\ell >0^\circ, b>0^\circ$ and $50<r/\kpc<60$, we obtain a \emph{total} MW+LMC mass $M(r<60\,\kpc) \approx 7.5^{+3.0}_{-2.2}\times 10^{11}\,\Msun$. Observationally, recovering the LMC and pre-infall MW mass profiles requires 6D samples at $r\gtrsim100$\,kpc, and a virial mass to within $\sim10-20\%$ requires $r\gtrsim120$\,kpc, both achievable with \textit{Gaia} DR5 astrometry for halo giants targeted by DESI Run~2 and 4MOST. Rather than treating the LMC-induced disequilibrium as an obstacle, our framework exploits it to weigh both galaxies.
\end{abstract}

\keywords{}

\section{Introduction}
\label{sec:intro}

The present day virial mass of the Milky Way (MW) is a critical benchmark in the calibration of cosmological scaling relations and the characterization of dark matter \citep[e.g.][]{Bland-Hawthorn2016, Cooper2010, Boylan-Kolchin2010, McGaugh2025}. In the current era of high-precision astrometry provided by \textit{HST} \citep[e.g.][]{HST2014, Sohn2026} and \textit{Gaia} \citep{Gaia_2016} in tandem with large spectroscopic surveys, the spatial distribution and kinematics of the MW's stellar halo are becoming increasingly well-constrained, even at large distances. Mass estimators that can leverage these data, such as those that use the Jeans equations \citep[e.g.][hereafter WEA10]{Watkins2010}, are critical tools to constrain the MW's mass profile. The goal of this study is to investigate the physical meaning of such MW mass estimates in the presence of its largest satellite, the LMC.

Despite great progress, virial mass estimates for the MW range from ${\approx 0.5-2\times 10^{12}\Msun}$ \citep[e.g., ][and references therein]{Callingham2019, Watkins2019, Wang2020, Vasiliev2021, Kravtsov2024, Strigari2025, Hayati2025, DESI_DR2KP_2026}. This range is sufficiently broad that the properties of the MW are consistent with either cold dark matter (CDM) or warm \citep[e.g.,][]{Bode2001, Lovell2014}, self-interacting \citep[e.g.,][]{Spergel2000, Tulin2018} or Fuzzy dark matter (DM) \citep[e.g.,][]{Hu2000, Hui_2021}. The magnitude of possible tensions with CDM scales with the MW's virial mass \citep{Bullock2017}. An accurate estimate of the MW's virial mass is thus critical to testing the CDM paradigm. 

Central to the vast majority of mass estimator methods is the assumption of a tracer population in dynamical equilibrium \citep[see e.g.][and \cite{Hayati2024} for related discussions]{McMillan2013, Hattori2021, Green2023, BeS2025, Li2025}. However, it is now clear that the observed kinematics of the MW's outer stellar halo ($r\gtrsim 30\kpc$) are perturbed by the recent infall of a massive LMC \citep{Garavito-Camargo2019, Erkal2019, Erkal2020, Petersen2021, Yaaqib2024, Bystrom2025, Chandra2025}. 

The LMC is likely just past its first approach towards the MW, entering within its virial radius 1-2 Gyr ago \citep[][]{Besla2007, Kallivayalil2013} -- for a recent review see \cite{Besla2026}. In this scenario, the LMC's cosmologically-expected infall mass is $\sim 1-3 \times 10^{11}$ M$_\odot$ \citep[$\sim$ 10-30\% that of the MW;][]{Behroozi2010, Guo2010, Moster2010, Moster2013}. In fact, several lines of evidence suggest an LMC mass in the range $\approx 1-2\times 10^{11}\Msun$ \citep[][]{Penarrubia2016, Erkal2019, Shipp2021, Watkins2024}, although these estimates still have significant uncertainties. The dynamical impact of such a massive LMC on the MW's halo is broadly described through two effects: 1) a local dynamical friction wake of MW stars and DM particles trailing the LMC's orbit \citep[][]{Garavito-Camargo2019, Belokurov2019, Conroy2021, Cavieres2025}; and 2) the global motion of the MW disc and inner stellar and DM halos in response to the gravitational pull of the LMC \citep[][]{Weinberg_1989, Gomez2015, Garavito-Camargo2019}. 

The LMC is on a nearly polar orbit about the MW, where the LMC is currently ``below'' the MW disc and is causing the MW disc to move ``downwards''. As such, the second effect produces a net ``upward'' motion of outer halo stars in the Galactocentric frame, which manifests as a global dipole in the radial velocities of stars for ${r\gtrsim 30\, \kpc}$ \citep[][]{Garavito-Camargo2019, Erkal2019, Cunningham2020, Erkal2021, Petersen2021, Yaaqib2024, Bystrom2025, Chandra2025, Brooks2026}. The observable consequence is generally referred to as a ``reflex" motion \citep{Erkal2019}, where the apparent motions of stars in the outer halo, as seen by an observer on Earth, are the reflex of our motion relative to them. 

The LMC-induced perturbations to the MW's DM and stellar halos vary as a function of time, radius and angle on the sky, in accordance with the LMC's orbit \citep{Garavito-Camargo2019, Erkal2020, Garavito-Camargo2021}. These perturbations drive the outer stellar halo out of equilibrium, complicating the inference of the MW mass distribution with equilibrium-based methods. The goal of this study is to understand how this disequilibrium manifests in the \citetalias{Watkins2010} mass estimators and whether the non-uniform nature of the LMC-induced perturbations can be exploited to uniquely constrain the present day combined MW+LMC mass, the MW mass pre-LMC infall, and the present day mass of the LMC.

The LMC's cosmologically-expected high infall mass ($\sim 10^{11}$ M$_\odot$) implies that the MW is still in the assembly process. Late accretion events of LMC-mass systems are relatively rare for MW-mass halos ($M_{vir} \sim 10^{12}$ M$_\odot$) in cosmological simulations \citep{Boylan-Kolchin2011, Busha2011, Patel2017}. This means that the average MW-analog in a large-volume cosmological simulation is more likely to have already merged with LMC-mass subhalos by z=0 \citep[e.g.,][]{Stewart2008}. To test for consistency between our MW Galaxy and CDM expectations, the appropriate virial mass with which to identify a cosmological MW analog is its virial mass {\it prior} to the LMC infall, rather than the present-day combined MW+LMC system. 

\citet{Erkal2020} (hereafter EBP20) applied the \citetalias{Watkins2010} mass estimator using the 3D velocity of stellar halo stars -- $M_v(r)$, Eq.~\eqref{eq:mass_est_v} below -- to simulations of the MW-LMC system in a first-infall scenario. They concluded that $M_v(r)$ estimates of the MW mass today will be higher than that of the MW pre-LMC infall by up to $\sim 50\%$, and are not uniform across the sky. Even compared with the simulated combined MW+LMC mass profile, \citetalias{Erkal2020} find that $M_v$ estimates using all-sky data can be biased high by up to $\sim$20\% at large radii, depending on the LMC mass (see their Fig. A1). Consequently, the mass returned by $M_v$ does not directly tell us the appropriate mass for cosmological MW analogs.  

Importantly, \citet{Garavito-Camargo2019} (hereafter GC19) demonstrate that the LMC's impact on the anisotropy profile of the MW halo varies across the sky. This means that in some regions, particularly where the LMC has not yet traveled (future orbit), the halo kinematics may not yet reflect the increase in mass from the LMC. \citetalias{Erkal2020} concluded that if the 3D motion of halo stars is corrected for the reflex motion as a function of radius and angle on the sky, the MW mass pre-LMC infall may be recovered within $\lesssim 5\%$ with the $M_v$ mass estimator for stars in the regions of the sky least impacted by the LMC.   
In this study, we expand on the work of \citetalias{Erkal2020} to explore a different approach to recover the combined MW+LMC mass profile and disentangle the mass profiles of the present-day LMC and MW pre-LMC infall. In Sec.~\ref{sec:sims} we summarize the set of MW+LMC simulations of \citetalias{Garavito-Camargo2019} used in this work. In Sec.~\ref{sec:mass_est}, we present the two mass estimators proposed by \citetalias{Watkins2010}, including the $M_v$ estimator used by \citetalias{Erkal2020} and a second one, $M_{v_r}$, based on only the radial velocity. In Sec.~\ref{sec:results} we show how well these mass estimators can recover the \emph{total MW+LMC} mass profiles, with or without corrections for reflex motion. We further propose a method to recover the LMC mass by exploring the reflex motion imprinted in different regions of the sky, and use this method to estimate the MW's pre-LMC infall mass. In Sec.~\ref{sec:desi} we use a sample of Blue Horizontal Branch (BHB) stars from the DESI Milky Way Survey DR1 \citep{Bystrom2025} to estimate the total MW+LMC mass within $r<60$ kpc. We discuss our results in Sec.~\ref{sec:discussions} and conclude in Sec.~\ref{sec:conclusions}.

\section{Simulations}
\label{sec:sims}

We use the N-body simulations presented by \citetalias{Garavito-Camargo2019} -- see their Sec. 3 and Sec. 2.1 of \cite{Garavito-Camargo2021} for details. Here, we summarize the properties of the simulations that are most relevant to our analysis. The SMC is not included in this study as its mass is expected to be $\approx 10\%$ of the LMC's halo mass \citep{Rathore2025} and so it makes a negligible contribution to large-scale perturbations to the MW's halo. 

The complete suite of N-body simulations in \citetalias{Garavito-Camargo2019} consists of eight MW–LMC simulations and two isolated MW models. The MW model consists of both stars and dark matter (DM) particles. The stellar component consists of a bulge with a Hernquist \citep{Hernquist1990} density profile (mass $1.4 \times 10^{10}\,\mathrm{M}_{\odot}$ and scale length of 0.7 kpc), and an exponential disk (mass $5.78\times 10^{10} \mathrm{M}_{\odot}$, and scale length and height of 3.5 and 0.5 kpc, respectively). The MW DM halo follows a Hernquist profile with total mass of $1.57\times10^{12}$M$_{\odot}$ and scale length of 40.82~kpc. These parameters are the same for all simulations.

\begin{figure}
\centering
\includegraphics[width=0.9\columnwidth]{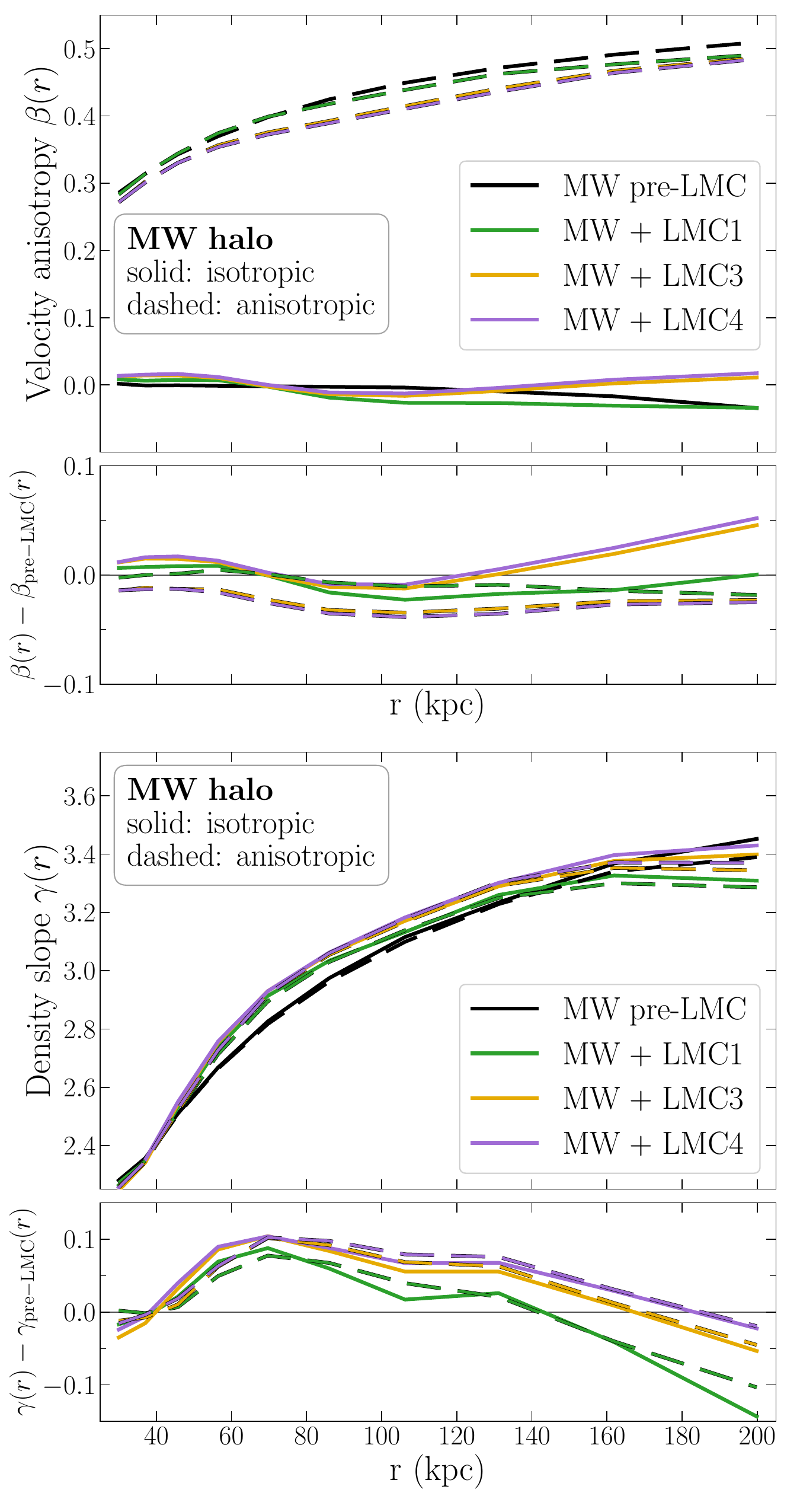}
\caption{Velocity anisotropy (top) and density slope (bottom) as a function of Galactocentric radius for MW DM particles (used as tracers) at present day in all simulations used in this study. Sub-panels show differences with respect to the equivalent MW pre-LMC model. The MW DM kinematics deviates most strongly from the pre-LMC models for the most massive LMCs (LMC3, LMC4). The density slope is higher than the pre-LMC case over 50-150 kpc, owing to the dynamical friction wake. Models where the MW DM velocities are initially isotropic (anisotropic) are shown with solid (dashed) lines. MW pre-LMC models are shown in black and represent the initial conditions for the MW halo in the MW+LMC models. Colored lines represent models with an LMC mass: $8\times 10^{10} \Msun$ (LMC1, green), $1.8\times 10^{11}\Msun$ (LMC3, gold) and $2.5\times 10^{11}\Msun$ (LMC4, purple).}
\label{fig:alpha_beta_gamma_all}
\end{figure}

The simulations include two realizations of the MW DM halo's velocity anisotropy, 
\begin{equation}
    \beta\equiv 1 - (\sigma_{\theta}^2 +
  \sigma_{\varphi}^2)/(2\sigma_{r}^2),
  \label{eq:beta}
\end{equation}
where $\sigma_i$ is the velocity dispersion: 1) \emph{isotropic kinematics}, $\beta=0$; or 2) \emph{anisotropic kinematics}, with ${\beta(r)=-0.5+ 0.2 \gamma(r)}$ where ${\gamma\equiv - \dd\ln \rho(r)/\dd\ln r}$
is the slope of the DM halo density profile. Thus, in the second case, the MW DM velocities are radially biased, and $\beta(r)$ is a monotonically increasing function of radius, as obtained in cosmological simulations. The two different kinematics provide more general conditions to probe the MW halo's response to the LMC arrival. The two isolated MW models, one for each kinematic model, are referred to as \textbf{MW pre-LMC} and serve to calibrate the impact of the LMC. Fig.~\ref{fig:alpha_beta_gamma_all} shows the MW DM particles' velocity anisotropy and density slope for the models used in this work, in the present-day snapshot.

The LMC halos follow a Hernquist profile, each with different mass and scale lengths. Although \citetalias{Garavito-Camargo2019} ran simulations for four different LMC models, we use only three in this work: LMC1, LMC3, and LMC4, with a total mass of 
$[0.8, 1.8, 2.5] \times 10^{11} \mathrm{M}_{\odot}$, and scale lengths $[10.4, 20, 25.2]$ kpc respectively. LMC3 is our fiducial LMC model, corresponding to the standard LMC model adopted in many numerical simulations of the Clouds \citep[e.g.][]{Foote2026, Rathore2025c, Kratos2024, Lucchini2020, Pardy2018, Besla2012}. LMC4 is included to illustrate the impact of a higher-mass LMC that is still consistent with cosmological expectations for its infall mass \citep[see][]{Garavito-Camargo2021}. LMC1 is also included as it represents a low-mass LMC model from \citetalias{Garavito-Camargo2019}, thus providing a lower bound for its impact on mass estimates of the MW. We exclude the LMC2 case, intermediate between LMC1 and LMC3, because results based on it would not provide additional physical insight.

The scale radius for each LMC halo model is chosen such that the enclosed mass at 8.7~kpc is 1.7$\times10^{10}\Msun$, consistent with measurements of \cite{VanderMarel2014}. Including the isolated MW models and the two MW DM kinematic models described above, a total of eight simulations are analyzed in this work.

The simulations were run with the N-body code Gadget-4 \citep{Springel2022}. Both the MW and the LMC halos were run in isolation for 2~Gyr to guarantee dynamical equilibrium before running the combined MW--LMC simulation \citepalias[][]{Garavito-Camargo2019}. 

For the combined MW--LMC simulation, at $t=0$ Gyr the LMC is placed $\sim$300 kpc from the center of the MW. The `present-day' is defined as the time at which the simulated LMC lies within the observed $2\sigma$ constraints in position and velocity. This occurs 2.2~Gyr after the time of infall of the LMC into the MW halo ($t=0$ Gyr). See Sec. 3.3 in \citetalias{Garavito-Camargo2019} for details. The present-day snapshots are the ones primarily analyzed in this work. For the MW pre-LMC simulations, the MW halos were evolved in isolation until the same time as the present-day MW--LMC ($\sim$4.2 Gyr). Hereafter, the {\bf fiducial simulation} refers to MW+LMC3 in an isotropic MW halo. 

\subsection{Halo tracers}
\label{sec:tracers}
There is no explicit stellar halo in the MW models. In this work, we assume that the MW possesses a stellar halo that reasonably traces the MW's DM halo. Thus, we use MW DM halo particles as ``tracers" throughout the paper. These tracers are used to estimate the mass profile of the MW's DM halo using two mass estimators (see next section). Note that the LMC's DM particles are excluded from the tracers.

\begin{figure}
\centering
\includegraphics[width=0.9\columnwidth]{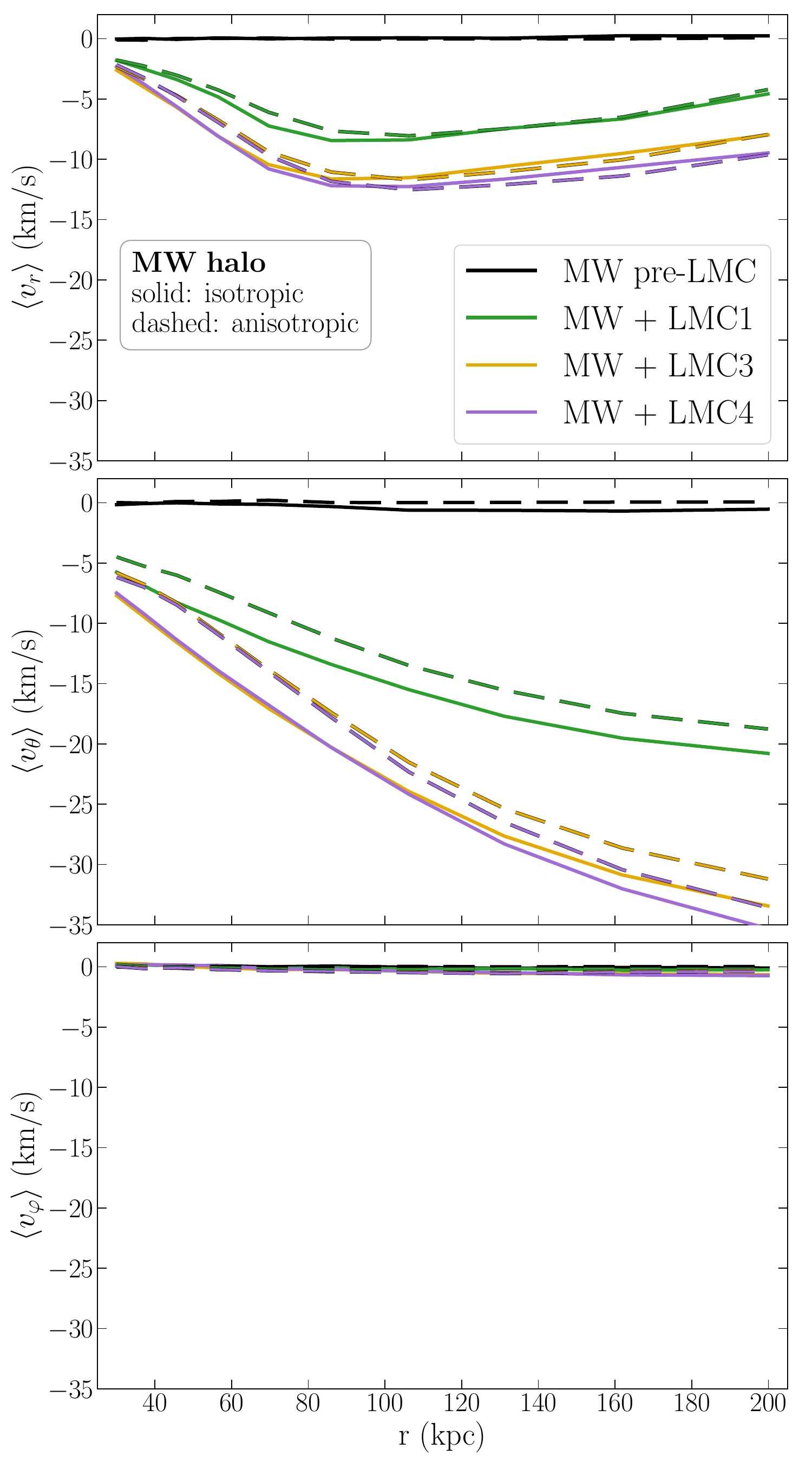}
\caption{Mean spherical velocity components of the MW DM particles, as a function of Galactocentric radius, for all simulations. For the MW+LMC simulations (colored lines), non-zero values are induced by the LMC infall. The corresponding reflex motion is largest in $\langle v_\theta\rangle$ owing to the polar nature of the LMC's orbit. The component $\langle v_\varphi \rangle$ is negligible for all simulations, and $\langle v_r\rangle$ reaches only $\approx -12$ km/s. This suggests that $v_r$-based mass estimators may be less biased by the disequilibrium induced by the LMC. Models with isotropic (anisotropic) MW DM halo velocity distribution are represented by solid (dashed) lines. Black lines denote the MW pre-LMC, with $\langle {\bf v}\rangle=0$ as expected.}
\label{fig:mean_v_all}
\end{figure}

Fig.~\ref{fig:mean_v_all} shows the present-day Galactocentric mean velocities in the MW pre-LMC (isolated) and MW+LMC models. In the latter, changes in the mean velocity reflect the reflex motion of the MW DM particles induced by the LMC infall. The velocity is computed in spherical coordinates, where $\langle v_r\rangle$, $\langle v_\theta\rangle$ and $\langle v_\varphi\rangle$ are the radial, polar and azimuthal components, respectively. In the MW pre-LMC models, all three mean velocities are zero. Models with increasingly massive LMCs show increasingly large reflex motions. 

Note, however, that while the reflex motion in $\langle v_\theta \rangle$ can reach up to $-35$ km/s, $\langle v_\varphi \rangle$ is essentially zero for all simulations and $\langle v_r\rangle$ only reaches up to $\sim -12$ km/s. This is a consequence of the LMC's orbit being nearly polar, producing a net ``upward'' motion of outer halo stars, as seen from the Galactic center\footnote{While the galactic latitude $b$ is defined ``from the origin to pole'', the spherical galactocentric latitude $\theta$ is defined ``from pole to the origin''. Thus, the negative $v_\theta$-values correctly represent ``upward'' motion of halo tracers associated with the ``downward'' motion of the disc and inner halo due to the pull by the LMC.}.

The true MW stellar halo contains substructure that can cause its velocity anisotropy to vary locally. As discussed in \citetalias{Garavito-Camargo2019}, this affects the stellar halo kinematics over a few kpc \citep{Cunningham2019, Cunningham2020}. Such spatial scales are much smaller than the large-scale perturbations caused by the LMC studied in this work. Indeed, the reflex motion caused by the LMC has been detected in the MW's stellar halo \citep{Petersen2021, Erkal2021, Chandra2025}, where observations are consistent with predictions from DM only simulations \citepalias{Garavito-Camargo2019}.  

The goal of this study is to demonstrate a methodology to disentangle the estimates of the LMC and MW mass profiles. The method is largely agnostic to the choice of halo tracer. Instead, the primary assumption is that the tracer responds to the LMC similarly to the MW's DM halo response. 

\section{Mass estimators}
\label{sec:mass_est}
We use two mass estimators proposed by \citetalias{Watkins2010}. These estimators assume a tracer population in equilibrium and deliver the total mass within a sphere of radius $r_\mathrm{out}$ by averaging over certain quantities inside the radial range of the data [$r_\mathrm{in}$, $r_\mathrm{out}$], assuming $r_\mathrm{out} \gg r_\mathrm{in}$.

The first estimator is given by \citepalias[Eq. (24) of][]{Watkins2010}:
\begin{equation}
  \label{eq:mass_est_v}
  M_v(r_\mathrm{out}) = \frac{(\alpha +
  \gamma - 2\beta)}{(3-2\beta)} \frac{r_\mathrm{out}^{1-\alpha}}{G}\langle v^2r^\alpha\rangle,
\end{equation}
where $v$ is the 3D speed of a tracer, $r$ is its galactocentric distance, $\alpha\equiv 1 - \dd\ln M(r)/\dd\ln r$
is related to the slope of the \emph{total enclosed mass} profile,
and $G$ is the gravitational constant. For a population that self-consistently produces the total mass\footnote{In other words, if a population itself sources the potential, i.e. its density $\rho$ is related to $\Phi$ via Poisson’s equation, $\nabla^2\Phi = 4\pi G\rho$.}, $\alpha = \gamma -2$, but self-consistency is not required by the method. Eq.~\eqref{eq:mass_est_v} is used by \citetalias{Erkal2020} to estimate the present-day MW mass.

Motivated by the small reflex motion in $\langle v_r\rangle$ seen in Fig.~\ref{fig:mean_v_all}, we additionally explore the second estimator proposed by \citetalias{Watkins2010} -- their Eq. (16):
\begin{equation}
  \label{eq:mass_est_vr}
  M_{v_r}(r_\mathrm{out}) = (\alpha +
  \gamma - 2\beta) \frac{r_\mathrm{out}^{1-\alpha}}{G}\langle v_r^2r^\alpha\rangle.
\end{equation}
Note that both Eqs.~\eqref{eq:mass_est_v} and \eqref{eq:mass_est_vr}
require 6D coordinates for all tracers. 

The application of these estimators to observational data requires an
educated guess for the total mass density slope $\alpha$ 
and the assumption that $\alpha$, $\gamma$
and $\beta$ are constants in the radial range of the data. 
As described in Sec.~\ref{sec:tracers}, in this study we use MW DM particles in the simulations as tracers of the gravitational potential. With 6D coordinates for all DM particles, we measure $\alpha$, $\gamma$, and $\beta$ explicitly in each radial bin and apply the above two estimators to each bin to recover the simulated MW's mass profile.

In what follows, we refer to the particles' coordinates in the frame centered in, and co-moving with, the MW disc's origin (Galactocentric). For analyses involving galactic coordinates, we place the Sun at $x=-8.122\kpc$, following the default value in astropy \citep[][]{astropy:2013, astropy:2018, astropy:2022}.

\section{Simulation Results: Disentangling the Mass Profile of the MW and LMC}
\label{sec:results}

In this section, we apply the mass estimators to simulations of the MW pre-LMC and of the combined MW+LMC at present day. In Sec. \ref{sec:MW+LMC}, we recover the MW+LMC total mass profile from 30-200 kpc, and partially confirm and reinterpret the results presented by \citetalias{Erkal2020}. In Sec.~\ref{sec:octants} we show the best regions of the sky to constrain the present day MW+LMC mass profile. We then illustrate a novel method to disentangle the LMC and MW mass profiles, recovering the simulation mass profiles for the present-day LMC (Sec. \ref{sec:LMC}) and MW {\it pre-LMC infall} (Sec. \ref{sec:MW}). 

\subsection{Estimating the combined MW+LMC mass at present day}
\label{sec:MW+LMC}

We analyze MW DM particles (no LMC particles) as tracers of the total mass profile, focusing on particles at $30<r/\kpc<200$, where the impact of the LMC is expected to be maximized. Radial bins are defined such that $r_\mathrm{out}$ is sufficiently larger than $r_\mathrm{in}$, but bin widths are small enough for the exponents $\alpha$, $\gamma$ and $\beta$ to be approximately constants in each bin. We follow \citetalias{Erkal2020} and define bin widths that are $19\%$ of $r_\mathrm{out}$ in the interval $30 < r/\kpc \lesssim 200\kpc$, such that the innermost bin has radial range $(r_\mathrm{in},r_\mathrm{out})=(24.3,30)$~kpc and the outermost range is $(162,200)$~kpc.

We calculate $\alpha$, $\beta$ and $\gamma$ for MW particles in each bin, as illustrated in Fig.~\ref{fig:alpha_beta_gamma_all}. The tracer density slope $\gamma$ and total mass slope $\alpha$ are calculated via numerical differentiation of the binned mass densities (with second-order central finite differences, as implemented in numpy.gradient).
While $\beta$ and $\gamma$ use only MW DM particles as tracers, $\alpha$ is based on the total potential produced by all MW particles, but in practice the MW disc or bulge contributions are negligible in this radial range. Thus, effectively, $\alpha = \gamma -2$, since the DM self-consistently generates the total mass in those radii. Note, however, that in this approach, when dealing with observational data, one has to assume some value for $\alpha$.

Fig.~\ref{fig:alpha_beta_gamma_all} (upper panel) shows the
velocity anisotropy for the MW DM particles. As the LMC mass increases, the MW's anisotropy deviates more significantly from the MW pre-LMC model. The density slope (bottom panel) is higher for the MW+LMC models than the pre-LMC model for $50\lesssim r/\kpc \lesssim 150$. That is, the MW DM density profiles get steeper and then shallower at larger radii as the halo responds to the LMC.

\begin{figure}
\centering
\includegraphics[width=0.9\columnwidth]{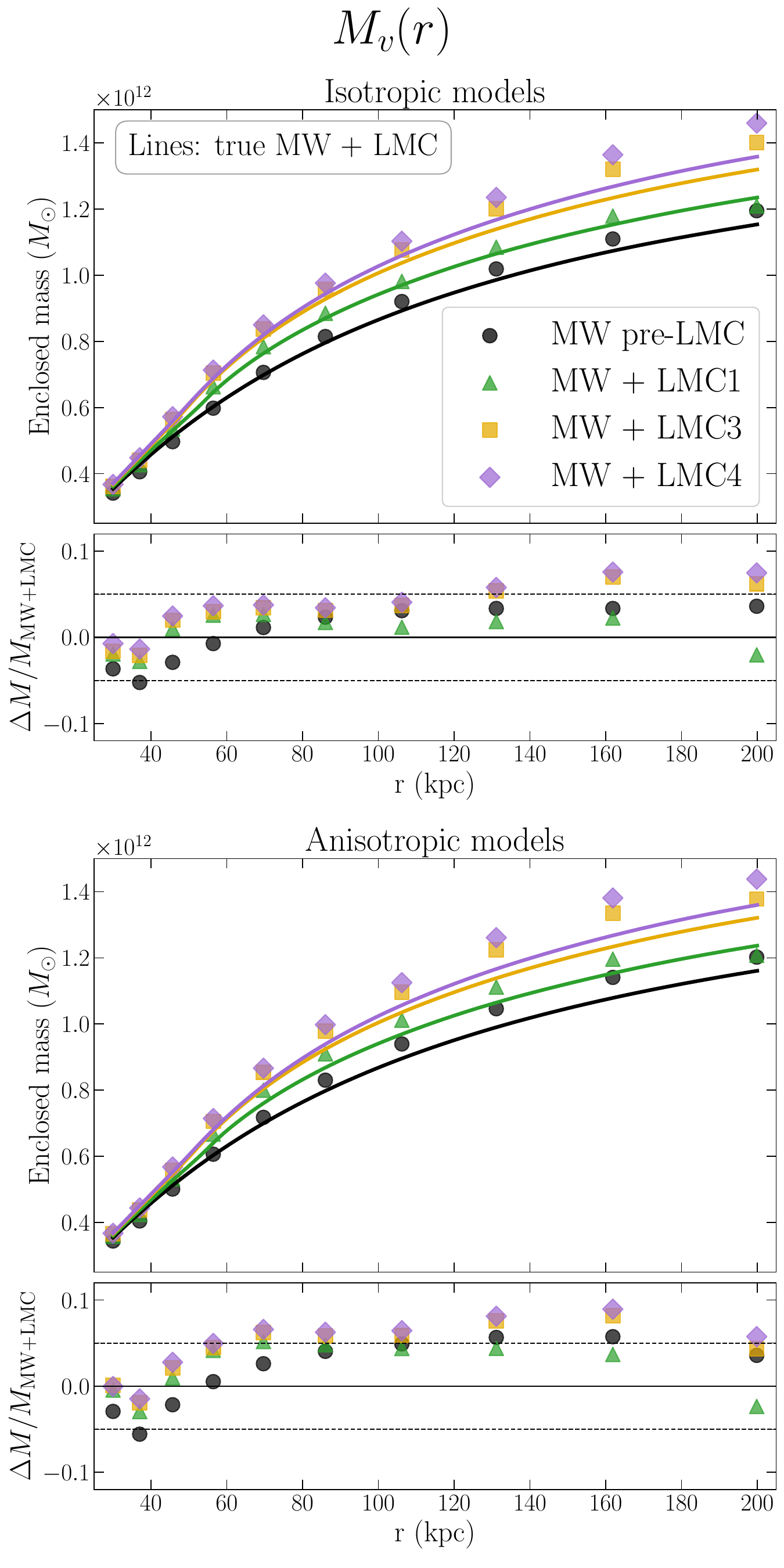}
\caption{Mass profile of the MW pre-LMC and combined MW+LMC derived from Eq.~\eqref{eq:mass_est_v} (dots) taking MW DM particles as tracers (Fig. \ref{fig:alpha_beta_gamma_all}), compared to the true simulated mass (lines). Models with isotropic (anisotropic) MW DM halos are shown in the top (bottom). Sub-panels show the fractional difference of the recovered and true mass profiles, $\Delta M/M_{MW+LMC}$. In all cases true mass profile is recovered in 30-120 kpc to within $\pm$7\%. For $r\gtrsim 120$ kpc, discrepancies 
 can reach 10\% for more massive LMC models, driven by the LMC-induced reflex motion, as pointed out by \citetalias{Erkal2020}.}
\label{fig:M_total_all_v}
\end{figure}

Having $\alpha$, $\beta$, and $\gamma$ in each bin, we estimate the total mass $M_v(r_\mathrm{out})$, Eq.~\eqref{eq:mass_est_v}, setting $r_\mathrm{out}$ as each bin's outer edge. Fig.~\ref{fig:M_total_all_v} shows that, for each MW+LMC model, the estimator recovers the true simulated total mass profiles within 30-120 kpc to within $\pm7\%$. The recovered mass is typically larger than that of the true simulated total MW+LMC mass. Note that even in the MW pre-LMC model, the recovered mass is still biased high by $\lesssim 5\%$, which may be due to a lack of perfect equilibrium in the MW pre-LMC models, or to the estimator intrinsic uncertainties produced e.g. by the power-law assumption over a radial range.

At larger distances ($>$120 kpc), the mass estimates are biased high, and increasingly so for heavier LMC models. This is similar to the conclusion of \citetalias{Erkal2020} -- see their Fig. A1. Although, for an LMC as massive as our LMC4 model, they find discrepancies from the combined MW+LMC mass up to $20\%$, while ours is $\lesssim$10\%.

As discussed in \citetalias{Erkal2020}, the mass estimator assumes a tracer population in equilibrium. The discrepancy with respect to the true LMC+MW mass profiles in the simulation is thus produced by the LMC's infall, which drives the MW DM halo out of equilibrium. A significant source of disequilibrium is the displacement of the center of mass of the inner MW halo with respect to its outer halo. The consequent apparent velocity of the halo as seen from the Galactic center, the reflex motion, increases with radius
\citep{Garavito-Camargo2021}. 

\begin{figure}
\centering
\includegraphics[width=0.9\columnwidth]{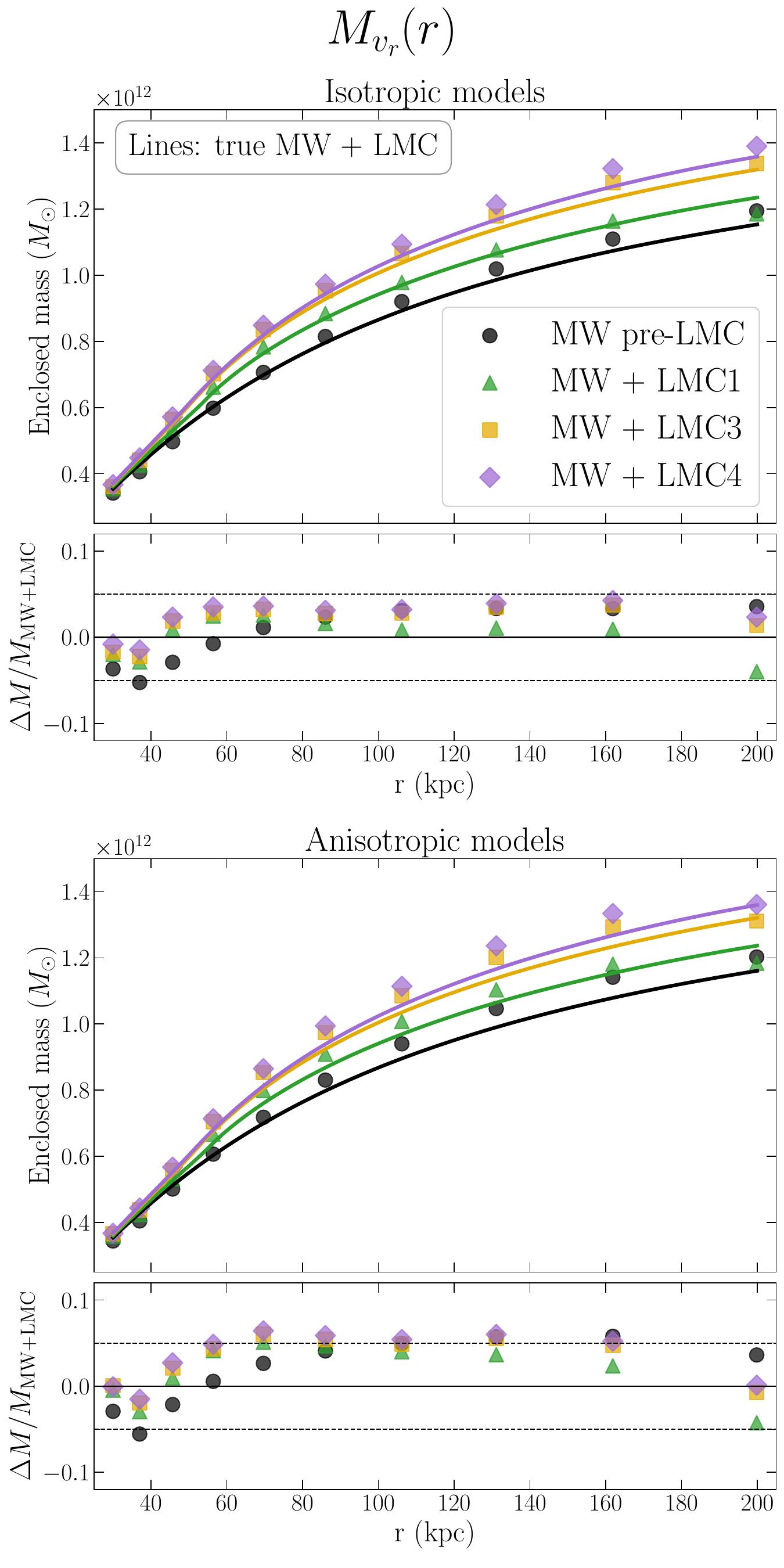}
\caption{Similar to Fig.~\ref{fig:M_total_all_v}, but using the $M_{v_r}$ estimator, Eq.~\eqref{eq:mass_est_vr}. Due to the small reflex motion in the radial direction (Fig.~\ref{fig:mean_v_all}), the true total MW+LMC mass is recovered within $\approx 5\%$ for all simulations out to $200\kpc$.}
\label{fig:M_total_all_vr}
\end{figure}

The small variations in $\langle v_r\rangle$ seen in Fig.~\ref{fig:mean_v_all} motivates us to use the $M_{v_r}$ mass estimator, Eq.~\eqref{eq:mass_est_vr}, which is based only on $v_r$. Fig.~\ref{fig:M_total_all_vr} shows that the resulting bias is significantly reduced, and the true MW+LMC mass is recovered within $\approx 5\%$ up to $200\kpc$ for all simulations.

\begin{figure}
\centering
\includegraphics[width=0.9\columnwidth]{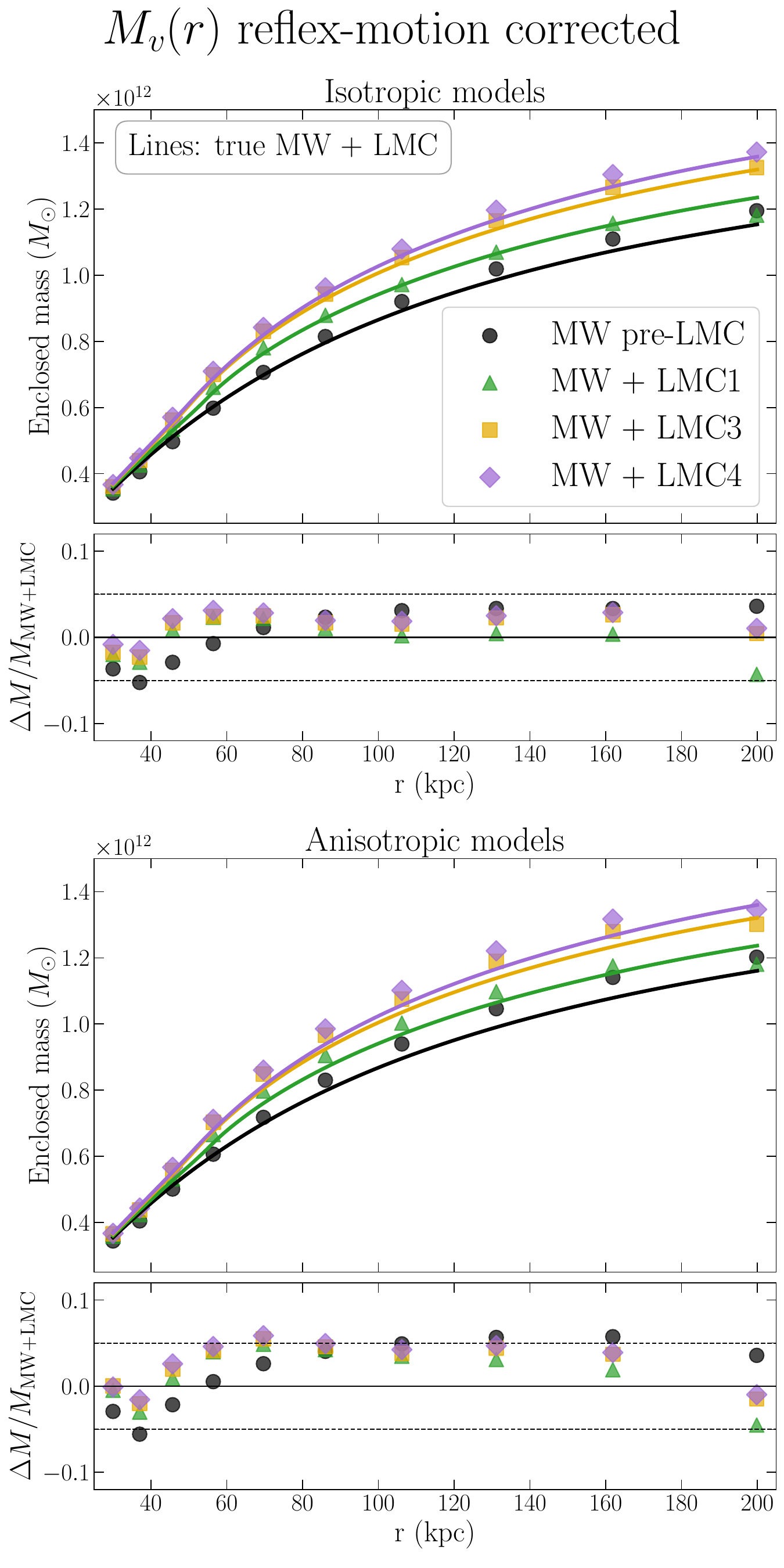}
\caption{Similar to Fig.~\ref{fig:M_total_all_v}, but with MW halo velocities in the MW+LMC simulations corrected for reflex motion, Eq.~\eqref{eq:mass_est_v_corrected}. The true total MW+LMC mass is now recovered within $\approx 5\%$ for all simulations, similar to the accuracy of the $M_{v_r}$ estimator with no reflex-motion correction (Fig.~\ref{fig:M_total_all_vr}).}
\label{fig:M_total_all_v_reflex}
\end{figure}

To strengthen the argument that the bias in the mass estimate is driven by the reflex motion, we follow the methodology outlined in \citetalias{Erkal2020}, but to demonstrate that the \emph{total} MW+LMC mass is recovered more accurately once the reflex motion is corrected for in the velocities of halo particles. Thus we re-estimate the mass with $M_v$, but instead of Eq.~\eqref{eq:mass_est_v}, we now use
\begin{equation}
  \label{eq:mass_est_v_corrected}
  M_v(r_\mathrm{out}) = \frac{(\alpha +
    \gamma - 2\beta)}{(3-2\beta)}
  \frac{r_\mathrm{out}^{1-\alpha}}{G}\langle (\boldsymbol{v} - \langle \boldsymbol{v}\rangle)^2r^\alpha\rangle,
\end{equation}
where $\langle \boldsymbol{v}\rangle = (\langle v_r\rangle, \langle v_\theta\rangle, \langle v_\varphi\rangle)$ is the mean reflex velocity in each radial bin (Fig.~\ref{fig:mean_v_all}). 
In contrast to \citetalias{Erkal2020}, who apply Eq.~\eqref{eq:mass_est_v_corrected} to tracers in different octants, each with its own $\langle \boldsymbol{v}\rangle$, here we apply Eq.~\eqref{eq:mass_est_v_corrected} to the all-sky spherical bins. In Sec.~\ref{sec:M_octs_reflex_motion_corrected}, we also correct for the reflex motion in each octant (Fig.~\ref{fig:M_octs_MWLMC5_b0_reflex}).
Fig.~\ref{fig:M_total_all_v_reflex} shows that, indeed, these reflex-motion corrected estimates recover the true MW+LMC mass within $\approx 5\%$ for all simulations.

This exercise highlights that the $M_{v_r}$ estimator -- Eq.~\eqref{eq:mass_est_vr},
Fig.~\ref{fig:M_total_all_vr} -- is similarly accurate to the reflex-motion corrected $M_v$ -- Eq.~\eqref{eq:mass_est_v_corrected}, Fig.~\ref{fig:M_total_all_v_reflex} -- with the added benefit of avoiding the need for a reflex-motion correction. Furthermore, a similar reflex-motion correction applied to $M_{v_r}$ improves the mass estimates only slightly.

\subsection{Estimating the MW+LMC mass with partial sky coverage}
\label{sec:octants}

We now investigate the impact on mass estimates of data usage only in parts of the sky, as opposed to the control tests of Sec.~\ref{sec:MW+LMC} which assume all-sky coverage. Given the LMC's past orbit and current location, the MW halo response is highly asymmetric, with the reflex motion magnitude varying across different regions of the sky \citepalias[see Fig. 9 and Fig. 1 in][respectively]{Garavito-Camargo2019, Erkal2020}. Thus, we expect a different mass estimate for each region, where some regions provide more accurate estimates, as shown by \citetalias{Erkal2020}.

We split the simulation data into eight regions according to their galactic longitude ($\ell$) and latitude ($b$)\footnote{Note that \citetalias{Erkal2020} define octants with only four regions in $(\ell, b)$ and splitting in $x>0$ or $x<0$.}. 
We then calculate the velocity anisotropy $\beta(r)$ and density slope $\gamma(r)$ for MW DM particles in the intersection of each radial shell with the angular aperture of the respective octant, as illustrated in Fig.~\ref{fig:geom_los}. The calculation of the volume of that region is detailed in the appendix \ref{sec:geom_los}.

\begin{figure}
\centering
\includegraphics[width=0.9\columnwidth]{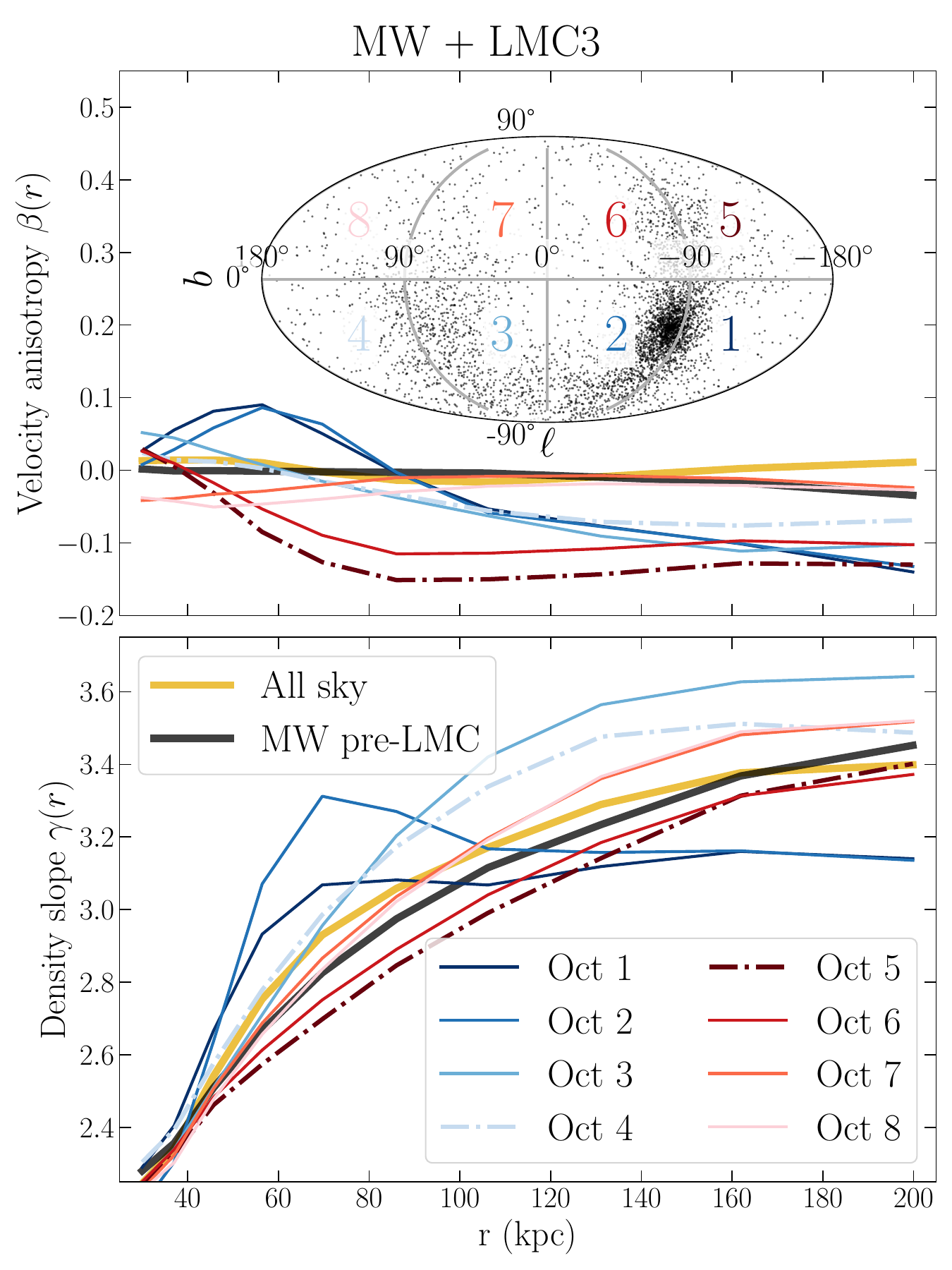}
\caption{Velocity anisotropy and density slope of MW DM particles for our fiducial model as a function of radius, splitting into octants in galactic coordinates ($\ell$,$b$) -- see the inset plot, with a sample of LMC DM particles. The LMC's past orbit nearly follows octants $8/7 \rightarrow 4/3\rightarrow 2/1$, whereas LMC's future orbit will take it through 5/6. Lines are color-coded by the respective octants, with 4 and 5 highlighted with dot-dashed lines for later reference in Sec.~\ref{sec:LMC}. Thick gold lines show all-sky quantities, and black lines represent the MW pre-LMC model (same colors as in Figs.~\ref{fig:alpha_beta_gamma_all}-\ref{fig:M_total_all_v_reflex}). Owing to the recent LMC infall, the profiles vary significantly across the sky, but are similar to those for the MW pre-LMC when averaged across the sky.}
\label{fig:beta_gamma_octs_MWLMC5_b0}
\end{figure}

Fig.~\ref{fig:beta_gamma_octs_MWLMC5_b0} shows the velocity
anisotropy and density slope in each octant for our fiducial model (isotropic MW+LMC3). The inset shows the Mollweide projection of ($\ell, b$) for a sub-sample of LMC DM particles at present day, which trace the orbit of the LMC. The octant definitions \citepalias[same as
in][]{Garavito-Camargo2019} are indicated by colored numbers, with lines colored accordingly. The LMC's past orbit approximately follows octants $8/7 \rightarrow 4/3\rightarrow 2/1$, 
whereas the LMC's future orbit will take it through octants 5/6. While $\gamma(r)$ and $\beta(r)$ vary significantly from octant to
octant, the all-sky quantities are close to those of the MW pre-LMC model, where prominent perturbations have been washed out in the spherical average. We verified that this is approximately the case for all simulations, including the higher LMC mass models.

In terms of $\beta(r)$, octants 1\&2 and 5\&6, the closest ones to the LMC's current location, are the most perturbed. This is consistent with the large velocity dispersions seen for these octants in Figs. 9-11 of \citetalias{Garavito-Camargo2019}. In terms of $\gamma(r)$, the Southern Galactic octants (blue) are significantly more perturbed than the ones in the North (red). The most prominent feature in $\gamma(r)$ is the profile steepening for octants 1 \& 2 near $\approx 70\kpc$, which is produced by the MW DM wake locally trailing the LMC. Octants 3 \& 4, although also significantly perturbed, show more regular density slopes that are closer to those of the all-sky results.
Octants 7 \& 8 are the least perturbed in terms of $\beta(r)$ and $\gamma(r)$.

\begin{figure}
\centering
\includegraphics[width=0.9\columnwidth]{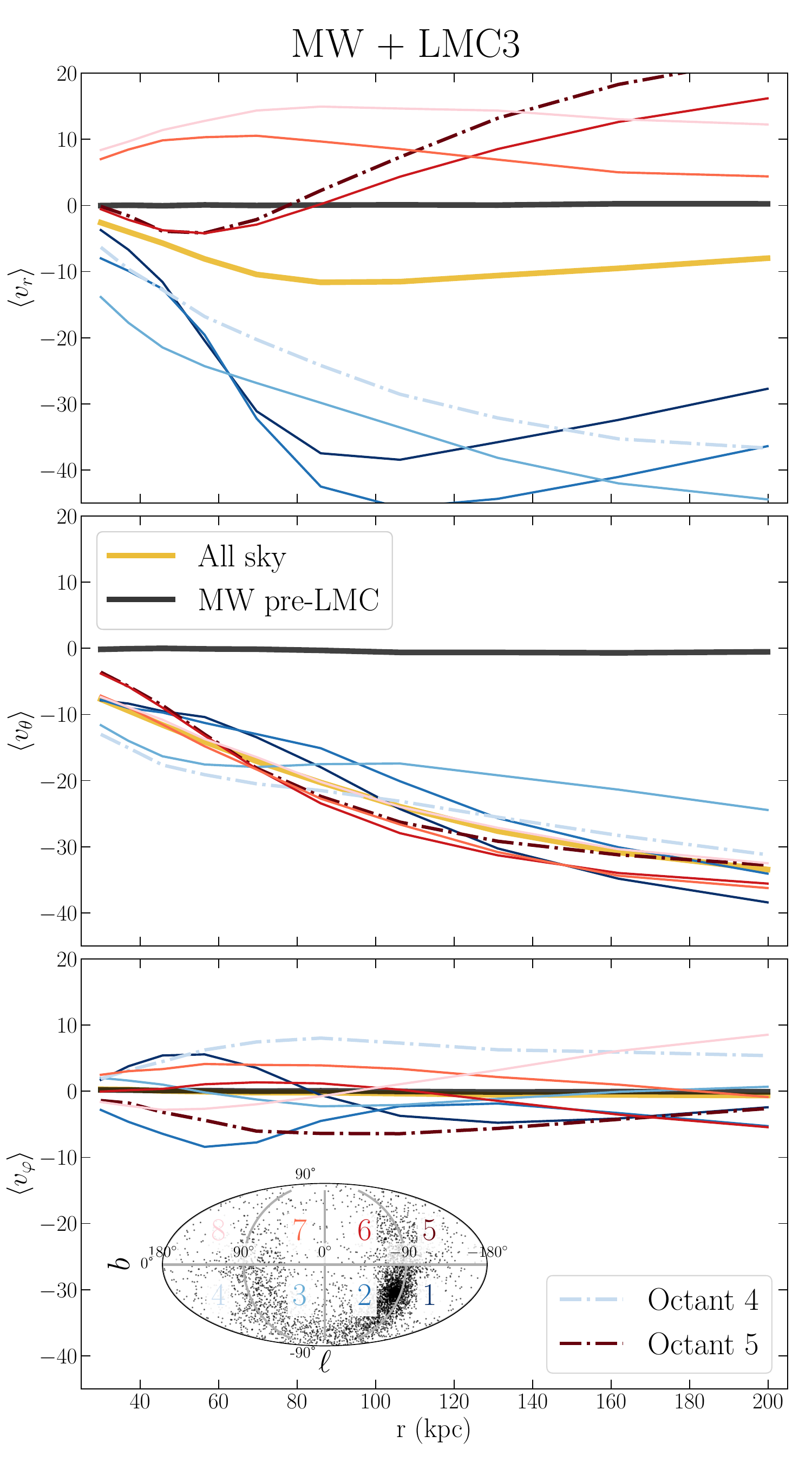}
\caption{Mean velocities of MW DM particles for our fiducial model, split into octants (see inset plot, which includes a sample of LMC DM particles for reference). Lines are color-coded by the respective octants. Octants 4 and 5 are highlighted with dot-dashed lines; these octants exhibit opposite behavior in every velocity component with respect to the mean.  Thick gold lines represent all-sky measurements, and black lines show the  MW pre-LMC model (same colors as in Figs.~\ref{fig:alpha_beta_gamma_all}-\ref{fig:M_total_all_v_reflex}). While the ``upward'' motion ($\langle v_\theta\rangle$) is coherent in all octants, $\langle v_r\rangle$ varies significantly across octants, being typically positive (negative) for those in the North (South).}
\label{fig:mean_vr_vtheta_vphi_octs_MWLMC5_b0}
\end{figure}

Fig.~\ref{fig:mean_vr_vtheta_vphi_octs_MWLMC5_b0} shows the mean velocities as a function of radius for each octant as defined in
Fig.~\ref{fig:beta_gamma_octs_MWLMC5_b0}. We see that the large $\langle v_\theta\rangle$ previously obtained with all-sky coordinates (Fig.~\ref{fig:mean_v_all}) is due to a coherent ``upward'' motion of MW DM particles for all octants. The figure further reveals that the smaller all-sky $\langle v_r\rangle$ is actually produced by a partial cancellation of opposite trends in different octants \citep[in accordance with Fig. 5 of][]{Vasiliev2023}. In particular, Southern octants, with $\langle v_r\rangle$ reaching up to $\approx -45$ km/s, are again more strongly perturbed than those in the North, which reach only $\approx 20$ km/s.

We compute the parameters for the mass estimators using data for each octant separately, except for $\alpha(r)$, which is based on all-sky measurements of Sec.~\ref{sec:MW+LMC}. This is because $\alpha(r)$ is associated with the total mass distribution and not with local tracers. Furthermore, in this approach $\alpha(r)$ would not be observationally obtained from tracers and some value would have to be adopted. Alternatively, one could model $\alpha$ and other parameters in a DF-fitting modeling \citep[e.g.][]{Deason2012, Deason2021}.

\begin{figure}
\centering
\includegraphics[width=0.9\columnwidth]{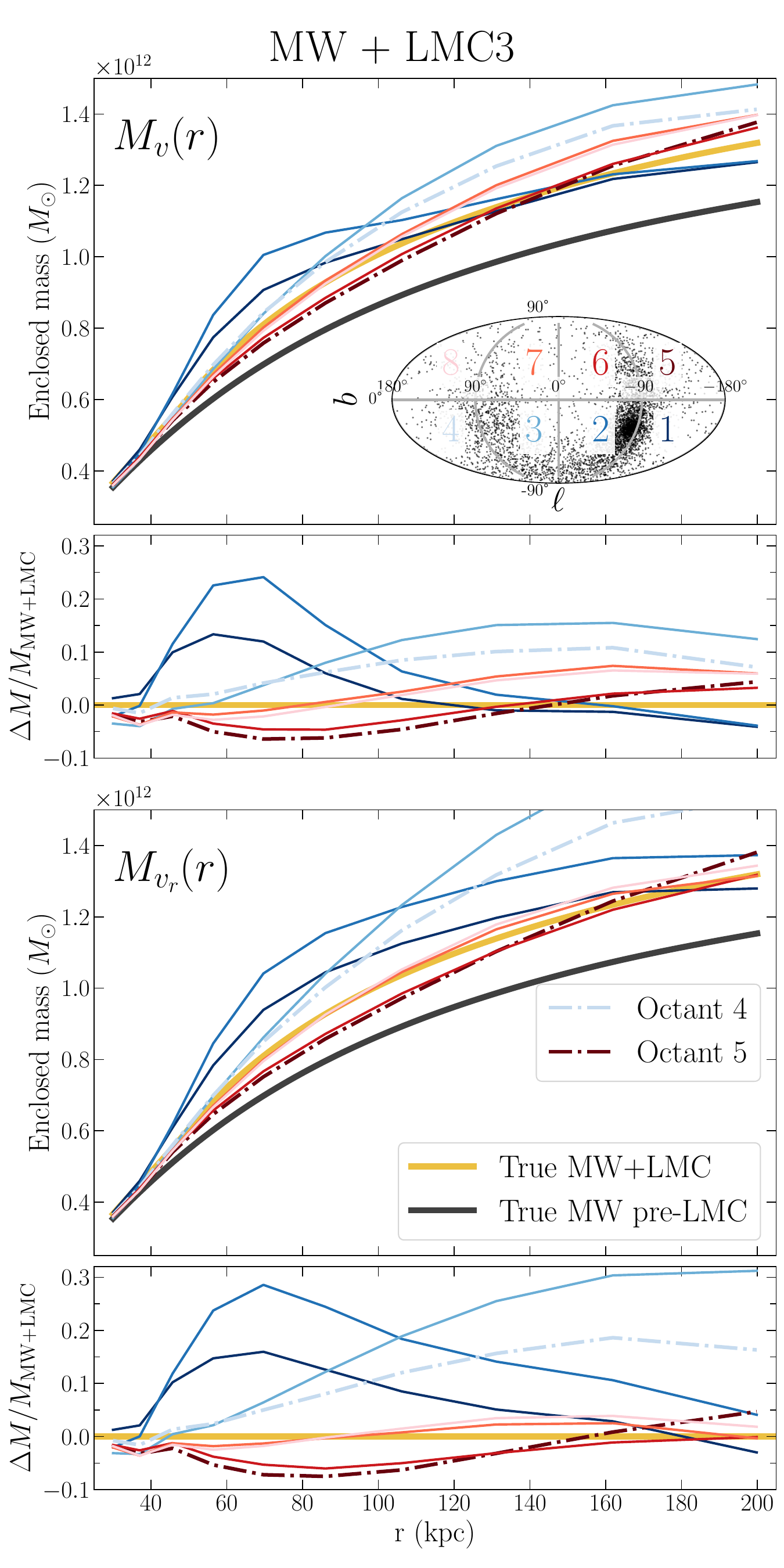}
\caption{Mass estimates for the fiducial model using data
  from separate octants (inset plot) and the estimator: $M_v$ (Eq.~\eqref{eq:mass_est_v}; top) or $M_{v_r}$ (Eq.~\eqref{eq:mass_est_vr}; bottom). Octants 4 and 5 are highlighted with dot-dashed lines. Thick gold lines show the true simulated MW+LMC mass; black lines represent the MW pre-LMC model (same colors as in Figs.~\ref{fig:alpha_beta_gamma_all}-\ref{fig:M_total_all_v_reflex}). Octants 1 and 2 are the most perturbed. In octants 7 and 8, $M_{v_r}$ reproduce the total MW+LMC mass across $30<r/\kpc \lesssim 200$ within $\lesssim 5\%$, while $M_{v}$ errors achieve $\sim 7\%$.}
\label{fig:M_octs_MWLMC5_b0}
\end{figure}

Fig.~\ref{fig:M_octs_MWLMC5_b0} shows the $M_v$ ($M_{v_r}$) mass estimates for each octant on the top (bottom) panel. For all MW+LMC simulations, we note that: 

\begin{itemize}
\item The Northern octants (red) provide the closest estimates of the total MW+LMC mass. This is likely because the LMC's orbital path is predominantly in the South, leading to pronounced perturbation in these regions of the sky causing local disequilibrium; 

\item The $M_{v_r}$ estimates for octants 7 and 8 are particularly accurate. This is because the reflex motion $\langle v_r\rangle$ is smaller for these octants (Fig.~\ref{fig:mean_vr_vtheta_vphi_octs_MWLMC5_b0}) and because the LMC traveled through them the longest time ago ($\sim 0.6$-1.2 Gyr ago, at Galactocentric distances $\sim 150$-200 kpc), giving more time for tracers in those regions to re-equilibrate to the new MW+LMC mass. From these octants, the total simulated MW+LMC mass 
is recovered within $\lesssim 5\%$ with $M_{v_r}$ across $30 \leq r/\kpc \leq 200$, which is within the intrinsic errors obtained with the MW pre-LMC model (Fig.~\ref{fig:M_total_all_vr}).
 
 \item  Octants 5 and 6 are the only ones with mass estimates below the true MW+LMC mass for almost all radii. These octants are in the region of the LMC's future orbit, and this region's tracers are the last to respond significantly to the LMC's arrival. As such, tracers in octants 5 and 6 can be seen either as biased tracers of the MW mass pre-LMC infall or as transitioning from being an MW mass detector to an MW+LMC mass detector.

\item Octants 1 and 2 are far from equilibrium and should not be used for this type of mass estimate. 
\end{itemize}

\begin{figure}
\centering
\includegraphics[width=0.9\columnwidth]{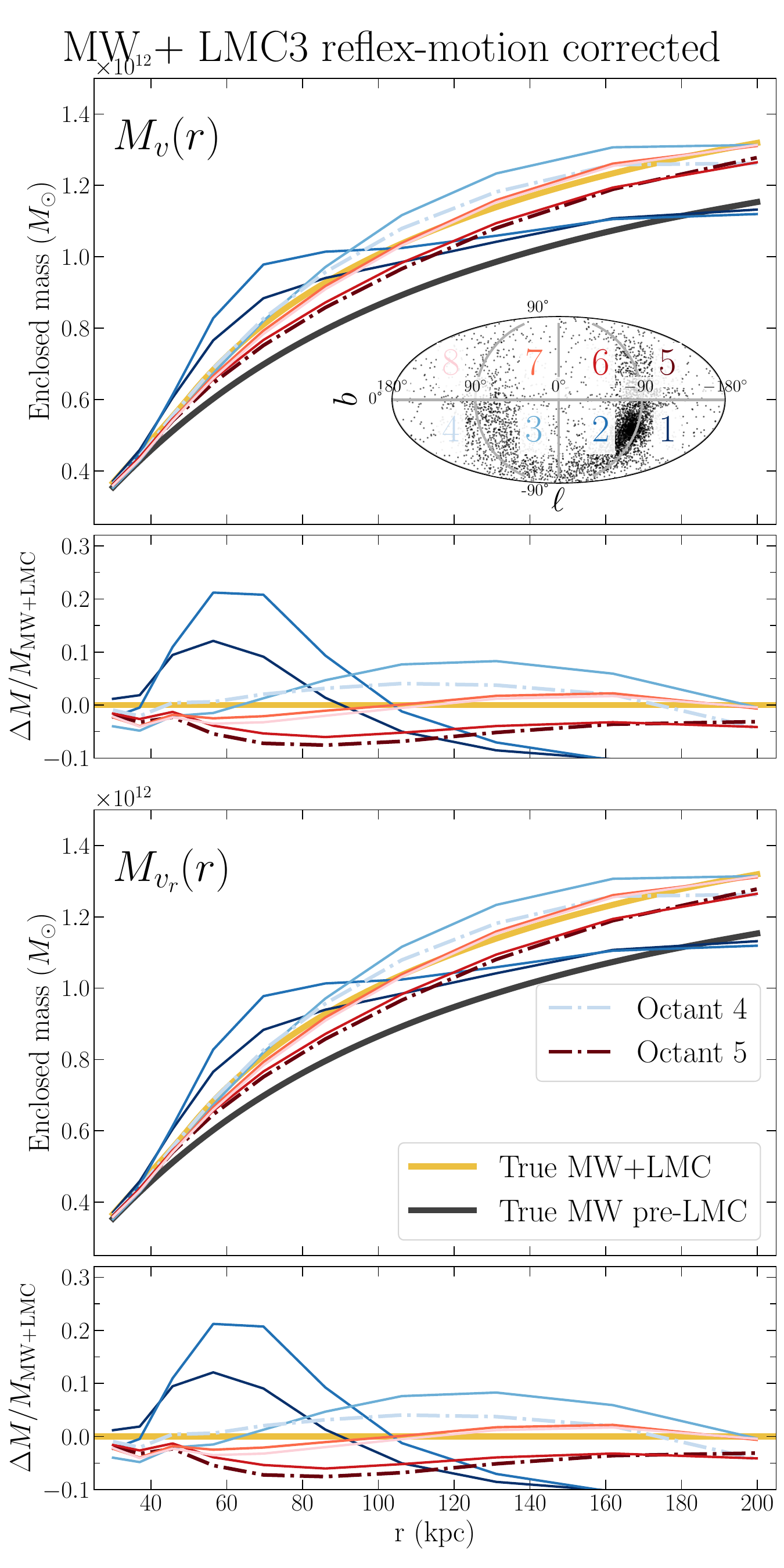}
\caption{Similar to Fig.~\ref{fig:M_octs_MWLMC5_b0}, but now
  correcting for the reflex motion in each radial bin and octant (as shown in Fig.~\ref{fig:mean_vr_vtheta_vphi_octs_MWLMC5_b0}). Now, octants in the past orbit of LMC approximately trace the total MW+LMC mass (8, 7, 4 and, to a lesser extent, 3), while octants in its future orbit (5 and 6) seem like biased tracers of the MW mass pre-LMC (overestimate) and of the combined MW+LMC mass (underestimate). Octants 1 and 2 still provide poor estimates of either mass profile.}
\label{fig:M_octs_MWLMC5_b0_reflex}
\end{figure}

\subsubsection{Reflex-motion-corrected mass estimates}
\label{sec:M_octs_reflex_motion_corrected}

Until now, we have not applied corrections for the LMC-induced reflex motion. We do so here, motivated by the work of \citetalias{Erkal2020}, i.e. using Eq.~\eqref{eq:mass_est_v_corrected} and, analogously, replacing $v_r$ by $v_r - \langle v_r \rangle$ in Eq.~\eqref{eq:mass_est_vr}. Fig.~\ref{fig:M_octs_MWLMC5_b0_reflex} shows the corresponding mass estimates corrected by the reflex motion computed in each octant. Comparison with Fig.~\ref{fig:M_octs_MWLMC5_b0} shows that this correction mostly improves $M_{v_r}$ estimates for Southern octants (blue), in agreement with their large $\langle v_r \rangle$ (Fig.~\ref{fig:mean_vr_vtheta_vphi_octs_MWLMC5_b0}, upper panel).

The most accurate estimates are achieved in octants 8/7, then 4 and, to a lesser extent, 3. This suggests that octants provide better reflex motion-corrected MW+LMC mass estimates the longer ago they were crossed by the LMC's orbit. The LMC has just entered octants 1 \& 2, leaving them significantly out-of-equilibrium. 

In contrast, octants 5 and 6 are in the path of the LMC's future orbit and can be described as providing biased estimates of the MW mass pre-LMC infall. Octants 5 and 6 are roughly the same octants \citetalias{Erkal2020} concluded provide the best reflex-motion corrected mass estimates of the MW pre-LMC infall. However, while their bias is $\lesssim 5\%$, our bias with respect to the MW mass pre-LMC can reach $11-12\%$ -- see Sec.~\ref{sec:compare_EB20}.

\begin{figure*}
\centering
\includegraphics[width=\textwidth]{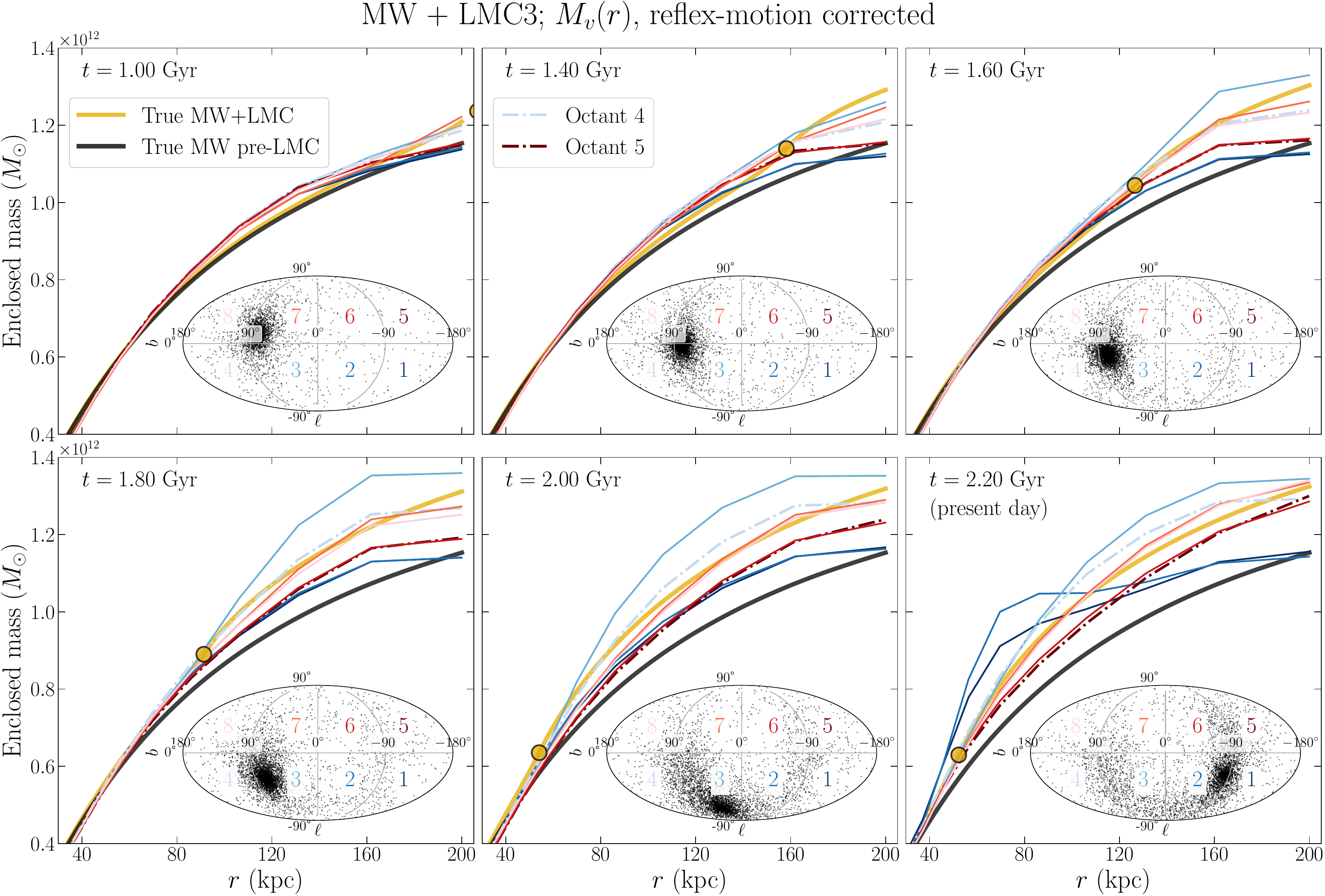}
\caption{Timeline of reflex-motion corrected mass estimates (using the total velocity; $M_v(r)$) in the fiducial simulation. Indicated in each panel are the times since the LMC first crosses the MW's virial radius (time of infall = 0). The lower right panel, $t=2.2$ Gyr after infall, corresponds to the present-day snapshot. The insets show LMC particles in galactic coordinates at the given time, with octants matching line colors (4\&5 are represented by dot-dashed lines). The gold dots indicate the LMC's galactocentric distances at the given time. The LMC's trajectory takes it first into octants 8/7 at infall (t=1.00, 1.40 Gyr), then to 4/3 (t=1.60, 1.80, 2.00 Gyr), to 2/1 at the present day (t=2.20 Gyr), and will eventually go to octants 5/6 in the future. Correspondingly, octants 7, 8 \& 4 are the first to respond to the LMC, returning mass profiles consistent with the total MW+LMC mass. Octants 5\&6 and 1\&2, in the latest stages of LMC's future orbit, are the last octants to begin to deviate from the MW mass; at present day, octants 5\&6 are expected to yield mass estimates that are biased low relative to the combined MW+LMC mass and high relative to the MW mass pre-LMC. Octants 1\&2 are the most disturbed at present day due to the very recent entry of the LMC in this region of the sky. A time-line video of these mass estimates is available \href{https://lberaldoesilva.github.io/videos/timeline_Mv_reflex.mp4}{here}.}
\label{fig:timeline_Mv}
\end{figure*}

Fig.~\ref{fig:timeline_Mv} confirms this evolutionary picture by showing a timeline of the $M_v$, reflex-motion corrected, estimates, Eq.~\eqref{eq:mass_est_v_corrected}, applied to each octant as the LMC infalls to its current location about the MW. The bottom-right panel shows the present-day snapshot, at $t=2.2$ Gyr after infall. The time evolution of the mass estimates in each octant is dictated by the LMC's orbital path. The first octants that begin to reflect the combined MW+LMC mass are octants 7, 8 \& 4, which is where the LMC enters the MW's halo ($t<1.4$ Gyr after infall). 

As the LMC approaches an octant, the estimated mass profile starts to deviate from the MW pre-LMC profile, approaching the combined MW+LMC profile. Correspondingly, octants in the future orbital path of the LMC return estimated masses between the MW pre-LMC  and combined MW+LMC. This is demonstrated by the mass profiles estimated from octants 5 \& 6 across $t = 1.4-2.2$ Gyr - these octants are always in the future orbital path of the LMC. This behavior is also seen in octants 1 \& 2 across $t=1.4-2.00$ Gyr, during which time frame the LMC had not yet entered these octants. 

When the LMC enters fully within an octant, the local tracers are significantly perturbed. The corresponding disequilibrium results in a pronounced overestimate of both the MW pre-LMC and combined MW+LMC mass profiles. This is illustrated in octant 3 at $t=1.80$ Gyr and octants 1 \& 2 at $t=2.20$ Gyr (present day).

After the LMC passes through an octant, the octant's tracers start to relax and achieve equilibrium; the mass estimates begin to decrease back towards the true combined MW+LMC mass profile. Correspondingly, mass estimates using octants 7 \& 8 (where the LMC enters the halo at the earliest times) reasonably recover the MW+LMC profile at almost all epochs illustrated. At $t=2.00$ Gyr after infall, mass profiles estimated in octants 3 \& 4 begin to decrease towards the MW+LMC profile, but have not reached equilibrium at present day. 

In summary, we have shown that the earliest perturbed octants 7 and 8 provide the best tracers of the total MW+LMC mass, with excellent $M_{v_r}$ mass estimates even without any reflex-motion correction (Fig.~\ref{fig:M_octs_MWLMC5_b0}). In addition, we have demonstrated that, because the LMC is now approaching octants 5 and 6 for the first time, these octants will return masses biased high relative to the MW mass pre-LMC-infall and low relative to the MW+LMC mass. This bias remains at the $\sim$10\% level with respect to the MW mass pre-LMC-infall even after the reflex motion of the octants are corrected. In what follows, we take advantage of the asymmetry created across the sky by the LMC's arrival to disentangle its mass from that of the combined MW+LMC system.

\subsection{Estimating the LMC mass}
\label{sec:LMC}

We now explore the idea that the LMC's induced reflex motion must carry information about the LMC's mass. The reflex motion originates on the MW's disc and inner halo motion with respect to the outer halo in response to the approach of the LMC \citep{Garavito-Camargo2021}. The speed of this motion scales with the mass of the LMC \citep{Gomez2015}. 
Thus, we focus on the $M_v$ estimator, Eq.~\eqref{eq:mass_est_v}, which encapsulates all three reflex motion components. This approach treats the reflex motion as having intrinsic value, rather than as a source of error or bias that must be corrected. 

Fig.~\ref{fig:mean_vr_vtheta_vphi_octs_MWLMC5_b0} shows that the mean radial velocities of DM particles exhibit a striking North-South asymmetry, with Northern (Southern) octants typically having positive (negative) $\langle v_r \rangle$. This radial velocity dipole scales with the mass of the LMC \citep[see Fig. 14 in][]{Garavito-Camargo2021} and it is to a large extent responsible for the different mass estimates obtained from different octants. Thus, one can expect that the difference between mass estimates based on a Northern and a Southern octant, which maximizes the radial velocity dipole, should provide information on the LMC mass. 

Furthermore, in Sec.~\ref{sec:octants} we argued that octants 5 and 6 are only starting to feel the impact of the LMC and thus carry most of the information on the MW mass pre-LMC infall. At the same time, octants in earlier phases of the LMC orbit (4, 7, 8) have had time to equilibrate and more accurately track the total MW+LMC mass. We thus expect the difference between mass estimates from octants (5,6) vs (4,7,8) to be most informative of the LMC mass. 

We combine the above arguments by choosing an octant in (5,6) and another in (4,7,8) that demonstrate maximal differences in the velocity components of the reflex motion. In Fig.~\ref{fig:mean_vr_vtheta_vphi_octs_MWLMC5_b0}, octants 4 and 5 (highlighted with dot-dashed lines in Figs.~\ref{fig:beta_gamma_octs_MWLMC5_b0}-\ref{fig:timeline_Mv}) produce almost exactly opposite signals in $\langle v_r \rangle$ and $\langle v_\varphi \rangle$ at all radii. Similarly, the trends in these octants are also opposite in $\langle v_\theta \rangle$ with respect to the all-sky average. Taken together, these arguments suggest that the difference in the mass estimates from octants 5 and 4 should be the most informative of the LMC mass.

\begin{figure}
\centering
\includegraphics[width=0.9\columnwidth]{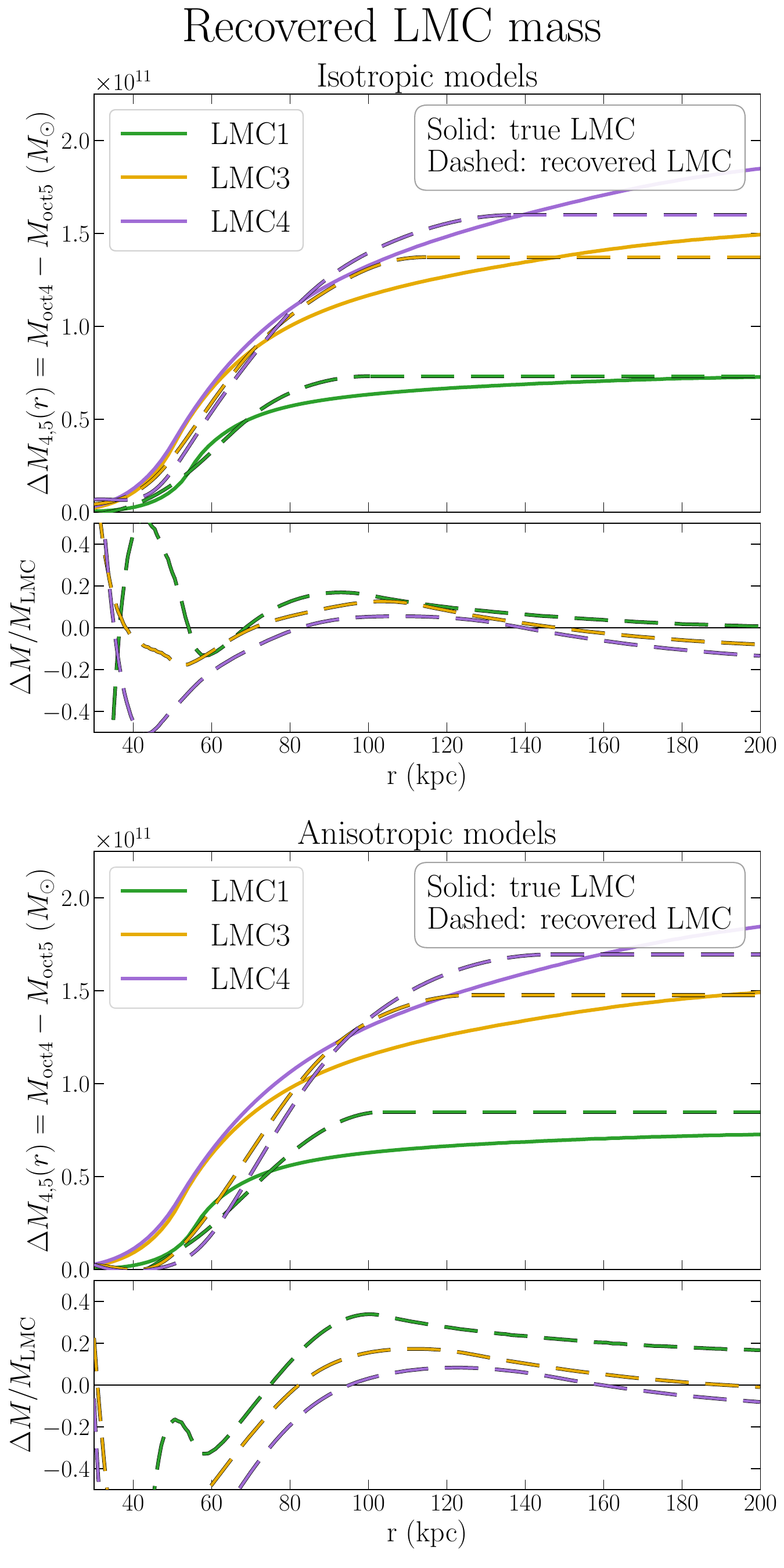}
\caption{The difference between the estimated mass profiles $M_v(r)$ from octants 4 and 5 (dashed lines; $\Delta M_{4,5}$) approximately reproduces the simulated LMC's mass profile (solid lines) with respect to the MW center at present day in all simulations. The top(bottom) pair of panels are for the Isotropic(Anisotropic) MW model. The properties of octants 4 and 5 in the fiducial simulation, Isotropic MW+LMC3, are illustrated in Figs.~\ref{fig:beta_gamma_octs_MWLMC5_b0}-\ref{fig:M_octs_MWLMC5_b0}, with these octants highlighted with dot-dashed lines. Sub-panels illustrate the errors of the inferred LMC profile relative to the true simulated LMC at present day. $\Delta M_{4,5}$ recovers the total LMC mass enclosed in Galactocentric $r < 200$ kpc with errors $\Delta M/M_\mathrm{LMC} \lesssim 20$\% if we use tracers at $r\gtrsim 100$ kpc.}
\label{fig:delta_M_octs_all}
\end{figure}

Fig.~\ref{fig:delta_M_octs_all} shows the true simulated Galactocentric cumulative LMC mass profiles (solid lines) compared to the reconstructed LMC mass (dashed lines). The LMC mass is reconstructed by first taking the difference in the $M_v(r)$ mass profiles for octants 4 and 5:  ${\Delta M_{4,5}'(r) = M_\mathrm{oct4}(r) - M_\mathrm{oct5}(r)}$. The resulting mass difference increases up to a certain radius, decreasing further out, in accordance with the relative behavior of these octants in Fig.~\ref{fig:M_octs_MWLMC5_b0} (top panel). We next fit a cubic spline to $\Delta M_{4,5}'(r)$. We finally define the recovered LMC mass profile, $\Delta M_{4,5}$, using the spline up to the radius of its maximum and assuming a fixed value at that maximal mass for larger radii. Despite significant deviations from the true LMC mass profile at $r\lesssim 100-120$ kpc, particularly in the anisotropic DM halos (bottom panel), the total LMC mass for $r\lesssim 200\kpc$ is recovered within $\approx$ 10-20\% errors.

\subsection{Estimating the Milky Way mass pre-LMC infall}
\label{sec:MW}
Summarizing our results so far, we have: i) recovered the total MW+LMC mass in $30<r/\kpc < 200$ with $\lesssim 5\%$ errors with the $M_{v_r}$ estimator using either all-sky data or data from octant 7 or 8; and ii) recovered the total LMC mass within Galactocentric $r< 200\,\kpc$ with $10-20\%$ errors, from tracers at $r\gtrsim 100-120$ kpc and by subtracting the $M_v$ estimate of octant 5 in the Northern hemisphere (LMC's future orbit) from that of octant 4 (oldest phase of LMC's past orbit in the South). The above results apply to all MW+LMC simulations used in this study (see Secs.~\ref{sec:MW+LMC}-\ref{sec:LMC}) and do not require a correction for the reflex motion induced by the LMC. With an estimate of the LMC's mass profile and the total MW+LMC profile, we now demonstrate how the MW's pre-LMC infall mass profile can be recovered. 

\begin{figure}
\centering
\includegraphics[width=0.9\columnwidth]{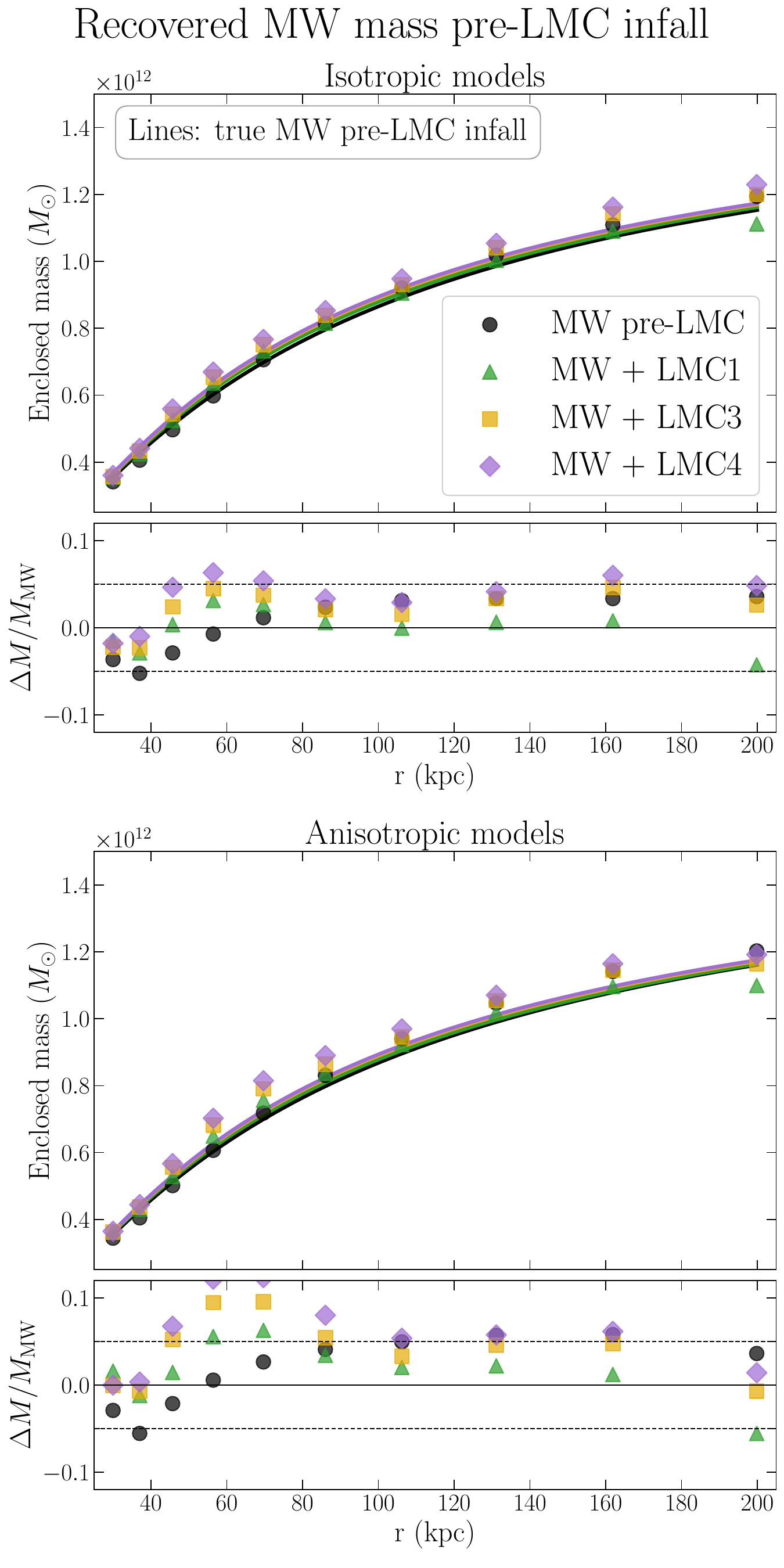}
\caption{The MW mass pre-LMC infall, recovered by subtracting the reconstructed LMC mass profile
  (Fig.~\ref{fig:delta_M_octs_all}), from the $M_{v_r}$ estimate of the total MW+LMC mass profile (Fig.~\ref{fig:M_total_all_vr}). Lines show the true MW mass pre-LMC infall. Colored symbols show the resulting enclosed mass. Sub-panels show the errors relative to the true MW mass pre-LMC infall, which is recovered within $\lesssim 5\%$ errors in all radii for the isotropic models (top). For the anisotropic ones (bottom), the same accuracy is only obtained for $r \gtrsim 100$ kpc. The virial radius of the simulated MW halo is $R_{200}\sim$225 kpc. As such, this method provides a pathway to the appropriate virial halo mass with which to compare MW cosmological analogs.}
\label{fig:M_MW_before_LMC_infall_all}
\end{figure}

Fig.~\ref{fig:M_MW_before_LMC_infall_all} shows the MW pre-LMC infall mass profile obtained by subtracting the recovered LMC mass profile (ii) from the recovered MW+LMC mass profile (i). For the isotropic models (top), the MW mass pre-LMC infall is recovered within $\lesssim 5\%$ though all radii $30<r/\kpc < 200$. For the anisotropic models (bottom), the MW mass pre-LMC infall is recovered with the same accuracy but only for tracers at $100<r/\kpc < 200$. Within $<$100 kpc in the anisotropic models, errors can be as large as the LMC mass, due to the poor recovery of the LMC profile in this region (Fig.~\ref{fig:delta_M_octs_all}).

We note that for these simulations the MW's virial radius ($R_{200}\sim225$ kpc) is close to the largest bin used for the mass estimates. As such, this method can estimate the real MW virial mass ($M_{200}$) nominally (i.e. with error-free data) within $\sim 5\%$. This provides a viable path forwards to the appropriate virial halo mass ($M_{200}$) which can be used to compare with MW cosmological analogs -- see Sec.~\ref{sec:virial} for a discussion on the extrapolation from inner mass estimates to $M_{200}$.

\subsection{Impact of fixing the total mass slope}
The previous subsections assumed full knowledge of the slope of the total MW mass profile $\alpha(r)$, which can be measured in the simulation. When dealing with observational data, one needs to assume some fixed value for $\alpha$, guided by results from simulations. In this subsection, we re-estimate the total MW+LMC mass, the LMC mass and the MW mass pre-LMC infall, but now fixing $\alpha=1$, in order to evaluate the impact of this choice. This value of $\alpha$ is expected for the outskirts of a Navarro-Frenk-White (NFW) halo \citep[][]{Navarro1997}.

\begin{figure}
\centering
\includegraphics[width=0.9\columnwidth]{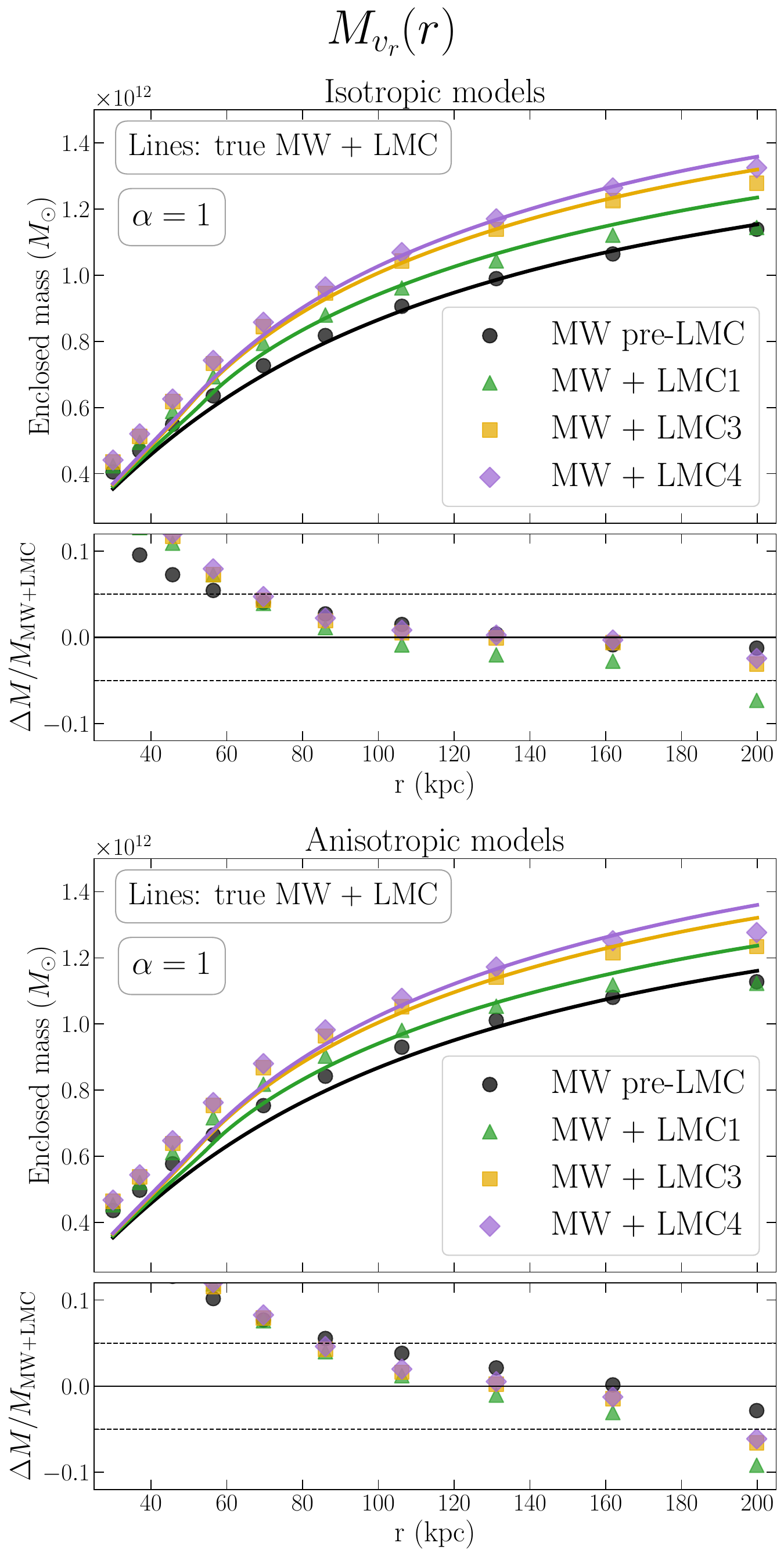}
\caption{Similar to Fig.~\ref{fig:M_total_all_vr}, but fixing
  $\alpha=1$ in Eq.~\eqref{eq:mass_est_vr}, as expected for an outer NFW halo. The true simulated total MW+LMC mass is recovered within $\approx 10\%$ for all simulations for $r\gtrsim 50\kpc$ and $\approx 5\%$ from $80-200\kpc$. We conclude that uncertainty in $\alpha$ is not a major source of error to the proposed method.}
\label{fig:M_total_all_vr_fixed_alpha}
\end{figure}

The total MW+LMC mass derived from the mass estimators, with a fixed $\alpha =1 $ is shown in Fig.~\ref{fig:M_total_all_vr_fixed_alpha}. We find that the total MW+LMC mass is recovered within $\approx 10\%$ errors at distances $\gtrsim 50\kpc$ and $\approx 5\%$ errors at distances from $80-200 \kpc$. At distances $\gtrsim 50\kpc$, the true $\alpha$ approaches the assumed value $\alpha=1$ (see Fig.~\ref{fig:alpha_beta_gamma_all}; bottom panel), and effectively $\alpha = \gamma -2$. 

The LMC mass obtained by subtracting estimates in octants is very similar to the previous case where $\alpha$ was extracted from the simulations instead of fixed, and is not shown here. Following our method, after subtracting the LMC mass from the total MW+LMC mass, we find that the MW mass pre-LMC infall is also recovered within $\approx 5-10\%$ errors at distances $\gtrsim 80\kpc$.

\section{Estimating the Combined MW+LMC mass with observations}
\label{sec:desi}

In this section, we consider real observational conditions for our proposed method to compute the combined MW+LMC mass estimates (outlined in Sec. ~\ref{sec:MW+LMC} and ~\ref{sec:octants}). We do this using data from the DESI Milky Way Survey \citep[DESI-MWS,][]{Cooper:2023}. DESI \citep{DESI_2016, DESI_2024, DESI_2026} is producing the largest spectroscopic catalogs to date, both in its cosmological survey and in its stellar surveys within the DESI-MWS \citep[][]{Cooper:2023, Koposov_2024, Koposov2026, Dey_2026}, with its first public data release providing spectra and stellar parameters for over 4 million stars. Additionally, the DESI-MWS DR1 has already published value-added catalogs for blue horizontal branch (BHB) stars \citep{Bystrom2025}, RR Lyrae stars \citep{Medina_2026}, giants \citep{Li_2025} and white dwarfs \citep{Swan2026}.

In Sec.~\ref{sec:desi_footprint} we assess the impact of the footprint of realistic spectroscopic surveys, and of DESI in particular, on our methodology by superimposing the DESI-MWS footprint over our fiducial simulation. In Sec.~\ref{sec:MW_LMC_mass}, we use our proposed methodology to estimate the total MW+LMC mass within $60$ kpc using the DESI-MWS BHB sample \citep[][]{Bystrom2025}.

\subsection{Effect of the DESI footprint}
\label{sec:desi_footprint}
\begin{figure}
\centering
\includegraphics[width=0.95\columnwidth]{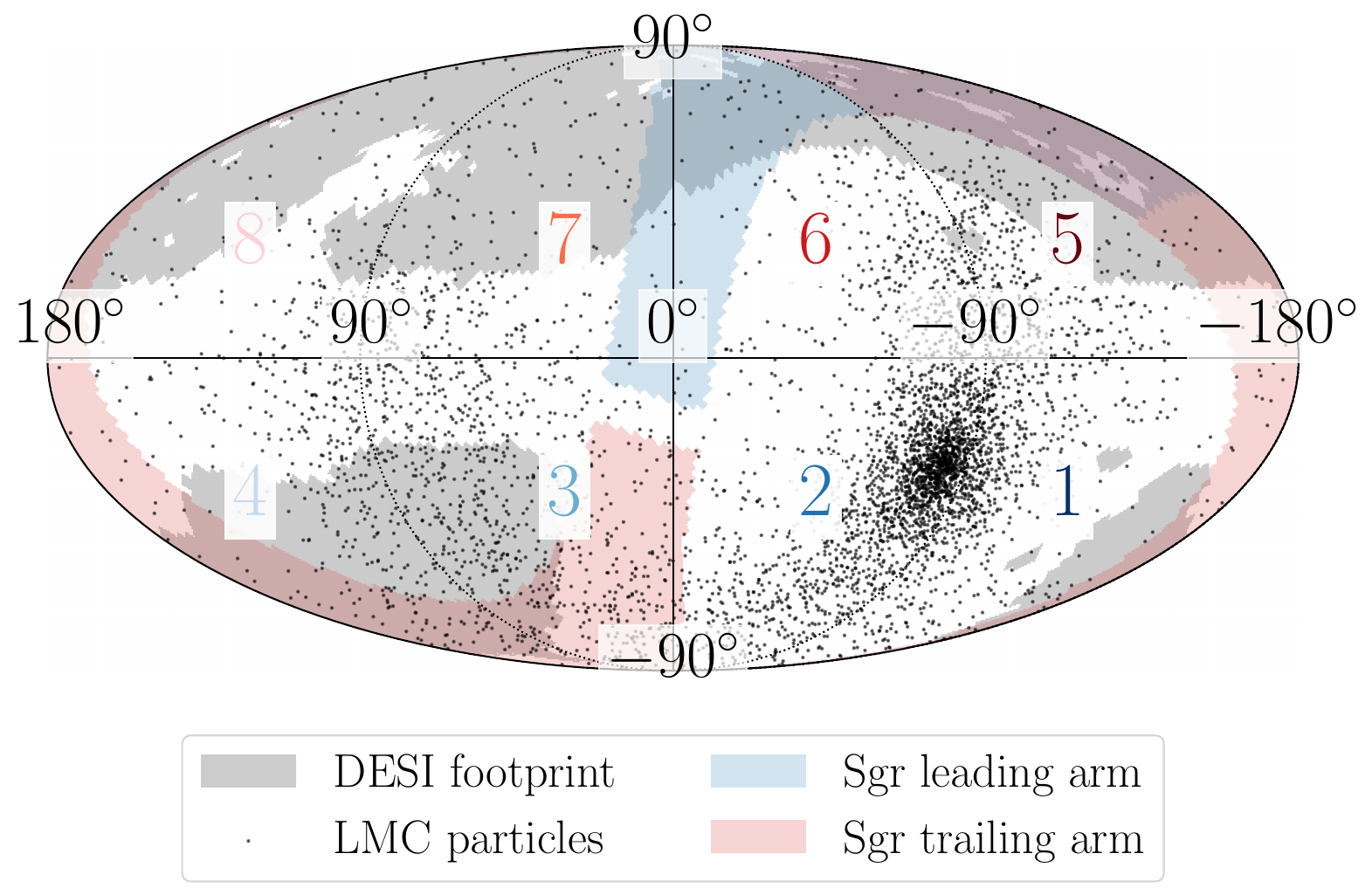}
\caption{Mollweide projection in Galactic coordinates showing the DESI footprint (grey). Scatter points show a random selection of LMC DM particles in the fiducial, isotropic simulation MW+LMC3. Overplotted are the on-sky regions of the Sgr stream, shown as $|\tilde{B}|<15^\circ$ in the Sgr stream coordinates $(\tilde{\Lambda},
\tilde{B})$ of \citet{Belokurov2014}, over the $\tilde{\Lambda}$ ranges spanned by the distance tracks of \citet{Hernitschek2017}, as provided by \citet{Bystrom2025} -- accessible at \href{https://github.com/abystrom/apex-likelihood-fitting}{https://github.com/abystrom/apex-likelihood-fitting}. The DESI footprint mostly covers regions of the Northern Galactic hemisphere, both in the early (octants 7 and 8) and future (octants 5 and 6) phases of the LMC's orbit.}
\label{fig:DESI_footprint}
\end{figure}

Fig.~\ref{fig:DESI_footprint} shows the DESI-MWS footprint, as shared by \cite{Bystrom2025}. Also plotted are a subset of LMC DM particles from our fiducial model along with octant definitions. We see that the DESI-MWS footprint mostly covers regions of the Northern Galactic hemisphere, both in the early (octants 7 and 8) and future (5 and 6) phases of the LMC's orbit.

In this section, as in Sec. ~\ref{sec:octants}, we split the data from the fiducial simulation into octants, but select only MW DM particles that reside within the DESI-MWS footprint. This allows us to isolate the effect of the survey footprint on the resulting mass estimates. In appendix \ref{sec:desi_foot_print_compare_sims} we show the results for the whole footprint and for all simulations considered in this study.

\begin{figure}
\centering
\includegraphics[width=0.9\columnwidth]{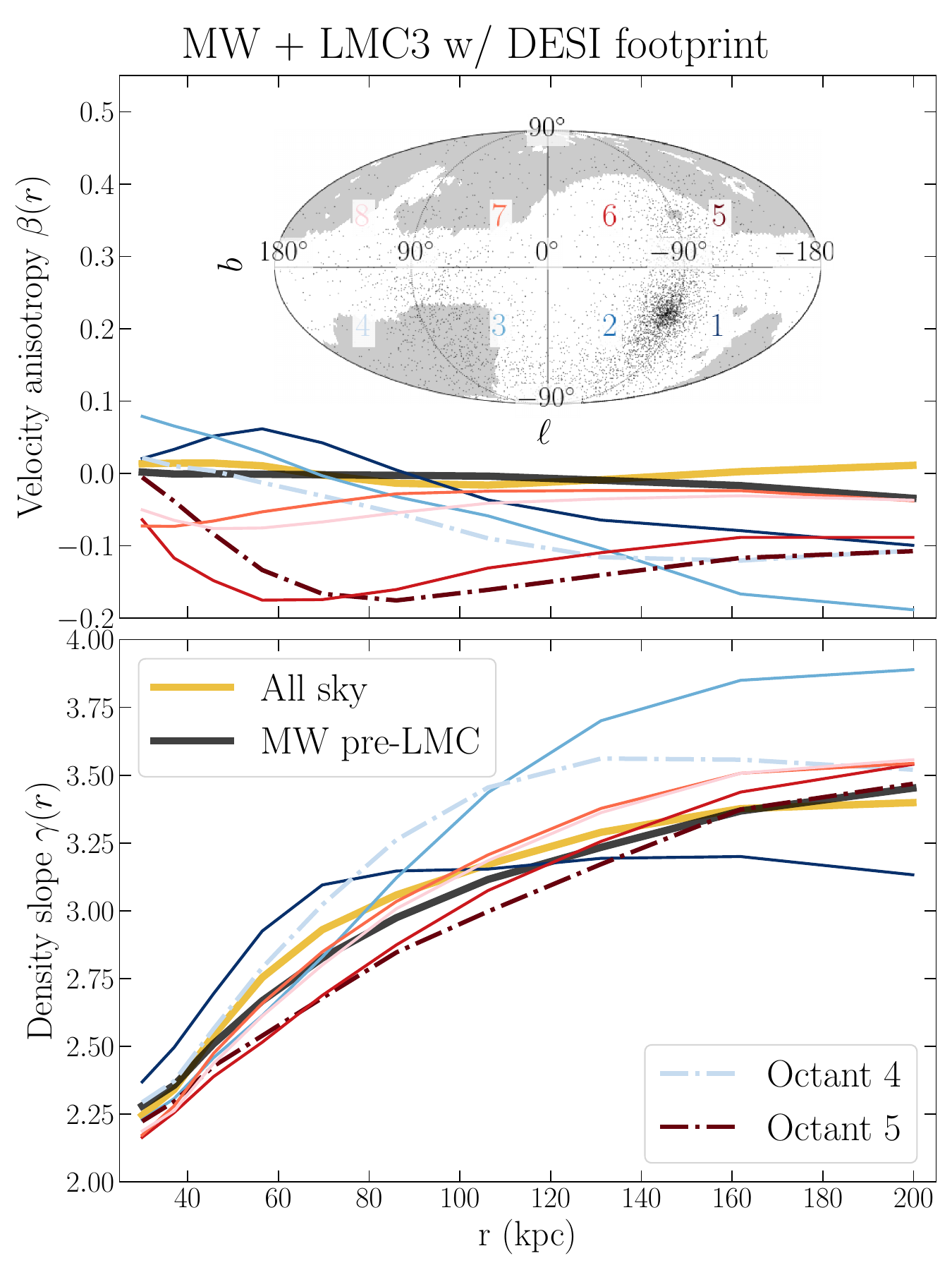}
\caption{Velocity anisotropy and density slope of MW DM particles within the DESI footprint (gray area in the inset plot) in the fiducial simulation. Northern octants, well covered by DESI, are not significantly different from Fig.~\ref{fig:beta_gamma_octs_MWLMC5_b0}. However, southern octants show significant differences.}
\label{fig:DESI_footprint_beta_gamma_octs_MWLMC5_b0}
\end{figure}

Fig.~\ref{fig:DESI_footprint_beta_gamma_octs_MWLMC5_b0} shows the velocity anisotropy and density slope for our fiducial model, for MW DM particles separated by octants and within the DESI-MWS footprint. For details on the calculation of the volume of these regions, see Appendix~\ref{sec:vol_desi}. Note the lack of a curve for octant 2, which is not covered by DESI. 

We find that the Northern octants are not significantly affected by the footprint, showing only minor differences with respect to Fig.~\ref{fig:beta_gamma_octs_MWLMC5_b0}. In the South, octant 3 now shows a density profile significantly steeper than if it had full coverage. Interestingly, even though octant 1 has limited coverage in DESI-MWS, the simulation data still reveals the larger steepness of the density profile over $50 \leq r/\kpc\leq 100$ associated with the DM wake trailing the LMC. This suggests that even with limited coverage, important features of MW's halo response to the LMC-MW encounter can be recovered. 

\begin{figure}
\centering
\includegraphics[width=0.9\columnwidth]{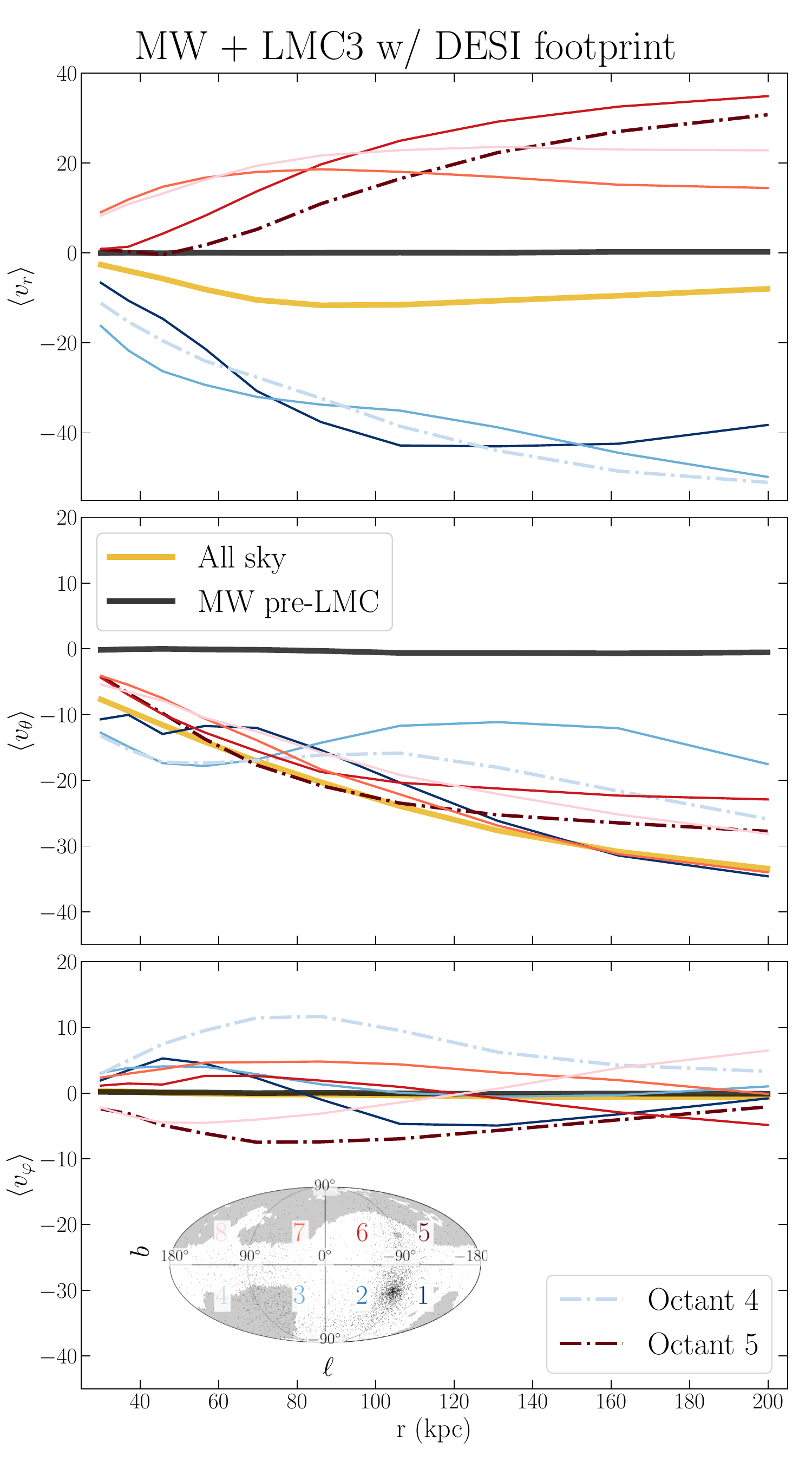}
\caption{Mean velocity components of DM particles selected within the DESI footprint (gray area in the inset plot) for the fiducial simulation. These velocities are somehow different from those in the absence of the footprint (Fig.~\ref{fig:mean_vr_vtheta_vphi_octs_MWLMC5_b0}), but the trends with radius are qualitatively similar.}
\label{fig:DESI_footprint_mean_vr_vtheta_vphi_octs_MWLMC5_b0}
\end{figure}

Fig.~\ref{fig:DESI_footprint_mean_vr_vtheta_vphi_octs_MWLMC5_b0} shows the mean velocities within the DESI footprint, separated by octants. The resulting velocities are qualitatively similar to the full-coverage case (Fig.~\ref{fig:mean_vr_vtheta_vphi_octs_MWLMC5_b0}), but with somewhat larger $\langle v_r\rangle$ amplitudes. Octants 7 and 8 still show the smallest overall $\langle v_r\rangle$.

\begin{figure}
\centering
\includegraphics[width=0.9\columnwidth]{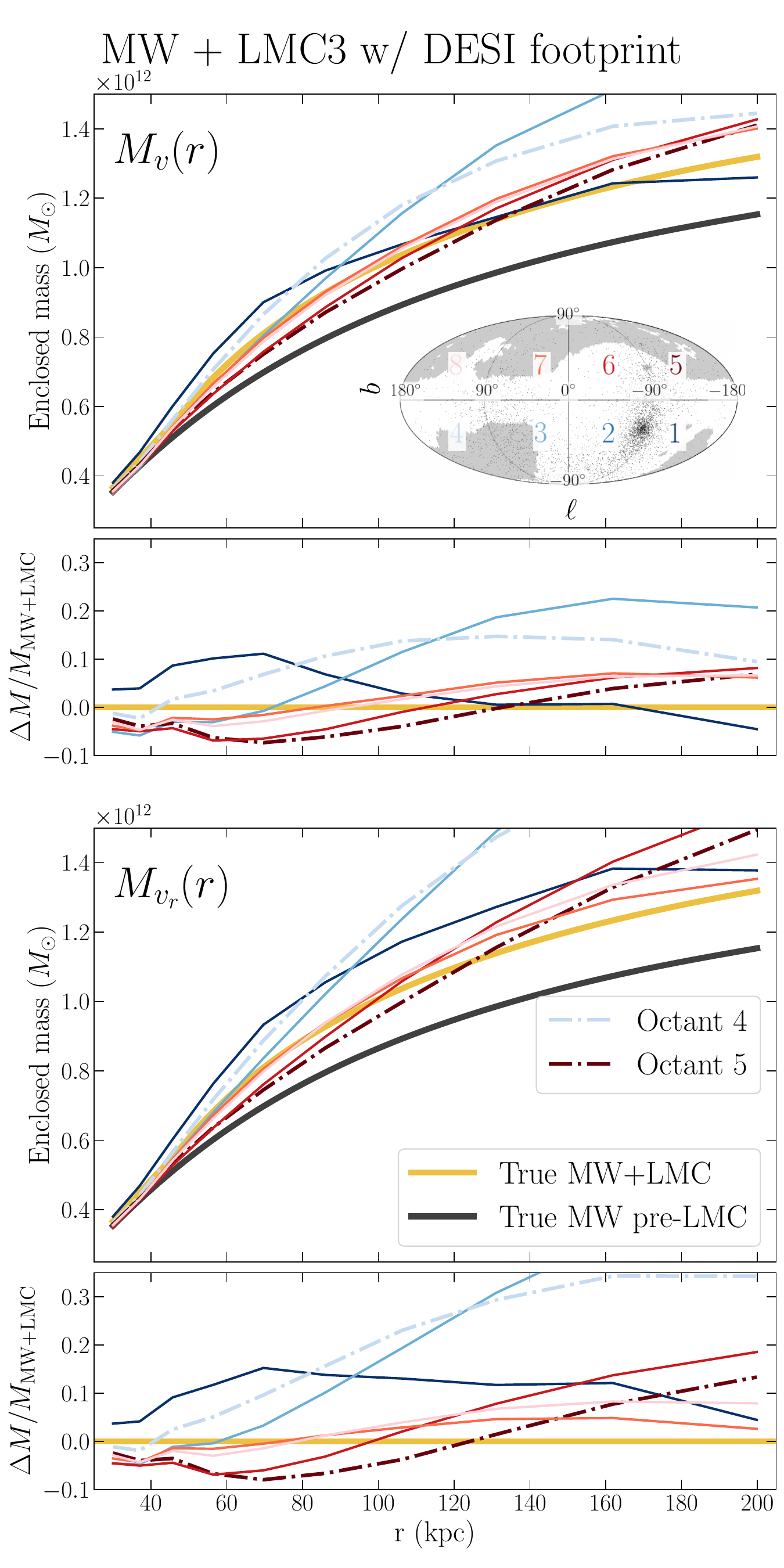}
\caption{Mass estimates for the fiducial simulation masked by the DESI-MWS footprint (grey region in the inset plot) and separated by octants, as in Fig. \ref{fig:M_octs_MWLMC5_b0}. Octants 7 and 8 still reproduce the total MW+LMC mass profile within $\sim 5\%$ for $r\lesssim 100$ kpc and within $\sim 10\%$ further out. That is also true for all simulations (not shown).}
\label{fig:DESI_footprint_M_octs_MWLMC5_b0}
\end{figure}

Fig.~\ref{fig:DESI_footprint_M_octs_MWLMC5_b0} shows the mass estimates for MW DM particles within the DESI-MWS footprint and separated by octants. Although some octants produce strongly-biased mass estimates, 7 \& 8 once again produce excellent estimates. In particular, their $M_{v_r}$ estimates still reproduce the total MW+LMC mass within $\sim 5\%$ for $r\lesssim 100$ kpc and within $\sim 10\%$ further out for all simulations. This exercise supports the validity of our proposed methodology (focusing on $M_{v_r}$ estimates in octants 7 \& 8 to recover the combined MW+LMC mass) using survey data with limited coverage. Having validated our method for the DESI-MWS footprint, in the next section we test the method using DESI-MWS data.

\subsection{The DESI-MWS sample of BHB stars}
\label{sec:BHB_sample}
In this section, we apply our proposed method to recover the combined MW+LMC mass using the DESI-MWS sample of BHB stars, made publicly available with its first data release \citep{Bystrom2025, Koposov2026}. The full sample contains 6,327 BHB stars, extending to Galactocentric distances of $\approx 100 \kpc$. To best recover the LMC and pre-LMC Infall MW mass, a sufficient sample of stars are needed at  distances larger than $100 \kpc$. As such, we limit our analysis to recovering the combined MW+LMC mass.  

From the full sample, we select 
4,161 BHB stars with \texttt{SURVEY\,=\,main}, \texttt{PROGRAM\,=\,bright}, \texttt{RVS\_WARN==0} and with measured \textit{Gaia} proper motions. We remove candidate members of the Sgr stream following the approach of \cite{Bystrom2025}: we use the stream coordinates ($\tilde{\Lambda}, \tilde{B}$) of \cite{Belokurov2014}, and select Sgr candidates as $|\tilde{B}| < 15^\circ$ along the stream and within $\pm 3\sigma$ of the stream, where $\sigma$ is the standard deviation of the distance to the stream at a given location, as identified by \cite{Hernitschek2017}. The Sgr stream region is shown as red/blue patches in Fig.~\ref{fig:DESI_footprint}. A video illustrating the Sgr region can be accessed \href{https://lberaldoesilva.github.io/videos/sgr_tube_spin.mp4}{here}. 

The primary targets of the DESI-MWS main survey (DESI bright) were selected from \textit{Gaia} in the apparent magnitude range $16\lesssim r \lesssim 19$, where the completeness is expected to be approximately uniform \citep[see Fig. 16 of][]{Koposov2026}. We focus on this magnitude range to avoid complicated selection effects. For an absolute magnitude $M\sim 0.6$ for BHB stars, the apparent magnitude range translates into heliocentric distances $12\lesssim d/\kpc \lesssim 50$, and given the Solar distance to the Galactic center $R_0\approx 8 \kpc$, the upper limit translates to Galactocentric distances $r\lesssim 60$ kpc. To avoid possible biases due to extra substructures, we also require $r\gtrsim 50$ kpc  -- see \cite{Bystrom2025}, and Table 1 in \cite{Amarante_2024} for a compilation of overdensities and their distances. This also avoids the highly anisotropic component associated with the Gaia-Sausage-Enceladus \citep{Belokurov2018, Helmi2018}, which is negligible beyond $\sim 30$ kpc and would require a kinematic modeling on its own \citep[][]{Lancaster2019}.

Thus, for the remaining analysis, we select BHB stars with $50\leq r/\kpc\leq60$, corresponding to 
245 stars before and  
131 stars after the Sgr removal. In this radial range, the Sgr-removed volume represents $\approx$11\% of the full footprint, $\approx$4\% for octants 7 \& 8, and $\approx$31\% for octants 5 \& 6 (the on-sky Sgr blue/red bands in Fig.~\ref{fig:DESI_footprint} appear larger because, as a projection, they show the stream at all distances, whereas only the portion at $r\simeq50$–$60\kpc$ removes volume from this shell).

To estimate the MW+LMC mass with Eq.~\eqref{eq:mass_est_v} or Eq.~\eqref{eq:mass_est_vr}, we need estimates of the total mass slope $\alpha$, the velocity anisotropy $\beta$, and the tracer density slope $\gamma$. In this analysis we set either $\alpha =1$ or $\alpha=0.5$ as reasonable choices, and estimate $\gamma$ and $\beta$ from the sample. While $\gamma$ is simply estimated via number counting and proper volume calculation (see appendix \ref{sec:geom_los} for details, including the Sgr removal), the velocity anisotropy, Eq.~\eqref{eq:beta}, requires estimates of velocity dispersions.

\subsection{Velocity anisotropy from error deconvolution}
\label{sec:beta_deconv}

The spherical Galactocentric velocity components $(v_r,v_\theta,v_\phi)$ of each BHB star are obtained from its sky position, heliocentric distance $d$, proper motions $(\mu_{\alpha*},\mu_\delta)$, and line-of-sight velocity $v_{\rm los}$. The dominant source of uncertainty is the proper motion: an error $\sigma_\mu$ at
distance $d$ propagates to a tangential-velocity error
\begin{equation*}
  \sigma_{v_t} \simeq 4.74\,
  \left(\frac{\sigma_\mu}{\mathrm{mas/yr}}\right)
  \left(\frac{d}{\kpc}\right)\ \mathrm{km/s} .
  \label{eq:vt_err}
\end{equation*}
Our sample stars typically have $\sigma_\mu\sim0.4\ \mathrm{mas/yr}$, yielding $\sigma_{v_t}\sim100\ \mathrm{km/s}$, comparable to the expected intrinsic dispersions. Evaluating the dispersions at face value therefore would inflate the tangential terms and strongly bias the velocity anisotropy towards strongly negative (tangential) values -- see \cite{Barbosa_2026} for a recent discussion, albeit focused on the effect of halo substructures.

To avoid this bias, we deconvolve the measurement errors with a maximum-likelihood estimator. For each star $j$ we build its $3\times3$ spherical Galactocentric velocities' covariance matrix $\mathbf{C}_j$: we draw ${N_{\rm samp}=1000}$ samples of $(\mu_{\alpha*},\mu_\delta)$ from the \textit{Gaia} error ellipse---including the correlation $\rho(\mu_{\alpha*},\mu_\delta)$---, transform each sample to $(v_r,v_\theta,v_\phi)$, and take the sample covariances. The errors on sky position, distance and $v_{\rm los}$ are subdominant in our radial range and are neglected in $\mathbf{C}_j$.

Within a radial bin, the spherical velocities are modeled as draws from a Gaussian of mean $\boldsymbol{\mu}$ and intrinsic velocity-dispersion matrix $\mathbf{S}$, assumed diagonal in the spherical frame,
$\mathbf{S}=\mathrm{diag}(\sigma_r^2,\sigma_\theta^2,\sigma_\phi^2)$, convolved with each star's measurement covariance: ${\mathbf{v}_j \sim
  \mathcal{N}\!\left(\boldsymbol{\mu},\,\mathbf{S}+\mathbf{C}_j\right)}$.
Thus the log-likelihood is
\begin{equation}
\begin{split}
  \ln\mathcal{L}(\boldsymbol{\mu},\mathbf{S}) = -\frac{1}{2}\sum_j
  \Big[&(\mathbf{v}_j-\boldsymbol{\mu})^{\!\top}
    (\mathbf{S}+\mathbf{C}_j)^{-1}
    (\mathbf{v}_j-\boldsymbol{\mu})\\
  &+
  \ln\big|2\pi(\mathbf{S}+\mathbf{C}_j)\big|\Big].
\end{split}
  \label{eq:loglike}
\end{equation}
We maximize Eq.~\eqref{eq:loglike} over $\boldsymbol{\mu}$ and $\mathbf{S}$ with the Nelder–Mead method, obtaining the deconvolved (intrinsic) dispersions $\mathbf{S}$ and, hence, a meaningful estimate of $\beta$. To estimate its uncertainties, we run the same procedure and maximize Eq.~\eqref{eq:loglike} for each of 1000 bootstrap realizations of the observed sample.

We obtain intrinsic dispersions ${\sigma_r=92.5\pm4.5\ \kms}$,
$\sigma_\theta=84.5\pm18.1\ \kms$, $\sigma_\phi=93.8\pm21.6\ \kms$
and finally $\beta = 0.07\pm0.29$, i.e.\ a near-isotropic halo.
This seems in disagreement with the more radially biased values of
\cite{Bird2021} using either BHBs or K giants, but compatible with
the closer-to-isotropic $\beta \sim 0.2$ measured by \cite{Lancaster2019} with BHBs at $r>30$ kpc or with values of \cite{Li2026} using K giants from
DESI DR2. The dispersion values can be compared with the face-values $\sigma_\theta\simeq132\ \kms$ and
$\sigma_\phi\simeq141\ \kms$, which are inflated well above the
radial value $\sigma_r\simeq95\ \kms$ purely by the proper-motion
errors.

\subsection{Observational constraints on the MW+LMC mass from DESI-MWS BHBs}
\label{sec:MW_LMC_mass}

Sec.~\ref{sec:beta_deconv} exposes the errors introduced by poorly constrained proper-motions at large radii. The same discussion applies to the mass estimates. The $M_v$ estimator, Eq.~\eqref{eq:mass_est_v}, is based on the full 3D velocity and is therefore prone to the same errors, while $M_{v_r}$ is only weakly dependent on proper motions since $v_r\approx v_\mathrm{los}$ at large radii. Furthermore, in Secs.~\ref{sec:results} and \ref{sec:desi_footprint} we demonstrated that smaller reflex motions in the radial direction, $\langle v_r\rangle$, result in better accuracy of $M_{v_r}$. Therefore, in the following analysis, we focus on the $M_{v_r}$ estimator. 

Having set $\alpha$, with $\gamma$ obtained from number counts and with $\beta$ estimated as described in Sec.~\ref{sec:beta_deconv}, the mass follows directly from Eq.~\eqref{eq:mass_est_vr}. We estimate its uncertainty by generating $N_{\rm boot}=1000$ Monte-Carlo realizations of the
observed sample, each one combining a bootstrap resampling of the stars with random draws from the distance and anisotropy error distributions.

In each realization, we proceed as follows: \emph{i}) We
bootstrap resample our full dataset (before the radial cut) with replacement, which captures the finite-sample (shot) noise of the tracer population and perturbs both the counts in the radial bin of interest and the set of stars entering the velocity moment $\langle v_r^2\,r^\alpha\rangle$. \emph{ii}) We fold in the dominant BHB distance uncertainty, which originates in the scatter of their absolute magnitude. A BHB distance is obtained from the distance modulus, $d = 10^{\,(m_g - M_g + 5)/5}$, in which the apparent magnitude $m_g$ is measured precisely but $M_g$ carries an intrinsic scatter $\sigma_{M_g}=0.133$ mag \citep{Bystrom2025}. This produces a relative uncertainty $\sigma_d/d \approx6\%$. We add this random Gaussian error to $M_g$ for each star, recompute its heliocentric distance and its Galactocentric radius, so that stars can migrate between radial bins as their distances scatter. \emph{iii}) We draw the anisotropy in each bin from $\mathcal{N}(\hat\beta,\sigma_\beta)$, with $(\hat\beta,\sigma_\beta)$ taken from the deconvolution of Sec.~\ref{sec:beta_deconv}. From the resampled, distance-perturbed sample, we recompute the tracer density slope $\gamma$ from the new counts, and estimate $M_{v_r}$, Eq.~\eqref{eq:mass_est_vr}.

In this procedure, the bootstrap resampling propagates the \emph{statistical} variance from the finite number of tracers and is the leading error source in $\gamma$; the distance scatter (step \emph{ii}) contributes secondarily. The Gaussian draws for the distances and the anisotropy (steps \emph{ii} and \emph{iii}) propagate the \emph{observational} uncertainties: for $\beta$ in particular the uncertainty is set almost entirely by the proper-motion errors encoded in $\mathbf{C}_j$ rather than by sample size, so it is drawn directly from its error distribution rather than re-estimated within each bootstrap. Line-of-sight velocities are held fixed throughout, since $v_{\rm los}$ is measured to high precision ($\langle \sigma_{v_\mathrm{los}}\rangle \approx 1.8 \kms$), thanks to the excellent DESI-MWS data, and is effectively noise-free relative to the tangential velocities. 

\begin{table}
\centering
\renewcommand{\arraystretch}{1.3}
\begin{tabular}{l c c c c c c}
\hline\hline
Octs. & Sgr & $N$ & $\beta$ & $\gamma$ & $\alpha$ &
$\frac{M_{v_r}}{10^{11}\Msun}$ \\
\hline
\multirow{4}{*}{All}
  & \multirow{2}{*}{incl.} & \multirow{2}{*}{245}
  & \multirow{2}{*}{$-0.47^{+0.38}_{-0.38}$} & \multirow{2}{*}{$5.11^{+0.46}_{-0.49}$}
  & $1$   & $6.00^{+1.05}_{-0.93}$ \\
  & & & & & $0.5$ & $5.83^{+1.09}_{-0.92}$ \\
\hhline{~------}
  & \multirow{2}{*}{excl.} & \multirow{2}{*}{131}
  & \multirow{2}{*}{$+0.09^{+0.30}_{-0.32}$} & \multirow{2}{*}{$3.39^{+0.69}_{-0.67}$}
  & $1$   & $5.20^{+1.41}_{-1.27}$ \\
  & & & & & $0.5$ & $4.82^{+1.44}_{-1.34}$ \\
\hline
\multirow{4}{*}{7\&8}
  & \multirow{2}{*}{incl.} & \multirow{2}{*}{53}
  & \multirow{2}{*}{$-0.73^{+0.80}_{-0.78}$} & \multirow{2}{*}{$5.79^{+0.96}_{-1.08}$}
  & $1$   & \cellcolor{gray!15}$6.91^{+2.16}_{-1.80}$ \\
  & & & & & $0.5$ & \cellcolor{gray!15}$6.84^{+2.22}_{-1.87}$ \\
\hhline{~------}
  & \multirow{2}{*}{excl.} & \multirow{2}{*}{44}
  & \multirow{2}{*}{$-0.76^{+0.91}_{-0.96}$} & \multirow{2}{*}{$5.69^{+1.22}_{-1.09}$}
  & $1$   & \cellcolor{gray!15}$7.58^{+2.90}_{-2.17}$ \\
  & & & & & $0.5$ & \cellcolor{gray!15}$7.49^{+2.97}_{-2.25}$ \\
\hline
\multirow{4}{*}{5\&6}
  & \multirow{2}{*}{incl.} & \multirow{2}{*}{156}
  & \multirow{2}{*}{$-0.79^{+0.75}_{-0.77}$} & \multirow{2}{*}{$5.34^{+0.61}_{-0.52}$}
  & $1$   & $4.95^{+1.48}_{-1.14}$ \\
  & & & & & $0.5$ & $4.88^{+1.50}_{-1.17}$ \\
\hhline{~------}
  & \multirow{2}{*}{excl.} & \multirow{2}{*}{51}
  & \multirow{2}{*}{$+0.59^{+0.36}_{-0.41}$} & \multirow{2}{*}{$0.06^{+1.20}_{-1.37}$}
  & $1$   & --- \\
  & & & & & $0.5$ & --- \\
\hline\hline
\end{tabular}
\caption{Combined MW+LMC mass within $r<60\kpc$ estimated from DESI-MWS DR1 BHB stars (selected with \texttt{SURVEY\,=\,main}, \texttt{PROGRAM\,=\,bright} and \texttt{RVS\_WARN==0}) in $50\leq r/\kpc\leq60$ -- see Secs.~\ref{sec:BHB_sample}-\ref{sec:MW_LMC_mass} for details. Column~1 indicates the on-sky region (see Fig.~\ref{fig:DESI_footprint}); column~2 whether the Sgr stream region is included in or excluded \citep[selection as in][]{Bystrom2025}; column~3 the sample size; columns~4--5 the deconvolved anisotropy and density slope; column~6 the assumed ${\alpha\equiv1-\dd\ln M(r)/\dd\ln r}$; column~7 the $M_{v_r}$ mass estimate, Eq.~\eqref{eq:mass_est_vr}. Highlighted in gray are our most reliable estimates. The lower mass estimates obtained from octs. 5 \& 6, compared with those from octs. 7 \& 8, support a first-infall scenario. The oct.~5\&6 / Sgr-excl.\ rows yield unphysical estimates ($\gamma\approx0$, $\beta>0$) because the Sgr removal is too drastic, leaving a strongly radially biased subsample.}
\label{tab:BHB_results}
\end{table}

Table \ref{tab:BHB_results} shows the $16$th/$50$th/$84$th percentiles of $\beta$, $\gamma$ and $M_{v_r}(r<60\, \kpc)$ across realizations. After removing likely Sgr stream members, we are left with 131 stars in $50 < r/\kpc < 60$, for which we measure a density slope 
$\gamma = 3.39^{+0.69}_{-0.67}$, marginally compatible with recent measurements \citep{Medina_2024, Amarante_2024}, though somewhat shallower. Setting $\alpha=1$ ($\alpha=0.5$) our total MW+LMC mass estimate is $M_{v_r}(r<60\,\kpc)=6.00^{+1.05}_{-0.93}\times10^{11}\,\Msun$ ($5.83^{+1.09}_{-0.92}\times10^{11}\,\Msun$). However, as mentioned before, some octants can be heavily affected by the Sgr removal, and we take this all-octants estimate with caution.

In Fig.~\ref{fig:DESI_footprint_M_octs_MWLMC5_b0}, we showed that at this radius, octants 7 \& 8 should provide mass estimates within $5\%$ or better for error-free data, even considering the DESI footprint. We also showed that the Sgr removed volume fraction is only $4\%$ of the footprint in these octants (see Fig.~\ref{fig:DESI_footprint}), so they provide our most reliable MW+LMC mass estimates. Restricting our (Sgr-cleaned) BHB sample to these octants 
(44 stars), we estimate a total MW+LMC mass 
$M_{v_r}(r<60\, \kpc) = 7.58^{+2.90}_{-2.17}\times10^{11}\,\Msun$
($7.49^{+2.97}_{-2.25}\times10^{11}\,\Msun$) for $\alpha=1$ ($\alpha=0.5$). These values, highlighted in gray in Table~\ref{tab:BHB_results}, remain consistent with those based on all octants within uncertainties.

\section{Discussions}
\label{sec:discussions}

\subsection{The meaning and accuracy of the MW+LMC mass estimates}
\label{sec:meaning}

In Sec.~\ref{sec:results}, we demonstrated that the \citetalias{Watkins2010} mass estimators, applied to all-sky MW tracers of an MW+LMC simulation, recover the \emph{total} MW+LMC mass of all simulations, with varying degrees of accuracy. Due to the relatively small $\langle v_r\rangle$ reflex motion, the estimator $M_{v_r}$, Eq.~\eqref{eq:mass_est_vr}, recovers the total MW+LMC mass with $\lesssim 5\%$ relative errors for $30 \leq r/\kpc \leq 200$ without any reflex-motion correction. This accuracy is also achieved by the $M_v$ estimator once it is corrected for the reflex motion as proposed by \citetalias{Erkal2020}, Eq.~\eqref{eq:mass_est_v_corrected}. This is the same accuracy returned by the method using the MW pre-LMC infall simulations, indicating that the reported accuracy is at the level of the method's intrinsic uncertainties.

When applied to tracers in different octants on the sky ($\ell, b$), the mass estimators provide wildly different accuracies due to the highly asymmetric perturbation of the LMC, in line with the conclusions of \citetalias{Erkal2020}. In Southern octants, the accuracy of $M_{v_r}$ is worse than that of $M_v$, since the $\langle v_r \rangle$ reflex motion is larger in those octants. The small $\langle v_r \rangle$ determined in the all-sky calculation is 
due to a partial cancellation between Northern and Southern octants. However, we found that in the Northern octants 7 or 8, $M_{v_r}$ recovers the total MW+LMC mass with $\lesssim 5\%$ errors for $30 \leq r/\kpc \leq 200$, once again {\it without} any reflex-motion correction (Sec.~\ref{sec:octants}).

We interpret this octant-dependent accuracy as being due to the timeline of the LMC's particular trajectory (Fig.~\ref{fig:timeline_Mv}). Before the LMC arrival, tracers in all octants trace the MW mass. As the LMC approaches the MW, all octants start providing mass estimates that are higher than the MW mass for Galactocentric radii larger than LMC's instantaneous distance. Octants 7\&8 are crossed by the LMC the longest ago, when the LMC was still far from the MW center and perturbing the MW outer halo nearly adiabatically. As such, octants 7\&8 are the first to trace the total MW+LMC mass and thus provide the most reliable measurements of the MW+LMC mass at present day.

Octants in the future LMC's trajectory (such as 5\&6 throughout the full simulation and 1\&2 for $t\leq 2$ Gyr) are out-of-equilibrium tracers of the MW mass. They provide mass estimates intermediate between the original MW and MW+LMC. Once the LMC enters an octant, tracers are strongly perturbed and provide poor MW+LMC mass estimates while the LMC is still in that region of the sky. This is the case of octant 3 for $t\geq 1.6$ Gyr, which is currently re-equilibrating as a tracer of the MW+LMC mass. At present day, the LMC resides in octants 1\&2, which are consequently strongly perturbed and should not be used for mass estimates. 

Although our results are based on the specific mass estimators of \citetalias{Watkins2010}, the coherent picture just described suggests that qualitatively similar conclusions would be obtained by any equilibrium-based mass estimator. This picture provides physical intuition for the dynamical state of the MW's stellar halo across the sky, shedding light on the meaning of mass estimates obtained with MW halo tracers in different directions and clarifying a possible source of reported discrepancies between different MW mass estimates \citep[see e.g.][]{Wang2020}. 

While we agree with \cite{Vasiliev2023} that quantitative agreement of mass estimates with the MW+LMC mass profile depends on the LMC orbit, our results and theoretical picture frame this as a consequence of the LMC's trajectory, rather than just a coincidence (which would be the case in a scenario of complete ignorance about the LMC path). In this work, we have assumed that the LMC is on first infall \citep{Besla2007}. The trajectory of the LMC through the Southern hemisphere is agnostic to this assumption. Gaia/HST proper motions require the LMC to travel from octants (present day) $1>2>3>4$ (past) direction over roughly the same timescale \citep{Besla2026}. If the LMC is on its second pericentric approach to the MW \citep[e.g.][]{Vasiliev2024}, our conclusion that octants 7 \& 8 will reasonably recover the MW+LMC combined mass would still be unchanged, as those octants would still have had enough time to equilibrate after the LMC passed through them. 

In a second passage scenario, the LMC will have previously traveled through octants 5\&6, which could in principle result in a different mass estimate than a first infall scenario. However, in a second passage scenario the previous pericentre occurs 5--10 Gyr ago, and the LMC subsequently travels to an apocenter of 300--400 kpc \citep{Vasiliev2024}, such that octants 5\&6 are expected to have had time to re-equilibrate. Indeed, \citet{Vasiliev2024} find that the all-sky density and line-of-sight velocity perturbations of the MW halo are nearly indistinguishable between first- and second-passage simulations, with any signature of the earlier passage having dissipated or been obscured by the most recent one. We therefore expect our proposed methodology to use octants 5\&6 to recover the present day LMC mass profile to also be valid for either LMC orbit scenario.

Interestingly, this means that we generically expect the mass estimate recovered in octants 5\&6 to be lower than that in octants 7\&8.
In Table \ref{tab:BHB_results}, we estimated the mass using $M_{v_r}$ in octants 5\&6 using DESI-MWS DR1 BHB stars and do indeed find lower mass estimates than in octants 7\&8. This is consistent with our expectations because the LMC is currently moving towards octants 5\&6. Following Fig. \ref{fig:timeline_Mv}, the recovered mass estimate in octants 5\&6 should be between the pre-LMC Infall MW mass and the combined LMC+MW mass. However, this estimate does not exclude the Sgr stream region; when excluded, the estimator yields unphysical results. This analysis should be revisited with future data sets and improved algorithms for Sgr removal.

More testing is needed to determine if there could be signatures in the kinematics of stars in octants 5\&6 at large radii that could distinguish between a first or second passage scenario. Currently the strongest support for an LMC first infall scenario is the SMC's high gas content and proximity to the LMC at the present day. In a second passage scenario, MW tides will disrupt a binary LMC-SMC state at the previous pericentric passage and force the SMC to have originally been a long term MW satellite where it would be subject to environmental quenching \citep{Besla2026}. 

Finally, we stress that our mass estimates were carried out by assuming that the stellar halo is smooth throughout to 200 kpc. In the MW, this is certainly not the case beyond 50 kpc. Further work is needed to test the methodology developed in this work with more realistic, non-smooth stellar halos.

\subsection{Comparison with \citetalias{Erkal2020}}
\label{sec:compare_EB20}

\citetalias{Erkal2020} presented a new methodology to recover the pre-LMC Infall MW mass using the $M_v$ estimator and a correction for the reflex motion induced by the LMC. In our study, we expand on this work, exploring a similar range of LMC/MW mass models and orbits, but also adopting the $M_{v_r}$ estimator. In general, our findings using the $M_v$ estimator qualitatively recover the results of \citetalias{Erkal2020}, but we report some quantitative discrepancies. Further, in our work, we present a different framework to interpret what we are fundamentally learning from the output of mass estimators using tracers in different regions of the sky (see Sec. \ref{sec:meaning}). 

\citetalias{Erkal2020} concluded that $M_v$ mass estimates are biased high, with respect to the MW mass, by up to $50\%$ in the halo outskirts for models with their most massive LMC. They also concluded that this bias is mostly due to the reflex motion of the outer stellar halo as observed from the MW disc, in response to the arrival of the LMC. When the estimator is corrected for this reflex motion, Eq.~\eqref{eq:mass_est_v_corrected}, and applied to tracers with $\ell < 0^\circ, b>0^\circ$ (our octants 5 \& 6), \citetalias{Erkal2020} recover the pre-LMC infall MW mass with $\lesssim 5\%$ errors. They also conclude that the $M_v$ estimates (with no reflex-motion correction) can be biased even compared with the total MW+LMC mass, reporting a bias of up to $\sim 20\%$ in the halo outskirts of models with the most massive LMC. Finally, they argue that the reflex-motion-corrected mass estimates should not be interpreted as returning the MW+LMC total mass, but rather the MW mass pre-LMC infall.

In our work, we compare mass estimates with the \emph{total} MW+LMC mass profiles of six simulations, which combine two different MW halo velocity anisotropy profiles with three LMC masses. The errors of our $M_v$ estimates are $\lesssim 10\%$, i.e. smaller than the $20\%$ errors reported by \citetalias{Erkal2020} for the outskirts of the model with the most massive LMC. \citetalias{Erkal2020} modeled the MW and the LMC as single particles sourcing their respective potentials, with stellar halo particles tracing this combined potential. In our study, we employ the full $N$-body simulations of \citetalias{Garavito-Camargo2019}, using MW DM particles as tracers. This distinction in method likely accounts for our reported quantitative discrepancies. 

We follow \citetalias{Erkal2020} to correct the $M_v$ estimator for the reflex motion and recover the total MW+LMC mass profiles with $\lesssim 5\%$ errors. This improvement confirms the conclusion of \citetalias{Erkal2020} that the mass estimate bias is mostly due to the reflex motion. However, 
we interpret this as a bias with respect to the \emph{total} MW+LMC mass, rather than a bias on the pre-LMC infall MW mass.

We then used the $M_{v_r}$ estimator (with no reflex-motion correction) to recover the total MW+LMC mass based on all-sky particle data with $\lesssim 5\%$ errors. This reinforces the conclusion that the bias is mostly driven by the reflex motion, since $\langle v_r \rangle < \langle v_\theta \rangle$ for all-sky data. This result also highlights a significant advantage over $M_v$, where accurate results can be recovered without adding a reflex motion correction.

Also in agreement with \citetalias{Erkal2020}, we find that different regions on the sky provide mass estimates with varying biases, driven by differences in the local reflex motion. \citetalias{Erkal2020} estimate the MW mass with $\lesssim 5\%$ errors with the $M_v$ estimator applied to $\ell < 0^\circ, b>0^\circ$ and corrected for the reflex motion in this region. We also find that this region of the sky provides the closest estimate to the pre-LMC Infall MW mass. However, when following the same method as \citetalias{Erkal2020}, we only recover the MW mass pre-LMC within $\lesssim 12\%$ errors. We interpret this as a result of octants 5 and 6 being in the future LMC trajectory and in the process of responding to the LMC, but still not having relaxed to the combined potential due to the MW+LMC. In this sense, our result from octants 5 \& 6 better fit \citetalias{Erkal2020}'s interpretation of a biased tracer of the pre-LMW Infall MW mass (as opposed to the total MW+LMC).

Besides possible differences in the LMC orbit and halo kinematics with respect to our simulations, \citetalias{Erkal2020} did not model the full MW and LMC $N$-body dynamics, thus missing e.g. the effect of the LMC mass loss. We suspect that our $5-15\%$ differences with respect to \citetalias{Erkal2020} is likely due to second-order dynamical effects that may be missed by those simplifications.


\subsection{Extrapolating from the Inner Halo to $M_{200}$}
\label{sec:virial}
\begin{figure}
\centering
\includegraphics[width=0.9\columnwidth]{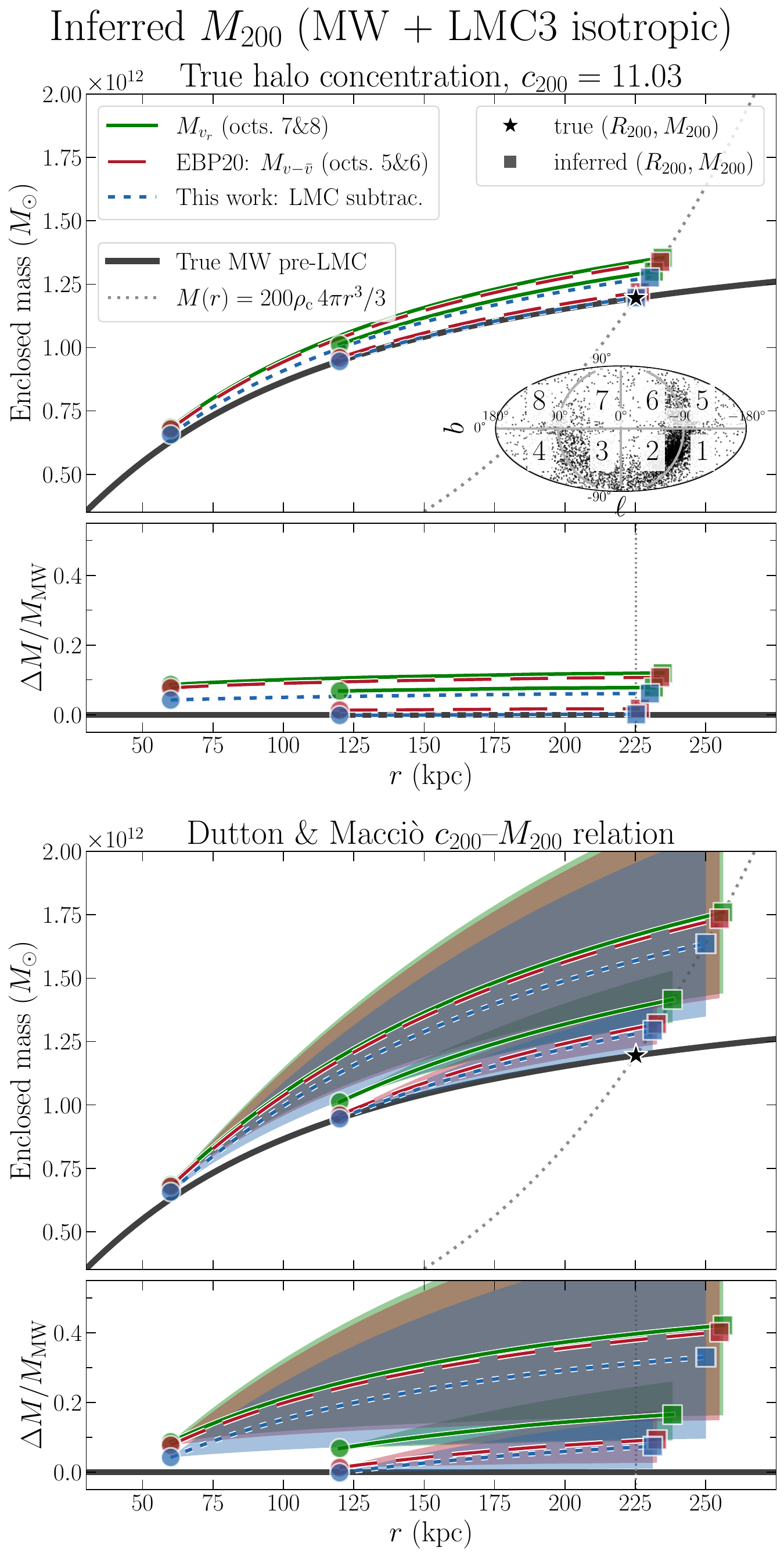}
\caption{Inferred $M_{200}$ (squares) for our fiducial simulation based on estimates of the pre-LMC infall MW mass at $r=60$ kpc or 120 kpc (circles). Each color and line style represents a different estimator (see main text). The method proposed in this work (blue), where the LMC mass is estimated and subtracted, provides the best estimates. This is closely followed by the method proposed by \citetalias{Erkal2020} (red), where the MW mass is estimated using $M_v$ with tracers in octants 5\&6 and corrected for the reflex motion. The extrapolation to $(R_{200}, M_{200})$ is done with the true $c_{200}$ (upper panels) or a $c_{200}-M_{200}$ relation \citep{Dutton2014} (bottom panels) with a corresponding 0.11 dex scatter (shaded areas). While the true $c_{200}$ approximately propagates the anchor mass errors to $M_{200}$, the use of a $c_{200}-M_{200}$ relation can significantly degrade the $M_{200}$ inference if the halo is even mildly atypical for its mass. Our simulated halo is $\sim 1.2\sigma$ off the assumed $c_{200}-M_{200}$ relation, extrapolating to $M_{200}$ with $\sim 37\%$ errors if anchored with our method at 60 kpc, but improving to $\sim 10\%$ if anchored at 120 kpc.}
\label{fig:extrapolate_virial}
\end{figure}

The MW virial mass is perhaps the most fundamental parameter to put the MW in context with cosmological simulations. We define the total enclosed mass
\begin{equation}
    M_{200} =200\rho_{\rm c}\,(4\pi/3) R_{200}^3,
    \label{eq:M200}
\end{equation}
 where $R_{200}$ is the radius of the sphere inside which the mean density is $200\rho_\mathrm{c}$, where $\rho_\mathrm{c} = 277.5 h^2\,\Msun\,\kpc^{-3} = 124.94\,\Msun\,\kpc^{-3}$ is the critical density of the universe (for $h=0.671$ in the Planck cosmology adopted in this section). For our MW pre-LMC infall models, we have\footnote{While we use $h=0.671$ and include the disc and bulge masses in the calculation of $R_{200}$, \citetalias{Garavito-Camargo2019} use $h=0.7$ and only consider the DM halo mass to get $R_{200}$. Consequently, our values for $R_{200}$ and $M_{200}$ differ slightly from those reported by \citetalias{Garavito-Camargo2019}.} $M_{200}=1.195\times10^{12}\,\Msun$ and $R_{200}=225.2$\,kpc -- black star and vertical line in Fig.~\ref{fig:extrapolate_virial}.

We infer $M_{200}$ by anchoring an analytic mass profile on a mass estimate at a given distance (60 or 120 kpc). We adopt the analytic Hernquist profile, as this was used to generate our controlled MW halos. If we have an estimate for the concentration ($c_{200}$) of the MW halo pre-LMC infall, then $c_{200}\equiv R_{200}/r_{-2}$, where $r_{-2}$ is the radius at which the logarithmic density slope equals $-2$ ($r_{-2}=a/2$ for a Hernquist profile of scale radius $a$, so that $c_{200}=2R_{200}/a$).

A Hernquist halo of total mass $M_{\rm h}$ encloses
\begin{equation}
    M_{\rm h}(<r) = M_{\rm h}\,\frac{r^2}{(r+a)^2},
    \label{eq:hernquist_mass}
\end{equation}
so that a single mass estimate $M(<r_{\rm out})$, with e.g. $r_{\rm out}=60$ kpc, fixes its normalization once $a$ is known:
\begin{equation}
    M_{\rm h} = \left[M(<r_{\rm out}) - M_{\rm baryon}\right] (r_{\rm out}+a)^2/r_{\rm out}^2,
    \label{eq:hernquist_norm}
\end{equation}
where $M_{\rm baryon}=7.19\times10^{10}\,\Msun$ is the baryonic mass (disc and bulge). Writing $a=2R_{200}/c_{200}$ in Eq.~\eqref{eq:hernquist_mass} with $r=R_{200}$, we get
$M_{\rm h}(<R_{200})/M_{\rm h} = [c_{200}/(c_{200}+2)]^2$, and from Eq.~\eqref{eq:M200}, we get $R_{200}$ as the root of
\begin{equation}
\begin{split}
    200\rho_{\rm c}\frac{4\pi}{3}R_{200}^3 = {}
    &\left[M(<r_{\rm out}) - M_{\rm baryon}\right]
     \left(\frac{c_{200}}{c_{200}+2}\right)^{\!2} \\
    &\times\left(\frac{r_{\rm out}+2R_{200}/c_{200}}
                      {r_{\rm out}}\right)^{\!2} + M_{\rm baryon}.
\end{split}
\label{eq:virial_solve}
\end{equation}
$M_{200}$ then follows from Eq.~\eqref{eq:M200}. 

We follow the above procedure to derive $M_{200}$ under two assumptions for the concentration, $c_{200}$. First, we use the known concentration of the pre-LMC Infall MW from the actual simulation $c_{200}\equiv R_{200}/r_{-2}=11.03$. Second, we assume $c_{200}$ is unknown and  
adopt a $c_{200}-M_{200}$ relation calibrated on
cosmological simulations. Fig.~\ref{fig:extrapolate_virial} illustrates the accuracy of the inferred $M_{200}$ (squares) that can be achieved from mass estimates derived by three different methods applied to a single radial bin, either 50-60 kpc or 100-120 kpc (circles anchored at 60 and 120 kpc), using the fiducial simulation. The thick black line shows the true cumulative MW mass pre-LMC infall. 

In the first method (green), we estimate the mass at the given radii by applying $M_{v_r}$ to tracers in octants 7\&8. In Sec.~\ref{sec:results}, we demonstrated that these octants provide good estimates of the \emph{total} MW+LMC mass, but here we naively interpret it as an estimate of the MW mass pre-LMC infall. In the second method (red), we estimate the mass with $M_{v}$ corrected for the reflex motion applied to octants 5\&6 (the ones providing estimates closer to the pre-LMC infall MW mass - Sec.~\ref{sec:octants}), as proposed by \citetalias{Erkal2020}. The third method (blue) is the one we propose in this work: using the $M_{v_r}$ estimate of octants 7\&8 we subtract the LMC mass profile, which was estimated using the difference between $M_v$ estimates of octants 4 and 5 (sec.~\ref{sec:LMC}). 

All these mass estimates assume $\alpha=1$. In Fig.~\ref{fig:extrapolate_virial}, each resulting analytic Hernquist profile is anchored at the circle (60 or 120 kpc) and terminates at the estimated ($R_{200}, M_{200}$, squares), where it intersects the dotted curve. Our proposed method (3) most closely matches the true MW pre-LMC infall mass, closely followed by the method (2) of \citetalias{Erkal2020} -- see subpanels for residuals and Table~\ref{tab:virial_extrapolation} (Column 3) for specific numbers.

Using the true $c_{200}$, the error of the inferred $M_{200}$ is only
mildly amplified with respect to the error of the MW mass pre-LMC infall estimate that anchors it: $+7.8\%\to+11.9\%$ ($+1.3\%\to+1.9\%$) for the \citetalias{Erkal2020} estimate at $60$ kpc ($120$\,kpc), and $+4.3\%\to+6.8\%$ ($-0.2\%\to+0.1\%$) for our method (Table~\ref{tab:virial_extrapolation}, Column 4). The error is amplified by $\simeq1.5$ ($\simeq1.2$) when extrapolating from $60$\,kpc ($120$\,kpc), and represents the irreducible cost of the extrapolation, with $c_{200}$ perfectly known.

\begin{table}
\centering
\begin{tabular}{lcccc}
\hline\hline
Method & $r_{\rm out}$ & $\frac{\Delta M}{M_{\rm MW}}$ & \multicolumn{2}{c}{$\frac{\Delta M_{200}}{M_{200}}$ (\%)} \\
\cline{4-5}
 & (kpc) & (\%) &true $c_{200}$&DM14\\
\hline
\multicolumn{5}{c}{\bf MW + LMC3 isotropic} \\
1) $M_{v_r}$, octs.\ 7\&8 & 60 & $+8.7$ & $+13.3$ & $+47.2^{+49.9}_{-29.4}$ \\
 & 120 & $+6.8$ & $+8.7$ & $+18.4^{+11.5}_{-8.4}$ \\
2) $M_{v-\bar{v}}$, octs.\ 5\&6 & 60 & $+7.8$ & $+11.9$ & $+45.1^{+48.8}_{-28.7}$ \\
 & 120 & $+1.3$ & $+1.9$ & $+10.4^{+10.1}_{-7.4}$ \\
3) LMC subtracted & 60 & $+4.3$ & $+6.8$ & $+36.9^{+44.5}_{-26.4}$ \\
 & 120 & $-0.2$ & $+0.1$ & $+8.2^{+9.7}_{-7.1}$ \\
\hline
\multicolumn{5}{c}{\bf MW + LMC3 anisotropic} \\
1) $M_{v_r}$, octs.\ 7\&8 & 60 & $+11.8$ & $+18.3$ & $+55.3^{+54.3}_{-31.7}$ \\
 & 120 & $+7.4$ & $+10.0$ & $+19.9^{+11.8}_{-8.6}$ \\
2) $M_{v-\bar{v}}$, octs.\ 5\&6 & 60 & $+11.3$ & $+17.6$ & $+54.2^{+53.7}_{-31.4}$ \\
 & 120 & $+1.4$ & $+2.6$ & $+11.1^{+10.2}_{-7.5}$ \\
3) LMC subtracted & 60 & $+7.7$ & $+12.2$ & $+45.4^{+49.0}_{-28.9}$ \\
 & 120 & $+0.8$ & $+1.8$ & $+10.2^{+10.1}_{-7.4}$ \\
\hline\hline
\end{tabular}
\caption{Accuracy of estimates of the simulated MW mass pre-LMC infall,
from estimates at $r_\mathrm{out}=60$ and $120$\,kpc, and of the $M_{200}$ inferred from each, for the MW+LMC3 simulations. Methods: \emph{i}) the $M_{v_r}$ mass estimator, Eq.~\eqref{eq:mass_est_vr}, applied to octants 7\&8 (previously shown to actually provide MW+LMC mass estimates, but here naively interpreted as estimating the MW mass pre-LMC infall); \emph{ii}) the $M_v$ estimator applied to octants 5\&6 and corrected for the reflex motion, Eq.~\eqref{eq:mass_est_v_corrected}, as proposed by \citetalias{Erkal2020}; and \emph{iii}) the method proposed in this work, where we estimate the LMC mass profile as described in Sec.~\ref{sec:LMC} and subtract it from the previously obtained $M_{v_r}$ from octants 7\&8. Column 3 shows the relative error of the estimated mass at $r_\mathrm{out}$. Column 4 shows the relative error of $M_{200}$ inferred by assuming the true halo $c_{200}=11.03$, and Column 5 shows the error based on the $c_{200}-M_{200}$ relation of \citet{Dutton2014}. The upper block is for the fiducial simulation, whereas the bottom one is for its anisotropic equivalent. See main text for details.}
\label{tab:virial_extrapolation}
\end{table}

In practice, the MW halo's concentration is not known, and one commonly adopts a $c_{200}-M_{200}$ relation calibrated on
cosmological simulations. We therefore repeat the exercise replacing the true
$c_{200}$ by the $c_{200}-M_{200}$ relation of \citet{Dutton2014}, obtained for NFW haloes at $z=0$ in a Planck cosmology,
\begin{equation}
    \log_{10}c_{200} = 0.905 - 0.101
    \log_{10}\!\left(\frac{M_{200}}{10^{12}h^{-1}\Msun}\right),
    \label{eq:cM}
\end{equation}
with a lognormal scatter of $0.11$\,dex. In this case, Eqs.~\eqref{eq:M200}, \eqref{eq:hernquist_norm} and \eqref{eq:cM} are solved simultaneously for $(R_{200}, M_{200}, c_{200}, M_{\rm h})$. For a trial $R_{200}$ we obtain $M_{200}$ from Eq.~\eqref{eq:M200}, $c_{200}$ from Eq.~\eqref{eq:cM}, $a=2R_{200}/c_{200}$, the halo normalization from Eq.~\eqref{eq:hernquist_norm}, and we require that the halo so
normalized, plus $M_{\rm baryon}$, enclose exactly $M_{200}$ within $R_{200}$. This $M_{200}$ inference is shown in the bottom panels of Fig.~\ref{fig:extrapolate_virial}, with the shaded bands showing the solution for $c_{200}$ displaced by $\pm1\sigma$ of the relation.

The inferred $M_{200}$ now degrades substantially: even the best mass estimate, which recovers $M(r<120\,\kpc)$ to $0.2\%$ with our method, implies $M_{200}$ with an error of $+8.2\%$, and an error $+36.9\%$ anchoring it at $M(r<60\,\kpc)$ (Table~\ref{tab:virial_extrapolation}, Column 5). It is worth noting that the relation's $\pm1\sigma$ band does not cover the true $M_{200}$ in any of the cases considered, although our method anchored at $120$\,kpc misses it by only $\sim1\%$. Comparing Columns 4 and 5 shows that this large error in $M_{200}$ is dominated not by errors in the mass estimates, but by ignoring the true concentration. Our true $c_{200}=11.03$ is $1.2\sigma$ above typical halos in cosmological simulations, i.e. Eq.~\eqref{eq:cM} at its true mass, implying $c_{200}=8.22$. Although this discrepancy is purely determined by the initial conditions of our simulations\footnote{Note that this discrepancy is not an artifact of the Hernquist parametrization, since the concentration $c_{200}\equiv R_{200}/r_{-2}$ does not depend on that parametrization.}, it illustrates a significant difficulty in inferring $M_{200}$ for the MW given our ignorance about its $c_{200}$. Unfortunately, it is not currently possible to directly measure $c_{200}$ for the MW at the required precision, with published values spanning $c_{200}\sim9$--$18$, with spread dominated by systematics rather than by statistical uncertainty \citep[e.g.][]{Cautun2020, Ou2024}. Columns 4 and 5 of Table~\ref{tab:virial_extrapolation} therefore bracket a realistic situation.

In summary, the accuracy with which the MW mass is measured within
$60$--$120$\,kpc does not translate into its virial mass accuracy. The extrapolation is dominated by an assumption on the concentration of the MW halo, and a MW that happens to be even mildly atypical
for its mass -- as our own simulated halo is -- will have $M_{200}$ in error by tens of per cent, in a way that no improvement of the kinematic estimator can remedy.

This calls for large stellar samples with full 6D information at $r\gtrsim120$\,kpc. At that distance, halo K giants observed by DESI-MWS \citep{Cooper:2023} sit at the survey faint limit, $G\simeq19$, where
\textit{Gaia} DR5 is expected to deliver tangential velocities accurate
to\footnote{From the predicted end-of-mission parallax error at $G=19$ for
\textit{Gaia} DR5, $\sigma_\varpi=149\,\mu$as, and the relation
$\sigma_\mu=0.27\,\sigma_\varpi$ quoted for DR5 at
\url{https://www.cosmos.esa.int/web/gaia/science-performance}.}
$\simeq23$\,km\,s$^{-1}$, well below the $\simeq85$\,km\,s$^{-1}$
tangential velocity dispersion of the outer halo
\citep[][]{Li2026}. Larger samples of such giants will come from DESI Run~2 and, in the South, from the 4MOST halo survey, which targets $\sim1.5$ million halo giants over $\gtrsim10^4$ deg$^2$ down to $G\simeq20$ \citep{Helmi2019}, about a magnitude fainter than the DESI-MWS main survey. At $120$\,kpc, BHBs and RR~Lyrae lie at $G\simeq21$, beyond the \textit{Gaia} astrometric limit and beyond current spectroscopic reach: the Rubin Observatory LSST \citep{Ivezic2019} will discover them and provide distances, while proper motions will require a multi-epoch \textit{Roman} survey \citep[such as proposed by][]{Han2026}, with joint \textit{Gaia}--\textit{Roman} astrometric solutions forecast to reach $\simeq0.1$\,mas\,yr$^{-1}$ at $G\simeq20-21$, i.e. $\simeq40-65$\,km\,s$^{-1}$ per star at $120$\,kpc, again below the halo dispersion.

\subsection{Advantages and limitations of the Method}
We showed that octants 7\&8 are expected to be nearly in equilibrium today because the LMC crossed this region of the sky sufficiently long ago. As such, octants 7\&8 provide excellent estimates of the \emph{total} MW+LMC mass. Fortunately, octants 7\&8 are the regions of the sky least contaminated by Sgr stream stars and the LMC itself (see Fig.~\ref{fig:DESI_footprint}). The removal of Sgr stream stars can significantly impact the kinematic studies and mass estimates in other regions. Further, octants 7\&8 are in the Northern Galactic hemisphere, 
which is better covered by current spectroscopic surveys than the South. Indeed we showed that the DESI footprint does not significantly impact the mass estimates based on octants 7\&8 (Sec.~\ref{sec:desi_footprint}). Thus, octants 7\&8 offer the best observational window to uncover the total MW+LMC mass.

Further, we showed that $M_{v_r}$, an estimator only based on the radial velocity, provides errors $\lesssim 5\%$ for all radial range $30 < r/\kpc < 200$. Such accuracy can only be achieved using the $M_v$ estimator if the stellar velocities are corrected for the reflex motion, which is a potential source of errors avoided by using $M_{v_r}$. 

By demonstrating the degree of disequilibrium in the stellar halo as a function of octant, our proposed methodology also sheds light on the meaning of the mass estimate recovered in different regions of the sky. This framework explains why stars in octants 5\&6 are biased tracers of the MW mass pre-LMC infall, as pointed out by \citetalias{Erkal2020}. Because octants 5\&6 lie in the future orbital path of the LMC, they have only just started to respond to the presence of the LMC. 

\citetalias{Erkal2020} propose to recover the MW mass pre-LMC infall using octants 5\&6 and correcting for their reflex motion. In this work we have identified a different pathway to recover the MW mass pre-LMC infall that exploits the reflex motion, rather than adding uncertainty by correcting for it.  We propose to first estimate the LMC mass by subtracting the $M_v$ estimates for octant 5 (in the future path) from that of octant 4 (in the South and in the past LMC trajectory). Then this LMC mass is subtracted from the MW+LMC mass recovered using $M_{v_r}$. This method provides slightly better accuracy when extrapolated to the virial radius (Table~\ref{tab:virial_extrapolation}, Column 3), with the advantage of avoiding any reflex-motion correction. Furthermore, our method delivers not only the MW mass pre-LMC infall, but also the LMC mass.

A caveat to the use of octants 5\&6 to constrain the MW mass is that they are heavily contaminated by the Sgr stream. While this affects both \citetalias{Erkal2020}'s and our method, ours can take advantage of independent estimates of the LMC mass \citep[e.g. from perturbations on the Orphan stream, or from LMC/SMC binarity, or from the kinematics of its globular cluster system; see][]{Erkal2019, Koposov2023, Besla2012, Watkins2024}, which cannot be done in \citetalias{Erkal2020}'s approach. Regardless, efforts to estimate the MW mass pre-LMC infall would strongly benefit from improvements on the identification and removal of Sgr members.

Finally, an intrinsic limitation of our method is that \citetalias{Watkins2010}'s estimators require the assumption of the total mass slope $\alpha$. Although we can make reasonable simulation-based choices, and results seem compatible for different choices within uncertainties (Sec.~\ref{sec:desi}), it would be desirable to expand our analysis to mass estimator methods that waive that assumption. 


\section{Summary and Conclusions}
\label{sec:conclusions}
In this paper we investigate the meaning of MW mass estimates in the presence of a massive LMC. As a key result, we show that stars in regions traversed long ago and perturbed nearly adiabatically by the LMC will trace the total MW+LMC mass. By contrast, stars in regions not yet crossed by the LMC path carry the most information on the MW mass pre-LMC infall, in agreement with \citetalias{Erkal2020}, who found stars at $\ell <0^\circ$ and $b>0^\circ$ to provide the least biased MW mass estimates (our octants 5\&6).

The response of the stellar halo to the orbital path of the LMC allows us to establish a strategy to recover the present-day combined MW+LMC mass profile and disentangle the LMC and MW mass profiles by exploiting the velocity perturbations imparted on stars in the outer stellar halo owing to the recent infall of the LMC. Critically, this method requires limited model assumptions to recover the MW mass profile pre-LMC infall, which is needed to identify appropriate cosmological mass-analogs of the MW in order to test CDM theory against the MW observables.  

Our strategy uses two mass estimators proposed by \citetalias{Watkins2010}, Eqs.~\eqref{eq:mass_est_v}-\eqref{eq:mass_est_vr}, which we apply to dark matter particles extracted from eight $N$-body simulations differing in the initial MW halo velocity anisotropy and LMC mass \citep{Garavito-Camargo2019, Garavito-Camargo2021}. 

Our conclusions are summarized below:

\begin{itemize}
\item For the MW+LMC simulations, we identify density profiles steeper than those of the isolated, MW pre-LMC infall models between $50\lesssim r/\kpc \lesssim 150$ (and shallower at larger radii), due to the MW halo response to the LMC (Fig. \ref{fig:alpha_beta_gamma_all}).

\item With the simulation all-sky data, we use the $M_v$ estimator, Eq.~\eqref{eq:mass_est_v}, which is based on the 3D velocity $v$, to recover the total MW+LMC mass with $\lesssim 5\%$ ($\lesssim 10\%$) errors for $r/\kpc \lesssim 120$ ($120 \lesssim r/\kpc\lesssim 200$) -- see Fig.~\ref{fig:M_total_all_v}. This contrasts with \citetalias{Erkal2020}, who obtained $M_v$ estimates higher by up to $20\%$ with respect to the total MW+LMC at large radii in their simulations (their Fig. A1). 

\item Due to a partial cancellation of the radial component of the reflex motion $\langle v_r \rangle$ of Northern and Southern galactic hemispheres (Fig.~\ref{fig:mean_vr_vtheta_vphi_octs_MWLMC5_b0}), the all-sky $\langle v_r \rangle$ reaches at most $\approx -12\kms$. This is significantly smaller than $\langle v_\theta \rangle$, which can reach $\approx -35 \kms$ (Fig.~\ref{fig:mean_v_all}) due to the LMC pull in its nearly polar orbit. Fortuitously, this allows the $M_{v_r}$ mass estimator, Eq.~\eqref{eq:mass_est_vr}, which is based on $v_r$ only, to recover the total MW+LMC with relative errors $\lesssim 5\%$ for all radii (Fig.~\ref{fig:M_total_all_vr}).

\item When corrected for the reflex motion as proposed by \citetalias{Erkal2020}, the $M_v$ estimator also recovers the total MW+LMC mass with errors $\lesssim 5\%$ for all radii using all-sky data (Fig.~\ref{fig:M_total_all_v_reflex}). This confirms \citetalias{Erkal2020}'s conclusion that biased mass estimates mostly result from the reflex motion, although they interpret the bias with respect to the MW mass pre-LMC infall, while we interpret this bias as being with respect to the total MW+LMC mass.

\item The reflex-motion corrected $M_v$ estimates are as accurate as those using $M_{v_r}$ with no correction. This means that the total MW+LMC mass can be recovered (with error-free data) to within $\lesssim 5\%$ errors without introducing a reflex-motion correction that would add significant uncertainty.

\item We confirm the findings of \citetalias{Garavito-Camargo2019} and \citetalias{Erkal2020} that some regions of the halo are more perturbed by the LMC than others. We find this degree of perturbation follows the orbital path of the LMC. Splitting the sky in octants based on Galactic longitude and latitude ($\ell, b$), the least perturbed regions correspond to octants 5\&6 ($\ell <0^\circ$ and $b>0^\circ$), where the LMC has not yet traveled, and octants 7\&8 ($\ell >0^\circ$ and $b>0^\circ$), which the LMC traversed sufficiently long ago that the system has re-equilibrated by present day. 
  
\item We estimate the MW mass with data for different octants. We show that octants 7\&8, due to their relatively small $\langle v_r \rangle$ reflex motion (Fig.~\ref{fig:mean_vr_vtheta_vphi_octs_MWLMC5_b0}), provide excellent $M_{v_r}$ estimates of the total MW+LMC mass. In the fiducial simulation, the total MW+LMC $M_{v_r}$ estimate has $\lesssim 5\%$ errors for all radii. In models with a more massive LMC, this error can reach $\lesssim 7\%$ for $r\gtrsim 120$ kpc.

\item We confirm the findings of \citetalias{Erkal2020} that octants 5\&6 provide the closest recovery of the pre-LMC infall MW mass. However, because the LMC is just now approaching octants 5\&6, these octants will return masses biased high relative to the MW mass pre-LMC infall and low relative to the MW+LMC mass. In contrast to \citetalias{Erkal2020}, we find that this bias remains at the $\sim$10\% level with respect to the MW mass pre-LMC infall, even after the reflex motion of the octants are corrected.

\item By exploiting the fact that the reflex motion contains information about the LMC mass, we demonstrate that the LMC mass can be recovered. We show that subtracting the $M_v$ estimate based on octant 5 (in the Northern hemisphere and in LMC's future orbit) from the one based on octant 4 (in the LMC's past orbit and the opposite location in ($l,b$), maximizing the reflex motion), we recover the LMC mass in all simulations within $\approx 20\%$-errors for tracers at $r\gtrsim 100$ kpc.
  
\item Subtracting the recovered LMC mass profile from the total MW+LMC mass profile estimated with $M_{v_r}$ (octants 7\& 8, or all-sky data), we recover the MW enclosed mass pre-LMC infall with $\approx 5-10\%$ accuracy up to $\approx 200\kpc$.

\item The method relies on prior knowledge of the slope of the MW mass profile, $\alpha$. Fixing $\alpha=1$ and taking tracers at
  $r\gtrsim 80\kpc$ still recovers the MW mass pre-LMC
  infall with $\approx 5-10\%$ accuracy.

\item Even with error-free data, extrapolation from precise mass estimates at $r=60$ kpc to $M_{200}$ can be wrong by several tens of percent due to our ignorance of the true concentration of the MW DM halo. Anchoring this extrapolation at 120 kpc can improve the $M_{200}$ estimates to within $\sim 10\%$.

\item Masking the simulation data with the DESI footprint, we still recover the total MW+LMC mass profile with the $M_{v_r}$ estimator for tracers in octants 7\&8, with errors $\lesssim 5\%$ for $r<100$ kpc and $\lesssim 10\%$ further out (Fig.~\ref{fig:DESI_footprint_M_octs_MWLMC5_b0}). This indicates that our method works even with limited coverage.

\item Besides tracing the MW+LMC mass, octants 7\&8 are also the least contaminated by the Sgr stream (Fig.~\ref{fig:DESI_footprint}), thus offering the best observational window to constrain the total MW+LMC mass.

\item Using the sample of BHB stars from DESI-MWS DR1 \citep[][]{Bystrom2025} in $50 < r/\kpc < 60$ and octants 7\&8, and after removing the Sgr stream, we estimate the total MW+LMC mass $M_{v_r}(r<60\, \kpc) = 7.58^{+2.90}_{-2.17}\times10^{11}\,\Msun$
($7.49^{+2.97}_{-2.25}\times10^{11}\,\Msun$) assuming a total mass slope $\alpha=1$ ($\alpha=0.5$). The DESI-MWS does not have sufficient numbers of stars at 100 kpc, and we do not test the proposed LMC mass and MW mass pre-LMC infall strategy with the DESI data. We note that other methods could recover the LMC mass profile \citep[e.g. from perturbation on the Orphan stream --][]{Erkal2019, Koposov2023}, which could then be subtracted from the total MW+LMC to recover the pre-LMC infall MW mass with DESI data. Precise proper motions from \emph{Gaia} DR5 and large spectroscopic samples of K giants at distances $>100 \kpc$ from DESI Run 2 and 4MOST will enable tests of our proposed method to recover the LMC and pre-LMC infall MW mass profiles.

\item The proposed method to recover the combined MW+LMC mass, LMC mass and pre-LMC Infall MW mass is agnostic to the orbital history of the LMC. The orbital path of the LMC in the Southern Hemisphere is dictated by the proper motion measurements. In the north, even in a second passage scenario octants 7\&8 should have sufficient time to achieve equilibrium. Although the LMC would pass through octants 5\&6 in a second passage scenario, the LMC would also travel to a large apocenter distance \citep[$\sim$300 kpc; see][]{Vasiliev2024}. Consequently mass estimates recovered within 200 kpc in octants 5\&6 should be similar to a first infall scenario. As such, we expect the recovered mass in octants 5\&6 to be lower than that in octants 7\&8, but higher than the true pre-LMC Infall MW mass (see Fig \ref{fig:timeline_Mv}). Using BHB stars from DESI-MWS DR1, we tentatively find that the mass recovered in octants 5\&6 is indeed lower than that in octants 7\&8. However, Sgr debris in octants 5\&6 strongly impacts this analysis. 

\item The DESI-MWS just recently published the results of its DR2 Key Project on measuring the MW mass profile with different stellar tracers \citep[][]{DESI_DR2KP_2026, Zhou2026, Medina2026, Riley2026}. These and other MW mass modeling methods would benefit from tests with MW+LMC simulations and from exploring the methodology proposed in this work.
\end{itemize}

Rather than reaffirming the disequilibrium produced by the LMC as an insurmountable problem, our results show that it is possible to take advantage of it to constrain the \emph{total} MW+LMC mass, the LMC mass, and then the MW mass pre-LMC infall. By making the meaning of these mass estimates clearer, we hope to shed light on how to estimate the MW virial mass, contributing to answering the century-long question of the nature and amount of dark matter in our Galaxy.

\section*{Acknowledgments}
LBeS acknowledges support from CNPq (Brazil) through a research productivity fellowship, grant no. [304873/2025-0]. MV \& LBeS acknowledge the support of NASA ATP award 80NSSC20K0509 and U.S. National Science Foundation AAG grant AST-2009122, and MV acknowledges support from NASA ATP award 80NSSC24K0938. (309723/2020-5). KJD \& LBeS acknowledge support from the Heising Simons Foundation grant \# 2022-3927. GB acknowledges support through NASA ATP award 80NSSC24K1225. The Theoretical Astrophysics Program (TAP) Dynamics Initiative at the University of Arizona provided resources to support this work. SK acknowledges support from the Science
\& Technology Facilities Council (STFC) grant ST/Y001001/1. CL acknowledges funding from the European Research Council (ERC) under the European Union’s Horizon 2020 research and innovation programme (grant agreement No. 852839) as well as funding from the Agence Nationale de la Recherche (ANR project ANR-24-CPJ1-0160-01). The authors acknowledge the use of Claude (Anthropic) and ChatGPT (OpenAI) for language editing, coding assistance, and bibliographic search during manuscript preparation; all content was reviewed and verified by the authors.
We respectfully acknowledge the University of Arizona is on the land and territories of Indigenous peoples. Today, Arizona is home to 22 federally recognized tribes, with Tucson being home to the O’odham and the Yaqui. 

\software{numpy \citep{numpy},
  scipy \citep{scipy}, matplotlib \citep{matplotlib},
  astropy.}


\appendix
\restartappendixnumbering

\section{Volume of intersection of radial shell, octants and DESI footprint}
\label{sec:geom_los}

\begin{figure}
\centering
\includegraphics[width=0.9\columnwidth]{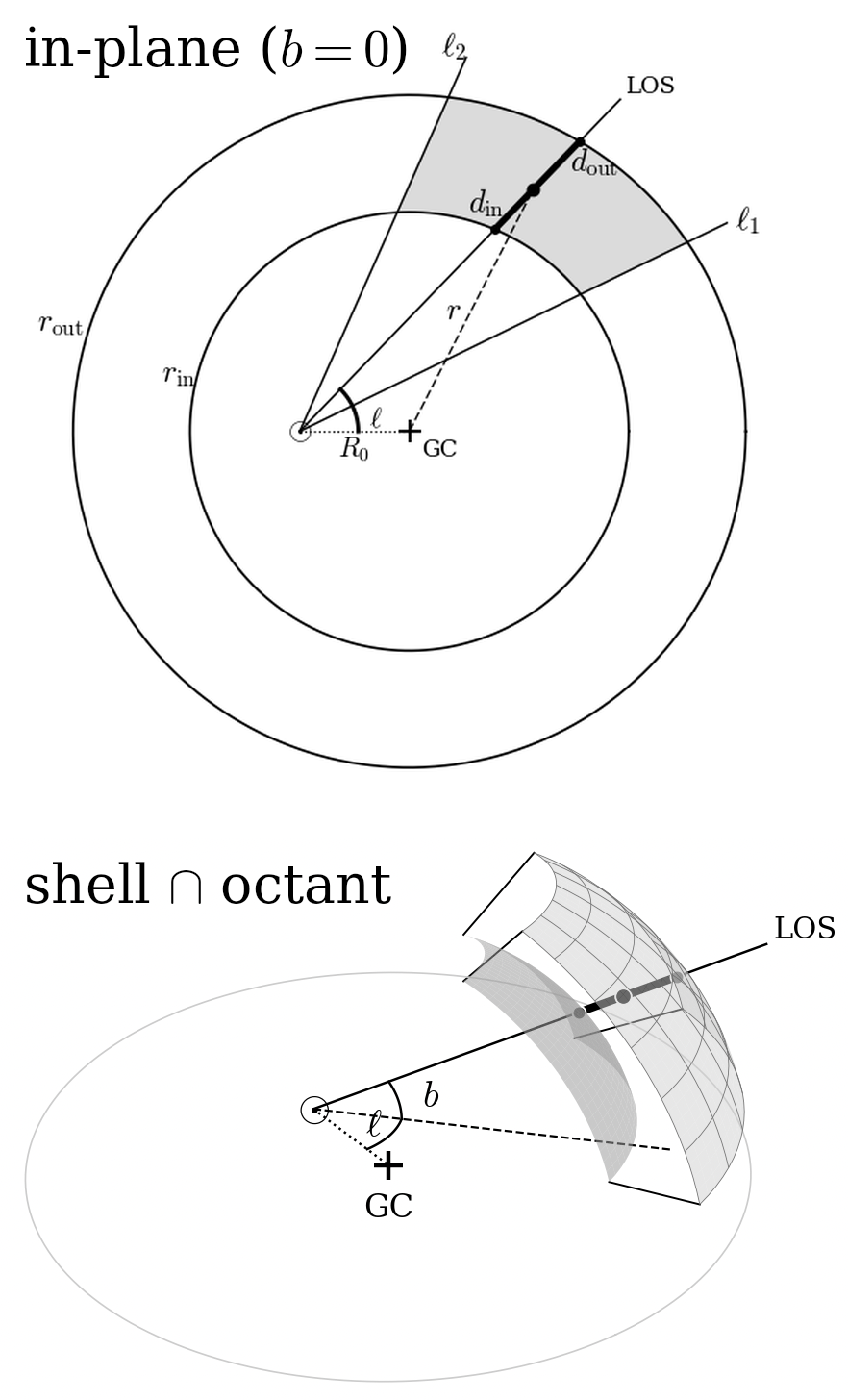}
\caption{Intersection of a radial shell with an octant (the region between the two spherical patches in the bottom panel), whose volume is required for the density calculation. The Sun is indicated by $\odot$, and GC is the Galactic center.}
\label{fig:geom_los}
\end{figure}

The estimate of the density slope $\gamma(r)$ in a given radial shell and in a given octant requires the proper calculation of the volume of that region, as illustrated in Fig.~\ref{fig:geom_los}.
For a star with heliocentric distance $d$ and galactic coordinates $(\ell,b)$, its galactocentric distance is $r^2(d;\ell,b)=d^2+R_0^2-2R_0 d \cos\psi$, where ${\cos\psi\equiv\cos b \cos\ell}$. The distance limits $d_\mathrm{in}$, $d_\mathrm{out}$
are the positive roots of ${d^2-2R_0d\cos\psi+(R_0^2-r^2)=0}$:
\begin{equation}
    d_\mathrm{in, out}(r)=R_0\cos\psi+\sqrt{\,r_\mathrm{in, out}^2-R_0^2\sin^2\psi\,},
    \label{eq:d_in_out}
\end{equation} 
with $\sin\psi = \sqrt{\sin^2 b + \cos^2 b\,\sin^2\ell}$. The radial part of the volume element $\mathrm{d}V=s^2 \mathrm{d}s \mathrm{d}\Omega$, where $s$ runs along the line-of-sight, integrates to 
\begin{equation}
I(\ell,b)\equiv\int_{d_{\rm in}}^{d_{\rm out}} s^2 \mathrm{d}s=\frac{d_{\rm out}^3-d_{\rm in}^3}{3}.
\label{eq:I_l_b}
\end{equation}
The volume of the intersection of the octant with each radial shell is then
\begin{equation*}
    V_{\rm shell \cap oct}=\int_{\ell_1}^{\ell_2}\!\!\int_{\mu_1}^{\mu_2} I\big(\ell,b(\mu)\big) \mathrm{d}\mu \mathrm{d}\ell,
\end{equation*}
where $\mu =\sin b$ and $[\ell_1,\ell_2)\times[b_1,b_2)$ defines each octant. This last integral is computed by a midpoint Riemann sum on a uniform
$n_\ell\times n_b=96\times96$ grid in $(\ell,\mu)$. We checked that the sum of these volumes over all octants equates the volume of the spherical shell. Furthermore, for the MW pre-LMC models, the density slopes calculated in each octant are identical to those based on the whole sky, as we would expect since these models are effectively spherically symmetric for $r>30 \kpc$.

\subsection{Intersection with DESI footprint}
\label{sec:vol_desi}
In this work we use the DESI-MWS footprint as provided by \cite{Bystrom2025}. This footprint is supplied as a HEALPix mask with $\text{NSIDE}=32$, such that the total number of equal-area pixels is $N_{\rm pix}=12\,\text{NSIDE}^2=12288$, each of solid angle $\Omega_{\rm pix}=4\pi/N_{\rm pix}$.

The intersecting volume of the footprint with each radial shell is calculated as 
\begin{equation}
    V_{\rm fp \cap shell}=\Omega_{\rm pix}\!\!\sum_{p\,\in\,\rm footprint}\!\! I(\ell_p,b_p),
    \label{eq:V_fp_oct}
\end{equation}
where $(\ell_p,b_p)$ is the center of each pixel in the footprint, and $I(\ell,b)$ is given by Eqs.~\eqref{eq:d_in_out}-\eqref{eq:I_l_b}. We checked that the fraction over the volume of a spherical shell, $V_{\rm fp \cap shell}/V_\mathrm{shell}$, varies from $\approx 0.33$ at $30 \kpc$ to $\approx 0.36$ at $200 \kpc$, approaching the sky fraction $\approx 0.363$ covered by DESI. This $r$-dependence is due to the footprint being an angular mask as seen from the Sun, while the shells are galactocentric.

The intersecting volume of the footprint with each radial shell and each octant is calculated similarly to Eq.~\eqref{eq:V_fp_oct}, but restricting to pixels within the octant:
\begin{equation*}
V_{\rm fp \cap shell\cap oct}=\Omega_{\rm pix}\!\!\sum_{\substack{p\,\in\,\rm footprint\\ (\ell_p,b_p)\,\in\,\rm oct}}\!\! I(\ell_p,b_p).
\end{equation*}

\subsection{Excluding the Sagittarius stream region}
\label{sec:sgr_volume}

As described in Sec.~\ref{sec:BHB_sample}, we remove likely Sagittarius (Sgr) stream members from the BHB sample following \cite{Bystrom2025}: a star is excluded if $|\tilde B|<15^\circ$ and its heliocentric distance lies within $3\sigma(\tilde\Lambda)$ of the mean Sgr distance track $\bar d(\tilde\Lambda)$, where $(\tilde\Lambda,\tilde B)$ are the Sgr stream coordinates \citep{Belokurov2014} and $\bar d(\tilde\Lambda)$, $\sigma(\tilde\Lambda)$ are the mean heliocentric distance and (Gaussian) line-of-sight depth of the stream from \cite{Hernitschek2017}. Since this cut removes a region of space -- all stars in that $(\tilde\Lambda,\tilde B,d)$ tube, not only genuine members -- the same region must be removed from the survey volume, so that the tracer density remains consistent.

Along each line of sight the Sgr cut amounts to the removal of a
heliocentric-distance interval. For a sightline $(\ell,b)$ we transform its direction into Sgr coordinates $(\tilde\Lambda,\tilde B)$ and define the excluded interval
\begin{equation}
  W(\ell,b)=
  \begin{cases}
    \big[\,\bar d(\tilde\Lambda)-3\sigma(\tilde\Lambda),\;
            \bar d(\tilde\Lambda)+3\sigma(\tilde\Lambda)\,\big],
      & |\tilde B|<15^\circ,\\[3pt]
    \varnothing, & \text{otherwise,}
  \end{cases}
  \label{eq:sgr_window}
\end{equation}
taken as the union of the leading- and trailing-arm intervals in the longitude range where both tracks are defined. The radial integral $I(\ell,b)$ of Eq.~\eqref{eq:I_l_b} is then replaced by its restriction to the part of the line-of-sight segment that survives the cut,
\begin{equation}
  \tilde I(\ell,b)\equiv\!\!\int_{[d_{\rm in},d_{\rm out}]\setminus W(\ell,b)}\!\!\!\!
  s^2\,\mathrm{d}s=\sum_k \frac{b_k^3-a_k^3}{3},
  \label{eq:I_tilde}
\end{equation}
where $[a_k,b_k]$ are the disjoint sub-intervals of $[d_{\rm in},d_{\rm out}]$ that remain after removing $W(\ell,b)$. The Sgr-corrected volumes follow by replacing $I(\ell_p,b_p)$ with $\tilde I(\ell_p,b_p)$ in Eq.~\eqref{eq:V_fp_oct} and in the octant expression above, e.g.
\begin{equation*}
  \tilde V_{\rm fp \cap shell\cap oct}=\Omega_{\rm pix}\!\!
  \sum_{\substack{p\,\in\,\rm footprint\\ (\ell_p,b_p)\,\in\,\rm oct}}\!\!
  \tilde I(\ell_p,b_p).
\end{equation*}
In the $50\leq r/\kpc\leq60$ shell used for our analysis, the removed volume
fraction is $\approx11\%$ for the full footprint, $\approx4\%$ for octants 7 \& 8, and $\approx31\%$ for octants 5 \& 6, the latter reflecting that the stream passes directly through those Northern octants.

\section{Effects of the DESI footprint on tracers' distribution and kinematics}
\label{sec:desi_foot_print_compare_sims}

\begin{figure}
\centering
\includegraphics[width=0.9\columnwidth]{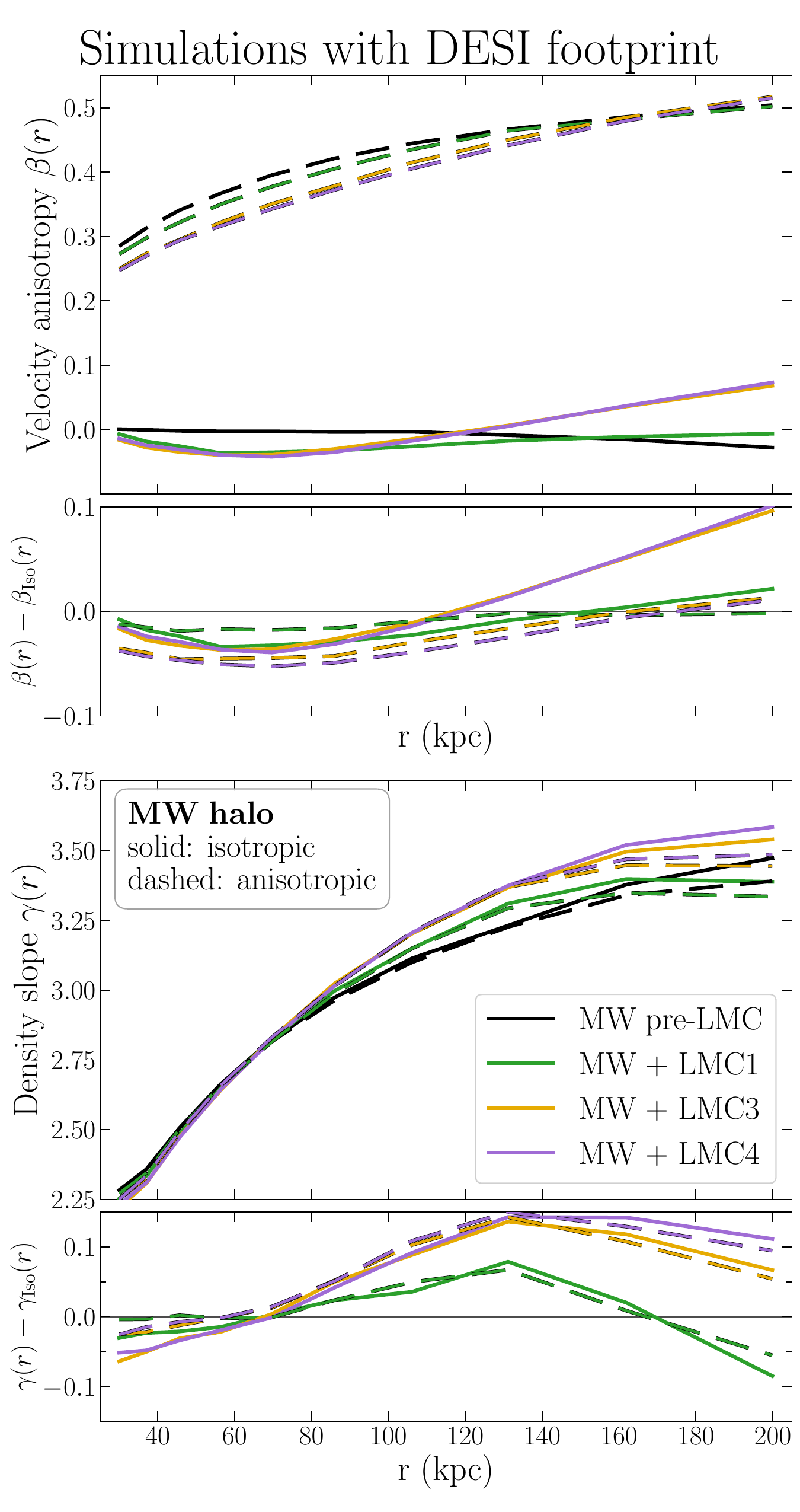}
\caption{Similar to Fig.~\ref{fig:alpha_beta_gamma_all}, but only for simulations MW DM particles in the DESI footprint. This introduces differences of up to $\approx 0.05$ in the velocity anisotropy and up to $\approx 0.1$ in the density slope. The footprint makes it harder to detect the larger steepness of the density profiles associated with the DM wake trailing the LMC in $50 \leq r/\kpc \leq 100$.}
\label{fig:DESI_footprint_alpha_beta_gamma_all}
\end{figure}

Similarly to Fig.~\ref{fig:alpha_beta_gamma_all}, in Fig.~\ref{fig:DESI_footprint_alpha_beta_gamma_all} we show the velocity anisotropy (top panel) and density slope (bottom panel) of DM particles for all simulations in the same radial bins used before, masked by the DESI footprint. For the MW pre-LMC Infall models, the density slopes are nearly identical to the whole-sky estimates in Fig.~\ref{fig:alpha_beta_gamma_all}, as expected since these models are nearly spherically symmetric at $r>30$ kpc. 

For all MW+LMC models, both $\beta(r)$ and $\gamma(r)$ show small differences with respect to the whole-sky estimates in Fig.~\ref{fig:alpha_beta_gamma_all}, due to the DESI footprint. The velocity anisotropy decreases by $\approx 0.05$ for $r\lesssim 100$ kpc, and increases by a similar amount for $r\gtrsim 100$ kpc with respect to the all-sky estimates. Similarly, the density slopes decrease by up to 0.1 in $50\lesssim r/\kpc \lesssim 100$ and can increase by the same amount for $r\gtrsim 100$ kpc. This is because the DESI footprint does not cover octant 2, and only barely covers octant 1, regions where simulations predict most of the DM wake to reside (Fig.~\ref{fig:DESI_footprint}). Thus, the DESI footprint makes it harder to detect the steep density slope associated with the DM wake trailing the LMC in $50\lesssim r/\kpc \lesssim 100$.

\begin{figure}
\centering
\includegraphics[width=0.9\columnwidth]{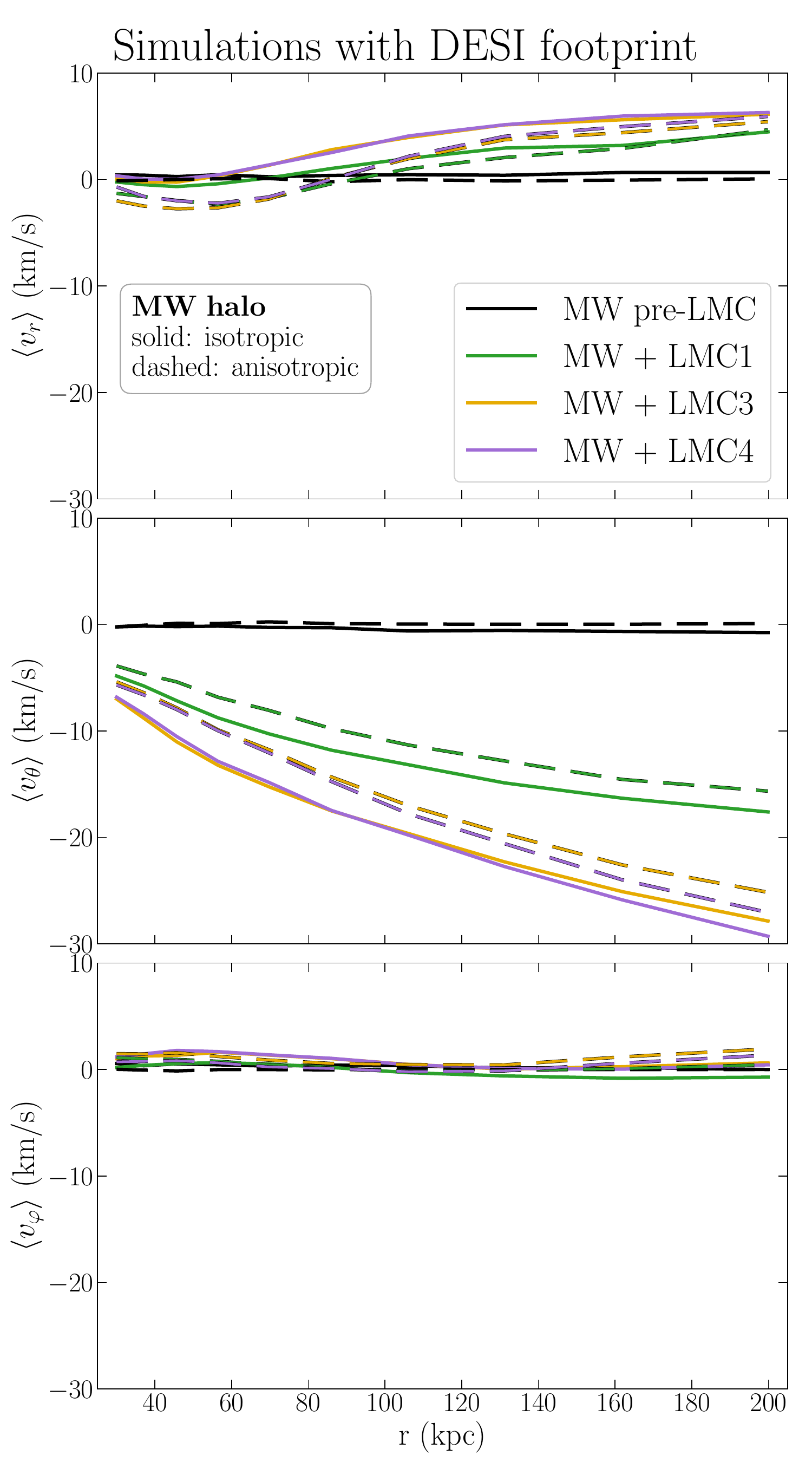}
\caption{Since $\langle v_\theta \rangle$ and $\langle v_\varphi \rangle$ are similar for all octants, these are not significantly affected by the DESI-MWS footprint, which preferentially selects on Northen octants. For $\langle v_r \rangle$, which varies strongly across octants, the footprint delivers different reflex motions, and we can see that $\langle v_r \rangle$ is a mixture of values for the Northen octants (red lines in Fig.~\ref{fig:mean_vr_vtheta_vphi_octs_MWLMC5_b0}).}
\label{fig:DESI_footprint_mean_vr_vtheta_vphi_all}
\end{figure}

Fig.~\ref{fig:DESI_footprint_mean_vr_vtheta_vphi_all} shows the three components of the reflex motion for all simulations, similarly to Fig.~\ref{fig:mean_v_all}, but now only for MW DM particles within the DESI-MWS footprint. The only significant differences with respect to Fig.~\ref{fig:mean_v_all} are for $\langle v_r \rangle$, which is now closer to zero, and varies between negative values for $r\lesssim 80$ kpc and positive values further out. This is because the DESI-MWS footprint mostly covers the Northern hemisphere, and $\langle v_\theta \rangle$ and $\langle v_\varphi\rangle$ are similar for all octants, while $\langle v_r\rangle$ strongly depends on the octant (Fig.~\ref{fig:mean_v_all}). Masked by the DESI footprint, $\langle v_r\rangle$ seems a mixture of the $\langle v_r\rangle$ we have seen for the Northern octants in Fig.~\ref{fig:mean_v_all}. 

\begin{figure}
\centering
\includegraphics[width=0.9\columnwidth]{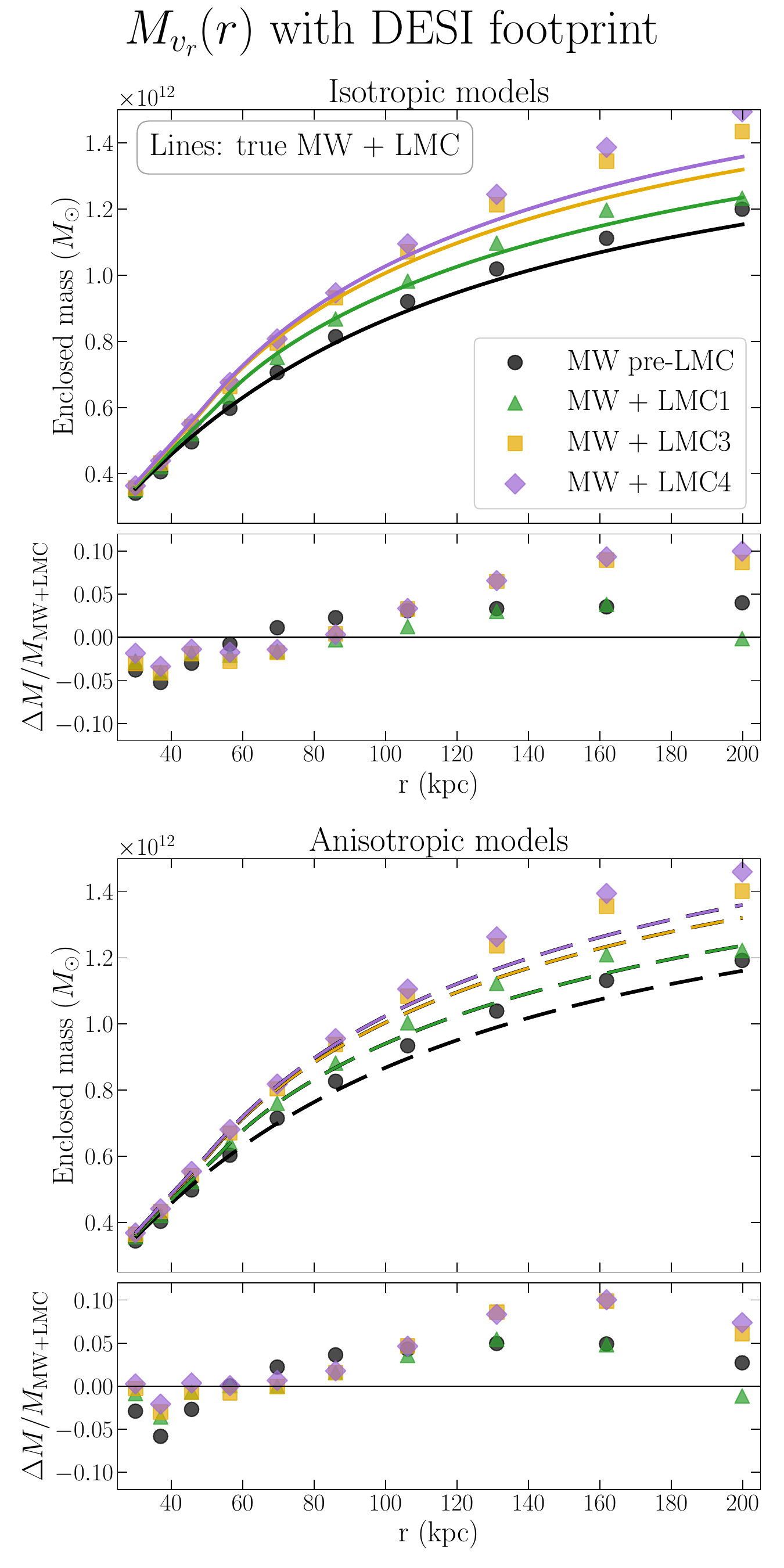}
\caption{Similar to Fig.~\ref{fig:M_total_all_vr}, but with simulation data masked by the DESI footprint. Due to the small reflex motion in the radial direction (Fig.~\ref{fig:DESI_footprint_mean_vr_vtheta_vphi_all}), the true total MW+LMC mass is recovered within $\lesssim 10\%$ for all simulations out to $200\kpc$.}
\label{fig:DESI_footprint_M_total_all_vr}
\end{figure}

Fig.~\ref{fig:DESI_footprint_M_total_all_vr} shows the $M_{v_r}$ mass estimates for all simulations masked by the DESI footprint. Similarly to the unmasked case (Fig.~\ref{fig:M_total_all_vr}), due to the small reflex motion in the radial direction (averaged over the whole footprint), the total MW+LMC mass is recovered to within $\lesssim 10\%$ errors out to $200\kpc$ for all simulations.

\bibliography{refs}{}
\bibliographystyle{aasjournal}

\end{document}